\documentclass[a4paper]{article}

\usepackage{endnotes}

\usepackage{setspace} 
\usepackage{wrapfig}
\usepackage{enumitem}
\usepackage{xr}
\usepackage{color}
\usepackage{amsmath,amsfonts,amssymb}
\usepackage{epsfig}
\usepackage{graphicx}
\usepackage{colortbl}
\usepackage{url}

\definecolor{grey}{rgb}{0.9, 0.9, 0.9}

\newcommand{\myparagraph}[1]{\vspace{-3mm}\paragraph{#1}}

\newcommand{\coout}[1]{[{\color{magenta} #1}]}
\newcommand{\co}   [1]{[{\it \color{red} #1}]}
\newcommand{\todo} [1]{[{\color{blue} #1}]}

\usepackage{amsmath,amsfonts,amssymb}
\usepackage{relsize}
\usepackage{ marvosym }

\newcommand{\sign}     {\mbox{sign}}

\newcommand{\const}{\mbox{\rm const.}}

\newcommand{\inv}{^{-1}}

\newcommand{\cov}{\mbox{\rm cov}}

\newcommand{\md}{{\rm d}}

\newcommand{\diag}{\mbox{\rm Dg}}

\newcommand{\e}   {\mbox{\rm e}}

\newcommand{\cv}{{\bf c}}
\newcommand{\vv}{{\bf v}}
\newcommand{\yv}{{\bf y}}

\newcommand{\kv}{{\bf k}}

\newcommand{\ev}{{\bf e}}
\newcommand{\xv}{{\bf x}}

\newcommand{\pv}{{\bf p}}

\newcommand{\sv}{{\bf s}}

\newcommand{\av}{{\bf a}}

\newcommand{\thetav}{{\boldsymbol \theta}}

\newcommand{\muv}{{\boldsymbol \mu}}

\newcommand{\Nmat}{\mathbf{N}}
\newcommand{\Imat}{\mathbf{I}}
\newcommand{\Cmat}{\mathbf{C}}
\newcommand{\Mmat}{\mathbf{M}}

\newcommand{\Emat}{\mathbf{E}}

\newcommand{\Amat} {\mathbf{A}}

\newcommand{\Bmat} {\mathbf{B}}

\newcommand{\Lmat}{\mathbf{L}}
\newcommand{\Kmat}{\mathbf{K}}
\newcommand{\Rmat}{{\bf R}}

\newcommand{\Sigmamat} {\mathbf{\Sigma}}

\newcommand{\trans}{^{\top}}

\DeclareMathAlphabet{\mathpzc}{OT1}{pzc}{m}{it}
\DeclareMathAlphabet{\mathcalligra}{T1}{calligra}{m}{n}

\providecommand{\esymbol}{e}

\newcommand{\steady}{^{\rm st}}

\newcommand{\vsteady}    {v\steady}

\newcommand{\csteady}    {c\steady}

\newcommand{\kcat}     {k_{\rm cat}}

\newcommand{\km}       {K_{\rm M}}
\newcommand{\keq}      {K_{\rm eq}}

\newcommand{\Euns}    {E}
\newcommand{\Eun}     {\mathbf{\Euns}}
\newcommand{\Eunc}    {\Eun_{\rm c}}

\newcommand{\Eunvlci} {{\Euns^{\rm v_{l}}_{{\rm c}_{i}}}} 

\newcommand{\Escs}    {\hat{E}}
\newcommand{\Esc}     {\mathbf{\Escs}}

\newcommand{\Escvlci} {\Escs^{\rm v_{l}}_{{\rm c}_{i}}} 

\newcommand{\Cvmat} {\Cmat^{\rm V}}

\newcommand{\RS}{R^{\rm S}}
\newcommand{\RJ}{R^{\rm J}}

\usepackage{etoolbox}

\newtoggle{bookversion}      \togglefalse{bookversion}
\newtoggle{shortbookversion} \togglefalse{shortbookversion}
\newtoggle{anhang}           \togglefalse{anhang}
\newtoggle{hidecomments}      \togglefalse{hidecomments}

\newcommand{\bookco}[1]{}

\definecolor{brown}{rgb}{0.9,0.69,0.34}
\definecolor{samoabrownlight}{rgb}{0.89,0.69,0.4}
\definecolor{samoabrowndark} {rgb}{0.5,0.3,0.15}
\definecolor{cbasamoabrown1}{rgb}{0.87,0.6,0.23}
\definecolor{cbasamoabrown2}{rgb}{0.87,0.6,0.23}
\definecolor{cbabrown1}{rgb}{0.87,0.6,0.23}
\definecolor{cbabrown2}{rgb}{0.87,0.6,0.23}
\definecolor{cbabrown3}{rgb}{0.87,0.6,0.23}
\definecolor{cbabrown4}{rgb}{0.87,0.6,0.23}
\definecolor{cbaecoblue1}{rgb}{0.8,0.8, 1.0}
\definecolor{cbaecoblue2}{rgb}{0.7,0.7, 1.0}
\definecolor{cbaecoblue3}{rgb}{0.87,0.6,0.23}
\definecolor{cbaecoblue4}{rgb}{0.87,0.6,0.23}
\definecolor{cbablue2}{rgb}{0.87,0.6,0.23}
\definecolor{cbapink}{rgb}{.99,0.92,0.75}
\definecolor{cbabeige1}{rgb}{0.86, 0.797, 0.625} 
\definecolor{cbabeige2}{rgb}{0.93, 0.812, 0.56}  
\definecolor{cbabeige3}{rgb}{1.0, 0.97, 0.88}  
\definecolor{cbalightgrey}{rgb}{0.95,0.95,0.95}
\definecolor{cbatablecolor1}{rgb}{0.86, 0.797, 0.625} 
\definecolor{cbatablecolor2}{rgb}{1,1,1}         

\newcommand{\myvalue}      {value}

\newcommand{\gain}         {gain}

\newcommand{{\fluxvalue}}  {flux \myvalue}
\newcommand{{\fluxgain}}   {flux \gain}

\newcommand{{\valueflow}}    {value flow}
\newcommand{{\Valueflow}}    {Value flow}

\newcommand{ {\flow}}        {flux profile}
\newcommand{ {\Flow}}        {Flux profile}

\providecommand{\esymbol}   {e}
\providecommand{\esymbolv}  {\mathbf{e}}

\providecommand{\prodrate}  {r}

\newcommand{\intprod}   {\prodrate^{\rm int}}

\newcommand{\rate}{\nu}

\newcommand{\ratelaw}{k}

\newcommand{\wsymbol}     {w}

\newcommand{\Nint}    {\Nmat^{\rm int}}
\newcommand{\Ntot}    {\underline{\Nmat}}

\newcommand{\Next}    {\Nmat^{\rm x}}
\newcommand{\NR}      {\Nmat^{\rm ind}}

\newcommand{\Deltar}{\Delta}

\newcommand{\fluxenzymecostl}  {\Delta \wsymbol_{\intprod_l:}}

\newcommand{\kM}{K_{\rm M}}

\renewcommand{\kM}{k^{\rm M}}
\renewcommand{\keq}{k^{\rm eq}}
\renewcommand{\kcat}{k^{\rm cat}}

\newcommand{\kX}{k^{\rm X}}
\newcommand{\kV}{k^{\rm V}}
\newcommand{\kA}{k^{\rm A}}
\newcommand{\kI}{k^{\rm I}}

\newcommand{\kcatpm}{k^{\rm cat}_\pm}
\newcommand{\kcatpml}{k^{\rm cat}_{\pm,l}}
\newcommand{\kcatp}{k^{\rm cat}_{+}}
\newcommand{\kcatm}{k^{\rm cat}_{-}}
\newcommand{\kcatpl}{k^{\rm cat}_{+l}}
\newcommand{\kcatml}{k^{\rm cat}_{-l}}
\newcommand{\kma}{q^{\rm ma}}
\newcommand{\DeltaR}{\Delta_{\rm r}}

\newcommand{\Nr}{{\bf N}_{\rm R}}
\renewcommand{\Next}{{\bf N}^{\rm ext}}
\renewcommand{\Nint}{{\bf N}}
\renewcommand{\Ntot}{{\bf N}^{\rm tot}}

\newcommand{\Ecun}{ {\bf E}_{\rm c}}
\newcommand{\Epun}{ {\bf E}_{\rm p}}

\newcommand{\CSun}{ {\bf C}^{\rm s}}
\newcommand{\CJun}{ {\bf C}^{\rm j}}

\newcommand{\zun}{{z}}
\newcommand{\etaun}{{\eta}}
\newcommand{\etasc}{{\eta}}

\newcommand{\Cun}{{C}}
\newcommand{\Csc}{\hat{C}}

\newcommand{\Run}{{R}}
\newcommand{\Rsc}{{R}}

\newcommand{\Ematsc}{\hat{\Emat}}

\newcommand{\RSdag}{{{\Rmat}^{\rm s}\dag}}
\newcommand{\Rmatun}{{\Rmat}}
\newcommand{\RSunp}{{\RS}_{\rm p}}
\newcommand{\RJunp}{{\RJ}_{\rm p}}

\usepackage{hyperref}
\definecolor{darkblue}{rgb}{0.05,0.,0.65}
\hypersetup{colorlinks=true,linkcolor=darkblue,citecolor=darkblue,urlcolor=darkblue}
\newcommand{\psfiles}                    {ps-files/}

\newcommand{\ecoliccmpsfiles}            {ps-files/ecoli_ccm/}
\newcommand{\hepatonetpsfiles}           {ps-files/hepatonet/}

\newcommand{\looppsfiles}                {ps-files/loop_model/}

\renewcommand{\Deltar}{\Delta_{\rm r}}

\renewcommand{\co}[1]{}
\renewcommand{\todo}[1]{#1}
\renewcommand{\myparagraph}[1]{}
\newcommand{\mysmallbreak}{\ \\}

\usepackage{lineno}

\renewcommand{\coout}[1]{}

\newcommand{\extralength}{0}
\newenvironment{adjustwidth}[2]{}{}

\begin{document}

\title{Structural thermokinetic modelling}
\author{Wolfram Liebermeister\\[3mm] 
Universit\'e Paris-Saclay, INRAE, MaIAGE, 78350, Jouy-en-Josas, France}
\date{}
\maketitle

\begin{abstract}
  \co{make sure changes from ``Metabolites'' version are copied to here}
  Translating metabolic networks into dynamic models is difficult if
  kinetic constants are unknown. Structural Kinetic Modelling (SKM)
  starts from a given metabolic state, defined by metabolite
  concentrations and fluxes, and replaces the reaction elasticities in
  this state -- the sensitivities of reaction rates to reactant
  concentrations -- by independent random numbers. Here I propose a
  variant that accounts for reversible reactions and thermodynamics:
  in Structural Thermokinetic Modelling (STM), correlated elasticities
  are computed from enzyme saturation values and thermodynamic forces,
  which are physically independent.  STM relies on a dependency schema
  in which basic variables can be sampled, fitted to data, or
  optimised, while all other variables are computed from them.
  Probability distributions in the dependency schema define a model
  ensemble, which leads to probabilistic predictions even if data are
  scarce.  STM highlights the importance of variabilities,
  dependencies and covariances of biological variables. By choosing or
  sampling the basic variables, we can convert metabolic networks into
  kinetic models with consistent reversible rate laws.  Metabolic
  control coefficients obtained from these models can tell us about
  metabolic dynamics, including responses and optimal adaptations to
  perturbations as well as enzyme synergies, metabolite correlations,
  and metabolic fluctuations arising from chemical noise. By comparing
  model variants with different network structures, fluxes,
  thermodynamic forces, regulation, or types of rate laws, we can
  quantify the effects of these model features. To showcase STM, I
  study metabolic control, metabolic fluctuations, and enzyme
  synergies, and how they are shaped by thermodynamic forces.
  Thermodynamics can be used to obtain more precise predictions of
  flux control, enzyme synergies, correlated flux and metabolite
  variations, and of the emergence and propagation of metabolic noise.
\end{abstract}

\co{make graphical abstract? evtl 2b und 6c}
\co{for 2nd submission: new abstract, following the outline below}


  \co{make sure changes from ``Metabolites'' version are copied to here}

\textbf{Abbreviations:} SKM: Structural kinetic modelling; MCT:
Metabolic control theory; FBA: Flux balance analysis; MoMA:
Minimization of metabolic adjustment.

\co{
\clearpage

  \textbf{Structure} (could also be a structure for the abstract / the graphical abstract)

  \textbf{Intro:}
  
  (1) What effects have thermodynamic forces? Equilibrium and strongly driven reactions; one may imagine many possible effects, e.g. a ``diode'' for propagating fluctuations (forward driven), plausible target for regulation (strongly driven); high heat dissipation and enzyme demand (equilibrium); to see how thermodynamic forces should be arranged, we need to show these effects. 

  How can we do this, since forces are not isolated, and metabolic behaviour reflects enzyme kinetics as a whole, not just the forces. That is, hwo can we show effects of forces IRRESPECTIVELY of, for example, enzyme saturation? this is a typical problem in uderstanding the cell! One possibility: study ensembles in which these other effects can assume various values, and check which properties persist, and depend on the forces. (that is, to show somethin independently of other variables, we construct an ensemble over these other variables)

  (2) Generally, how can we understand metabolic dynamics without having precise models, either because of missing knowledge or knowing that there is large variability in cells?  To do this, how can we build models systematically, from partial knowledge? For example, how can we build models systematically around known fluxes? Thermodynamic FBA allows us to obtain plausible states (consitent fluxes and metabolite levels), but how can we extend those to full kinetic models, while describing things as uncertain if they are? 

SKM does this; it links FBA and kinetic models (with irreversible rate laws); but in the meantime, thermodynamic laws have been used to improved models an both sides; STM account for this and links THERMODYNMIC FBA and thermodynamically consistent kinetic models (with consistent reversible rate laws).

\textbf{Methods:}

STM is based on a dependency schema, elasticity formulas, and the possibility of sampling. It  allows for ensemble simulations and  can be used to \\
o study broad model ensembles; demonstrate effects of model details (eg different patterns of forces in the network); requires statistical significance.\\
o build realistic models (with uncertainty in parameters, and allowing
for probabilistic forward simulation)

\textbf{Results:}

The elasticities themselves (even for a SINGLE model from the ensemble, eg one that has been built around kinetic data);
yield intersting insights about\\
o stability (see how forces influence the eigenvalue spectrum)\\
o effects of differential expression (UTP is just one ``crazy'' example) - but alos show a ``normal'' prediction\\
o synergies -  explain how this was studied by FBA so far, and what MCA can add to it\\
o (static) variance and correlations, describing (i) uncertainty or (ii) actual variability (e.g. between cells)\\
o fluctuations (and correlations), caused for instance by enzyme level fluctuations or chemical noise

In all this, we can now study computationally (and partially, analytically!) how forces are involved (discuss general expected effcts, now mathematicallly, knowing how forces enter the formulas for control coefficients etc etc; [explain briefly for each of the 5 effects described above])

\textbf{Discussion}

schema modelling as a General approach to  model construction

(ii) ensemble modelling, fitting, and optimisation

understanding relationships between biological variables: understanding of variation and correlations, e.g. between cells or across evolution.

retromodelling  is  also Helpful for ``engineering'' models, starting from a desired outcome}

\co{JA! statt ``schema-based'' lieber ``structured modelling
  approaches'' auch in RBA} \co{hier und model balancing und RBA
  symbolic: ``schema-based modelling'' erklaeren: systematsicher weg
  vom schema (plus network structure) to model; schritte koennen sein:
  waehlen, constrainen, sampling, optimieren. bsp PB, MB, STM, ECM,
  RBA construction}

\co{bei factorised rate law sagen, dass flussrichtung bekannt sein muss! (vorteil von modular rate laws: sie muss es nicht sein)} \co{What can we learn from network structure alone? Structure and function; buch (idee von vincent): stromnetz und telefonnetz haben die gleiche strukture -> sind aber sehr verschieden!}

\co{mention factorised rate laws: $v = e\,\sign(\theta)\,\kcat_{\rm ff}\,(1-\e^{-|\theta|})\,\eta^{\rm kin}(\cv)$ where ff means ``in flux direction'' (other formulation: EMC2)\\
  if $\eta^{\rm kin}=\const$, then $\Escvlci=-\diag(\frac{1}{1-\e_{|\theta|}})\,{\Ntot}\trans$, \\
  $\Eunvlci=-\diag(\vv)\,\diag(\frac{1}{1-\e^{|\theta|}})\,{\Ntot}\trans\,\diag(\cv)\inv$
}

\co{ o consider MDPI rules} \co{ o The name, version, corporation and
  location information for all software used should be clearly
  indicated. Please include all the parameters used to run
  software/programs analyses.}  \co{ o Citations and References in
  Supplementary files are permitted provided that they also appear in
  the reference list of the main text.}
\co{o neuen bayesian kram schon in STM-paper andeuten; ebenfalls dynamischen kram schon dort andeuten
      - aber nicht sagen, dass ich das mache! (noch korrektur noetig)}
\coout{retromodelling passt gut!! erwaehnen in abstract, intro, discussion; gilt dan auch fuer ECM, FCM; auch erwaehnen in FCM paper und CBA opt}
\co{alle matlab-grafiken durchgehen (auch SI): ueberall richtiges beispiel? font und farben ok? ergebnisse ergeben sinn? name of utp enzyme?}
\co{in SI add section on matlab code, showing commands for some
 typical analyses? lieber auf webseite!}

\co{reviewers: ralf steuer, vassily hatzimanikatis} \co{cover letter
  schreiben (alten teilweise wiederverwenden?)}  \co{anderes format
  für bibliograpie-referenzen im text?}   \co{nochmal ueberlegen
  (driving force or thermodynamic force?) - hier und MB gerade eher
  thermodynamic force} 

\co{JA! MIT BESTEHENDEM ABSTRACT VERBINDEN?\\
 ABSTRACT: Main question: how do thermodynamic forces and enzyme saturation influence metabolic dynamics? To study this precisely, we need to compare model variants in which these properties are individually varied (while other properties, e.g.~the flux distribution, are kept constant). Usual kinetic modelling starts from given enzyme concentrations and kinetic constants and determines metabolic (often steady) states, including all model variables. For our present question, it makes more sense to follow an SKM-like approach in which thermodynamic forces and saturation can be varied independently, and in a controlled way. I present a method for building models and model states exactly in this way, and (unlike the original SKM approach) accounting correctly for kinetics and thermodynamics. As a method for model construction, it generally has certain advantages (a systematic way to include data for specific model variables, and apply model assumptions and constraints.\\
 ARTICLE: first the theoretical basis (analysis of model variables
 and their dependencies); then the method for model construction; the
 building models; then analyse models to find statistical
 relationships and significant differences. This approach allows us
 to study the influences of network structure, metabolic state
 (including fluxes, concentrations, thermodynamic forces), and
 saturation systematically and precisely. The focus of the article
 is on the theroretical connections, not on precise modelling or pure
 parameter sampling. The formulae show explicitly how forces
 co-determine reaction elasticities (also for 2nd order). This also
 clarifies the relationships between 1st and 2nd order coefficients.
 I apply this to (i) control; (ii) prediction of effects of
 perturbations; (iii) synergies; (iv) (static) variation and
 (temporal) fluctuations. Other possible applications (model fitting
 and optimisation, model reduction for model embedding\co{explain
 somewhere!}) can be envisaged. STM works as a general practical
 modelling method, in particular when missing data need to be
 compensated by meaningful specific or fuzzy general modelling
 assumptions.}

\co{o welche allgemeinen aussagen koennte ich machen? BEWEISE!\\
1. schwaechere kraefte machen modelle tendenziell stabiler NICHT ZU SEHEN IN grossem ecoli-netz
1a Do varying enzyme concentrations induce correlated variations of their own reactant concentrations? 
2. Enzymes that
 have a higher influence on this target by themselves tend also to
 show larger synergies with other enzymes. NICHT ZU SEHEN IN grossem ecoli-netz\\
3. Enzymes that
 have a higher influence on this target by themselves tend also to
 show higher synergy degrees. STARKES ERGEBNIS IN grossem ecoli-netz\\
4. Synergies are correlated with the negative influence product (nachschauen!)\\
5. There is modular epistasis (BILD SOLLTE INS PAPER! aber es ist wohl nur sinnvoll, wenn es zumindest 
 einen vergliech mit FBA gibt)\\
6. Enzyme (individuell) haben tendenziell mehr negative als positive signifikante synergien (ueber threshold)
 d.h., die signifikanten synergien sind eher aggravating als buffereing. \\
7 Es gibt klare muster in den Zyklen (bsp in additional, fuer hefe,odell diskussionstext gibt es dort)\\
8 gute diskussion von signifikanten unterschieden und beweisbaren einfluessen. als beispiel evt das branch-modell (oder die drei glykolysemodelle)

o beweise fuer synergie-vorzeichen suchen. aber erstmal schauen, welche tendenzen in den daten zu sehen sind, bzw welche realistisch klingen}

\co{Update fuer webseite!! umbenennen in STM; alle modelldateien neu
 machen! text ``matlab code'' pruefen und nach github verschieben}

\co{neuere referenzen!} \co{From Ron: What is new in this paper? It
  will help to clarify exactly what is new. - Colormaps legends on
  some figure are needed. - Do we have one "punchline" of the
  analysis?}  \co{in intro + discussion: use term ``layered
  modelling'' or ``sequential modelling'' or ``sequential modelling of
  different layers of variables'' // ``Flux-scaffold''? konzept
  erklaeren, aber lieber schoeneres wort // Allgemeines wort fuer alle
  methoden, die von gegebenen fluessen ausgehen, so wie auch ECM und
  convex model bal.? ein adjektiv? (oder flux-first? / systematic?)
  mari fragen?}  \co{``TIPS'' aus caspars masterarbeit einbauen
  (checkliste in SI?)} \co{mention SBtab, github usw}

\co{WICHTIG! der artikel ohne gute anwendungen ist ein bisschen perlen
 vor die saeue, .. nochmal versuchen, was konkretes zu modellieren
 und hoeher zu publizieren. konstruktion und epistase fuer richtiges
 (grosses) netzwerk (zB dougies modell!), oder zumindest
 hepatozytenbeispiel?? und doch nochmal was mit differentieller
 expression versuchen? vielleicht jetzt mit besseren daten und
 dougies modell?? mit irgendjemandem drueber reden!} 
\co{show more results about uncertainties in the graphics}
\co{symmetric signs: allgemeine aussagen fuer
 synergien moeglich?} \co{put into SI: synergy cycles (from
 additional) } \co{symbole fuer
 elastizitaetsmatrizen?}

\co{TO DO:\\ 
o additional material und matlab-code durchsehen //
 schauen, ob andere beispiele in additional material relevant sind, 
 zb large e coli network - text from additional mat, copy to SI?\\
Alle bilder machen und anschauen!\\
o datenintegration doch nochmal probieren, mit realistischen konzentrationen!\\
o differential expression: branch point beispiel in SI einbauen! und als demo in repo\\
o was ist mit hefe-beispiel (oxygen)? nicht sehr gut!\\
o bei bayesian estimation lieber den neuen algorithmus implementieren und die 
 modelle/daten/texte dort wiederverwenden\\ 
o allgemeine ausrichtung klaeren // klaeren: welche themen werden behandelt? was in ornung bringen, was weglassen?\\
o weitere analysen (vorhersage von expressionsfolgen; modellfitten) durchsehen\\
o neue punkte; was wohin?\\
o gedankengang des artikels klarkriegen; einleitung + grosser plan \\
o alles in STM umbenennen? \\
o Welche beispiele: kleineres E. coli-modell als erstes beispiel in text? 
oder sogar ausschnitt aus grossen beispiel?\\
o ueberpruefen jetzt realistische konzentrationen und kinetik? samplen um KM-werte herum?
o automatisch erzeugte bilder schoenmachen / colorbar mit flag an- und ausschalten\\
o describe use of KM values (including beta distribution; already implemented in workflow script) in the paper examples?\\
o webseite, an workflow ausrichten; alle wichtigen infos (zb annotationen); einfacher ablauf sollte klar sein; mit matlab-doku abgleichen. phasen einheitlich numerierern\\
o demodateien : gute erklaerung aller optionen // statistik // synergien
o loesung fuer infeasible fluxes zur funktion machen, in workflow dokumentieren\\
o lactococcus glycolyse - neues modell von ralf?\\
o pmf in demobeispiel geht nicht - wieso? \\
o woher stammte domenicos lactococcus-netzwerk? bilder machen?\\
o weitere datensammlung: fluesse? e-coli-expression + fluesse zur vorhersage von fluss aenderungen??\\
o check: realistische konzentrationen + fluesse + km-werte + ergebnisse in allen beispielen? \\
o alle analysen nochmal laufen lassen / daten und reports nach html kopieren\\
o liste aller modelle in SI? groessen usw\\
o demo-beispiel: weitere fehlende methoden + infos ergaenzen! \\
o make everything useful in practice: ist beschreibung von funktionen
 und code ausreichend? fix code, documentation, and examples; describe
 SBtab workflow \\
o matlab-dokumentation in github auffrischen // Resourcen-Verzeichnis fertig // gith
ub auffrischen\\
o gutes bild fuer workflow, auch auf webseite (matlab-doku!)\\

Fix comments in text // auf l-lactis-paper eingehen // text auf papier korrigieren, abtippen //
tabelle mit methoden in matlab in SI? // fix text: mention large e
coli model o E-coli-Modelle in Artikel erwähnen o SBtab erwaehnen,
kurz matlab-workflow beschreiben o fluesse nicht stationaer, aber
thermodyn korrekt // bessere 1. abbildung // show all distributions
(e.g.~of synergies) for at least one example in SI // Fix all figures
// nochmal referenzen durchgehen

Fix abstract and title // Find keywords // fix github repo // update file on arxiv //
 submit to JTB?}

\coout{PLAN: alles in ordnung bringen, dann zu JTB}

\co{SCHREIBEN} 
 \coout{mention that predictions can be important for enzyme-flux
 relation (interpolation using the flux response coefficients of
 prototype fluxes} \co{The legends of the figures are not very
 complete.} \co{REFEREE My main criticism is therefore not so much
 about the scientific contents of the paper but its presentation. I
 propose you change this considerably. This work has great potential
 but you have to try to reach a broader audience than you do now by
 making it far more accessible. The presentation to a broader
 audience than the MCT and FBA/MOMA geeks is very poor. I think that
 this work can greatly improve if the author explains more what the
 aim of this study is and why this type of modelling is actually
 useful for answering particular metabolic control or dynamics
 problems. At the moment the paper is more a collection of
 illustrations, without coming anywhere to some clear conclusion.}

\coout{OTHER EXAMPLES / INSIGHTS} \coout{andere / groessere modelle? Hefe, E
 coli L lcatis, Hepatonet viele fluesse: Eq const, von elad mappen}
\coout{hepato-modell auf webseite?} \coout{doch grosses e-coli-beispiel
 bringen (aus additional material) und statistische tendenzen
 beschreiben?} \coout{natalies modell fuer sampling verwenden? ODER
 NEUES (dougie-) hefemodell?} 
\co{nochmal nach der phosphat/bas-geschichte schauen!} \co{Editor:
 some additional work with examples is needed to expand the claims
 being made} \co{REFEREE: Also of interest would be the application
 of the new approach to an important network / pathway including the
 detailed discussion of the new biological insights achieved.}

\co{CONCENTRATIONS}
\co{realistische metab. kinetics + kraefte} \co{wichtig: thermodyn!
 aus paar. bal? wie in Antrag, bei schlechten kraeften kompromisse //
 dougie-hefemodell als beispiel? auf webseite?} \co{irgendwelche
 testbaren vorhersagen?} \co{wichtig: realistischere konzentrationen
 und gleichgewichtskonstanten!}

\co{USEFULNESS / WORKFLOW}
\co{make sure it's all made to be useful in practice}
\co{kleine pdf-anleitung fuer toolbox. wie macht man ein modell und laesst ES laufen?}
\co{SBtab workflow?}
\co{conc und forces aus MDF oder PB (cite equilibrator)}
\co{alles schoen mit SBtab, webseite / github mit modellen und daten}
\coout{update settings files}
\coout{fuer software usw: plaene von zettel}

\co{yeast stanford / dougies version des jol-netzes // evtl auch grosses
  ecoli / l-lactis-modell zum laufen kriegen // beschreibung des
  grossen ecoli-modells ist schon in additional material (bilder in
  extra-supplement?) 1. Supplements fuer verschiedene modelle
  automatisch erzeugen // 2. Fuer Hefe zeigen: Study the impact of
  network structure, rate laws, reference fluxes, and regulation on
  control coefficients??}

\section{Introduction}

\coout{WO? For details about specific models and their construction, see
  section \todo{\ref{sec:SIexamplemodels}}. All models in the article
  were checked for stability of the reference state.}

\co{NACH ERSTEM REVIEW: einen parag vorschalten, in dem mit thermodyn
  kraeften angefangen wird und wie schwierig es ist, eine solche frage
  isoliert zu behandeln, damit schliessen, dass man eigentlich genaue
  modell braeuchte, aber man die erstens nicht hat und man zweitens
  fuer ALLGEMEINE aussagen tatsaechlich ensemblmodelle braeuchte

  What effects have thermodynamic forces? Equilibrium and strongly
  driven reactions; one may imagine many possible effects, e.g. a
  ``diode'' for propagating fluctuations (forward driven), plausible
  target for regulation (strongly driven); high heat dissipation and
  enzyme demand (equilibrium); to see how thermodynamic forces should
  be arranged, we need to show these effects.

  How can we do this, since forces are not isolated, and metabolic
  behaviour reflects enzyme kinetics as a whole, not just the
  forces. That is, hwo can we show effects of forces IRRESPECTIVELY
  of, for example, enzyme saturation? this is a typical problem in
  uderstanding the cell! One possibility: study ensembles in which
  these other effects can assume various values, and check which
  properties persist, and depend on the forces. (that is, to show
  somethin independently of other variables, we construct an ensemble
  over these other variables)}

\myparagraph{\mysmallbreak Metabolic models} The metabolic fluxes in
cells are shaped by network structure, reaction thermodynamics,
enzymatic rate laws, and enzyme regulation.  Flux, metabolite, and
protein data provide a detailed picture of metabolism, and
computational models can help us answer important questions.  For
example, how do enzyme-inhibiting drugs, enzyme overexpression, or
changes in nutrient levels affect the cell's metabolic states? Will
local enzyme perturbations have long-range effects on the fluxes, or
are they compensated by changes of nearby metabolite concentrations
\cite{fbrp:10}?  To address such questions, metabolic networks,
comprising thousands of reactions, have been built
\cite{fhrk:07,Henry2010}, and methods such as Flux Balance Analysis
(FBA) \cite{fesm:86,vapa:94}, MoMA \cite{sevc:02}, the principle of
minimal fluxes \cite{holz:04}, or ROOM \cite{shbr:04} predict
plausible flux distributions from heuristic assumptions \cite{scks:07}
or sampling \cite{prsp:04}.  Thermodynamic flux analysis
\cite{belq:02,kuph:06,hjbh:06,hohh:07} relates fluxes to metabolite
concentrations via equilibrium constants and thermodynamic
forces. However, it does not describe how metabolic fluxes arise
mechanistically. How can we predict the effects of enzyme
concentrations, external metabolite concentrations, or parameters like
temperature or the dilution rate on fluxes? If metabolite
concentrations were constant, reaction fluxes would be directly
proportional to enzyme concentrations. But in reality this is
different: the interplay between fluxes, concentrations, and enzyme
activities leads to stationary fluxes and metabolite concentrations,
which depend on enzyme levels in complicated ways.  To understand how
an enzyme inhibition changes metabolic fluxes, we need to consider its
effect on metabolite concentrations. Flux analysis cannot describe
this because rate laws, enzyme saturation, and regulation by effector
molecules, are not even considered. Kinetic models would allow us to
quantify the effects of enzyme perturbations (or other parameter
perturbations) on fluxes and metabolite concentrations in steady
state, but kinetic rate laws and rate constants are largely unknown,
especially in less well-studied organisms.

\coout{DONE: in vorigem abschnitt: Varianten von SKM zitieren: natalie-approach
 kurz erwaehnen // Murabito \cite{mvbb:14} so aehnlich wie natalie,
 aber mit zufallsparametern // andere referenzen (aus
 murabito-artikel)\\
 \cite{japa:08} Jamshidi and Palsson Mol Syst Biol. 4(2008)171 THEY
 ASSUME that the elast (``gradient'') matrix can be split into a
 diagonal ``kinetic'' matrix kappa and a ``thermodynamic'' matrix
 Gamma. \coout{We describe here a framework for building and
 analyzing such models. The mathematical analysis challenges are
 reflected in four foundational properties, (i) the decomposition
 of the Jacobian matrix into chemical, kinetic and thermodynamic
 information, (ii) the structural similarity between the
 stoichiometric matrix and the transpose of the gradient matrix,
 (iii) the duality transformations enabling either fluxes or
 concentrations to serve as the independent variables and (iv) the
 timescale hierarchy in biological networks. Recognition and
 appreciation of these properties highlight notable and challenging
 new in silico analysis issues.}\\
 \cite{japa:10} IS LIKE FITTING MASS ACTION KINETICS TO ELASTICITIES
 Jamshidi and Palsson Biophys J. 98(2010)175-185. Mass action
 stoichiometric simulation (MASS) models. \coout{Enzymes and their
 various functional states are represented explicitly as compounds,
 or nodes in a stoichiometric network, within this
 formalism. Analyses and simulations of MASS models explicitly show
 that regulatory enzymes can control dynamic states of networks in
 part by binding numerous metabolites at multiple sites. Thus,
 network functional states are reflected in the fractional states
 of a regulatory enzyme, such as the fraction of the total enzyme
 concentration that is in a catalytically active versus inactive
 state. The feasible construction of MASS models represents a
 practical means to increase the size, scope, and predictive
 capabilities of dynamic network models in cell
 and molecular biology.}\\
 \cite{trrl:08} Tran LM, Rizk ML, Liao JC (2008) Ensemble modeling of
 metabolic networks. Biophys J 95: 5606–5617. \coout{ Complete
 modeling of metabolic networks is desirable, but it is difficult
 to accomplish because of the lack of kinetics. As a step toward
 this goal, we have developed an approach to build an ensemble of
 dynamic models that reach the same steady state. The models in the
 ensemble are based on the same mechanistic framework at the
 elementary reaction level, including known regulations, and span
 the space of all kinetics allowable by thermodynamics. This
 ensemble allows for the examination of possible phenotypes of the
 network upon perturbations, such as changes in enzyme expression
 levels. The size of the ensemble is reduced by acquiring data for
 such perturbation phenotypes. If the mechanistic framework is
 approximately accurate, the ensemble converges to a smaller set of
 models and becomes more predictive. This approach bypasses the
 need for detailed characterisation of kinetic parameters and
 arrives at a set of models that describes relevant
 phenotypes upon enzyme perturbations.}\\
 \cite{trca:11} 27. Tan Y, Rivera JGL, Contador CA, Asenjo JA, Liao
 JC (2011) Reducing the allowable kinetic space by constructing
 ensemble of dynamic models with the same steady-state flux. Metab
 Eng 13: 60–75. }

\myparagraph{Linearised models and Metabolic Control Theory} Close to
a steady state, metabolic dynamics can be described by linearised
models.  Metabolic Control Theory \cite{hesc:96,hofm:01} (MCT, Figure
\ref{fig:elasticitiesthermoandcontrol}), considers local perturbations
(for example enzyme inhibitions), quantifies their direct effect on
reaction rates, and infers network-wide effects on the steady
state. MCT uses two types of sensitivity coefficients: reaction
elasticities describe how reaction rates change with changing
metabolite concentrations: they concern immediate effects on a fast
timescale (which depend only on rate laws of reactions
perturbed). Response and control coefficients, which depend on network
structure and elasticities, describe long-term, wide-range effects of
parameter perturbations on steady-state fluxes and concentrations (see
Supplementary Materials). 
MCT has various applications. Using response coefficients, we can
assess the effects of enzyme inhibition, differential enzyme
expression, varying external metabolite concentrations, enzyme
inhibition by drugs, or genetic modifications. A Taylor expansion
based on response coefficients can also describe more complicated
perturbations (e.g.~simultaneous activity changes of all enzymes),
optimal enzyme allocation in pathways \cite{klhe:99}, and parameter
uncertainty or variability \cite{likl:05}. Second-order response
coefficients \cite{hohe:93}, describing synergy effects, play a role
in predicting optimal differential expression \cite{lksh:04}.  In
systems with periodic \cite{inga:04,lieb:2005} or random
\cite{likl:05} parameter fluctuations, the fluctuations of fluxes and
metabolite concentrations depend on spectral response
coefficients. Importantly, all these phenomena can be described
without a full kinetic model: linearised models with the same network
structure and reaction elasticities as in the original model, suffices
to describe the dynamics of small fluctuations caused by static or
periodic parameter perturbations or chemical noise, and powerful
theory exists for optimal control and model reduction in linear models
\cite{libk:05}.  So all we need to know are network structure and
(first or second order) elasticities. But if only the network
structure is known, how can we obtain elasticities from few or no
other data?

\myparagraph{Structural kinetic modelling} Since elasticities are
unknown (unless a kinetic model is already available), Structural
Kinetic Modelling (SKM, \cite{sgsb:06,gsbh:07}), the ORACLE framework
\cite{wabh:04,waha:06a,waha:06b,somh:12}, and similar methods
\cite{trrl:08,japa:08,japa:10,trca:11} replace them by random
numbers. Reaction rates increase with the substrate concentrations and
decrease with the product concentrations, and this is encoded in the
signs of reaction elasticities: positive for substrates and
activators, and negative for products and inhibitors (where
``substrate'' and ``product'', in this case, refer to the flux
direction).  Assuming irreversible rate laws, SKM equates the scaled
elasticities to saturation values, numbers between 0 and 1 that
describe the fraction of catalysing enzyme molecules that are bound,
on average, to metabolites.  In SKM, saturation values are treated as
free variables and sampled from random distributions.  Possible bounds
can be derived from the rate laws (e.g.~the range $]0,1[$ for
substrate elasticities in mass-action rate laws). The resulting
elasticity matrix is sparse, reflecting network structure and
regulation arrows. Each row describes one reaction: substrates and
activators lead to positive matrix elements, while products and
inhibitors lead to negative elements.  To sample the elasticities,
each non-zero matrix element is replaced by a random number with the
required sign.  Each sampled elasticity matrix defines a linearised
kinetic model, whose Jacobian matrix determines the stability of the
reference state as well as the dynamic behaviour around it. Given
metabolite concentrations and fluxes, reaction elasticities can be
converted into kinetic constants: hence, instead of sampling
elasticities it is also possible to sample some kinetic parameters
directly and to compute the others \cite{slsk:13, mvbb:14}. Structural
kinetic modelling has been applied to various cell biological
questions \cite{rese:10,mssw:11,gigs:12,mura:13}.

\myparagraph{Uncertainties and variability} If elasticities are
variable, uncertain, or unknown, how will this translate into (actual)
variability or (or subjective) uncertainty?  To explore this
variability or uncertainty, we can study model ensembles in which some
model features are given (e.g.~network structure and flux
distribution), while others are varied (e.g.~metabolite concentrations
and kinetic constants). By sampling variable model features from
random distributions \cite{likl:05, miha:10} and translating the
linearised model back into a full kinetic model, a model ensemble can
be created. Given a set of sampled model instances, we can estimate
the probability distributions for an infinitely large ensemble. In
practice, by generating many model instances (each with its own
elasticity matrix), one can explore the possible dynamics allowed by
our network, and assess their probabilities. By computing
probabilities for model outputs or types of behaviour, we can see how
they depend on network structure, fluxes, or rate laws.  If most of
the model instances show a certain behaviour, this behaviour can be
attributed to model structure (or generally: to features that were
fixed during sampling), while varying properties may be attributed to
the features sampled. In the model ensemble, we can also screen for
models with a given property -- e.g.~stable oscillations -- and check
which model details (e.g.~inhibition arrows or specific elasticity
values) are overabundant in this subensemble, and thus potentially
causing these features \cite{sgsb:06}.  Finally, different model
variants can be compared (e.g.,~with different network structures,
regulation arrows, different fluxes, etc.): each model variant is
translated into a model ensemble and significant differences between
the ensembles can be attributed to differences in model structure.

\myparagraph{Reaction elasticities and thermodynamic forces} Despite
all its merits, ensemble modelling by SKM has one major drawback: it
ignores the fact that elasticities are interdependent due to basic
physical laws. The net flux in chemical reactions results from a
difference of one-way rates $v_{+}$ and $v_{-}$, the rates of
microscopic reaction events in forward and backward direction. In
chemical equilibrium states, the ratio of product and substrate
concentrations is given by the equilibrium constant: the net rate $v$
vanishes and the two one-way rates must be equal. More generally,
their ratio is given by $\frac{v_{+}}{v_{-}} = \e^{{\theta}}$ with the
dimensionless thermodynamic force $\theta=-\Deltar G/RT$, the reaction
Gibbs free energy $\Deltar G$ (given by the difference of chemical
potentials along a reaction), Boltzmann's gas constant $R$, absolute
temperature $T$ \cite{beqi:07}. The quantity
$A=-\Deltar G = RT \,\theta$ is called reaction affinity.  Depending
on their thermodynamic force (which depends on substrate and product
concentrations), reactions range between two extremes. In equilibrium
reactions, the force vanishes, the one-way rates $v_{+}$ and $v_{-}$
are equal, and the net flux $v$ vanishes.  Strongly driven reactions,
in contrast, have large thermodynamic forces, a negligible backward
flux $v_{-}$, and a net flux close to the forward flux $v_{+}$. By
tuning the one-way fluxes
($v_{+}= \frac{\e^{\theta}}{\e^{\theta}-1}\,v,
\frac{1}{\e^{\theta}-1}\,v_{-}= v$), the thermodynamic force shapes
elasticities: if a force is large, the backward rate will be
negligible, the net rate is does not dependent on the product
concentrations, and the product elasticities vanish.  Since the forces
themselves satisfy Wegscheider conditions (a zero sum over loops in
the network) and elasticities are dependent on them, elasticities may
be interdependent (and thus, correlated in reality) across the entire
network \cite{liuk:10}. Ignoring this fact, SKM leads to
thermodynamically inconsistent models \cite{liuk:10} (for an example,
see Supplementary Materials).

\myparagraph{Structural thermokinetic modelling} How can we solve this
problem?  Sampling correlated elasticities -- satisfying network-wide
thermodynamic constraints -- seems difficult, but it is actually easy
for certain rate laws: using formulae from \cite{liuk:10},
elasticities can be computed from thermodynamic forces and saturation
values, variables that can by independently varied. Below I use this
to construct model ensembles: I describe an algorithm for sampling
reaction elasticities, while fully accounting for dependencies due to
reversible rate laws.  I then show how thermodynamics and enzyme
saturation shape control properties, dynamic timecourses, enzyme
synergies, and fluctuations in metabolic systems. Importantly, also
enzyme synergies and other second-order effects are described by
closed formulae. Since it adds thermodynamics to SKM, the framework is
called Structural Thermokinetic Modelling.  Compared to the original
SKM, modelling is more formalised: all model variables (including the
reference metabolite concentrations and fluxes) are determined step by
step, by inserting known values, sampling, or optimisation. A
thermodynamically feasible metabolic state is a basis for constructing
thermodynamically feasible models. The workflow follows a schema that
describes the model variables, guarantees physical correctness, and
can be used to define probability distributions. By separating network
(defining the biological system) and schema (defining the types of
variables and their physical dependencies), models can be build
flexibly, and their mathematical relationships are easy to see. This
makes STM particularly suitable for automatic model construction.

\section{Materials and Methods}

\coout{REVIEWER: You hardly introduce the main
  terminology; i.e. elasticity coefficients, MCT, FBA, MOMA, response
  coefficient are not clear to a wider audience, those have to be
  introduced. In particular not to the audience that is interested in
  using this tool. I think your audience is primarily the experimental
  systems biologist working in the metabolism field with a basic
  understanding of enzyme kinetics. He/she is dealing with data and
  would like to develop predictive/insightful models by using data to
  infer a feasible set of models. This is in principle what your
  software achieves and you illustrate its applicability to a range of
  biologically relevant applications, where the examples are purely in
  silico (and do not deal with actual experimental data).}

\subsection{Constructing kinetic metabolic models}

\myparagraph{\mysmallbreak Construction of biochemically valid models}
To build metabolic models, STM relies on kinetic and thermodynamic
laws and on Metabolic Control Theory.  An overview of concepts and
formulae is given in Supplementary
Materials. 
To translate metabolic networks into kinetic models, we need to
integrate data about kinetics, thermodynamics, and metabolic
states. In general, model construction poses a number of challenges:
(i) finding realistic rate laws and kinetic constants; (ii) ensuring
consistent equilibrium states of the model, with metabolite
concentrations leading to vanishing fluxes; (iii) choosing a reference
state with realistic fluxes and metabolite concentrations; and (iv)
assessing variability and uncertainties in model parameters and
metabolic states.  Metabolic model can be parameterised automatically
with the help of generic rate laws including mass-action, power-law
\cite{sava:70}, linlog \cite{vihe:03}, or modular rate laws
\cite{likl:06a,liuk:10}.  The use of reversible rate laws with
consistent kinetic constants (satisfying Wegscheider conditions and
Haldane relationships) can guarantee consistent equilibrium states.

\co{cite and explain  thermodynamic-kinetic formalism \cite{edgi:07}}

\co{say here: assume that fluxes and conc are known; how can we get to
  the elasticities? ein possibility: EMC2 - formulae from blatt!
  requries keq and correct state; accounts for theta; ignores variable
  saturation effects by setting $\eta^{\rm kin}=const$; ie ignores
  kinetic effects, e.g., of lowering substrates and all allosteric
  effects. (but could be added as an extra term, like in teh forumlae
  used here .. ); but here we use other formulae, from modular rate
  laws.} 

\co{FN: (say that
    ``tacitly, invariant theta convention will be used'') insist that
    theta by definition refers to MOLECULARITIES, ie. h=1 (stoich
    coeff and molecularities must be the same!); das auch elad sagen
    und ueberall (in MB usw) betonen!}

  \myparagraph{Retromodelling} In kinetic models, enzyme levels and
  external metabolite concentrations are usually treated as parameters
  that determine the steady state. However, such model are difficult
  to fit to fluxes (whether measured or predicted by
  FBA). Retromodelling reverts this procedure: we start from a flux
  distribution and construct metabolite concentrations and rate laws
  around it. We can do this as follows. After choosing the fluxes, the
  MDF method \cite{nbfr:14} can be used to find feasible metabolite
  concentrations and thermodynamic bottlenecks (reactions with
  inevitably poor thermodynamic forces). Then, there are different
  ways to proceed.  First, one may define reversible rate laws, choose
  plausible metabolite concentrations, and compute the enzyme demands
  \cite{sssm:10,ldjs:10,slsk:13}, or optimise metabolite and enzyme
  concentrations simultaneously for a minimal enzyme cost
  \cite{fnbl:13}.  Second, one may first construct a full metabolic
  state (including fluxes and metabolic concentrations) and then find
  kinetic constants that can realise this state.

\subsection{Elasticities and their dependence on thermodynamic forces}

\begin{figure*}[t!]
  \begin{adjustwidth}{-\extralength}{0cm}
\begin{center}
\includegraphics[width=15.5cm]{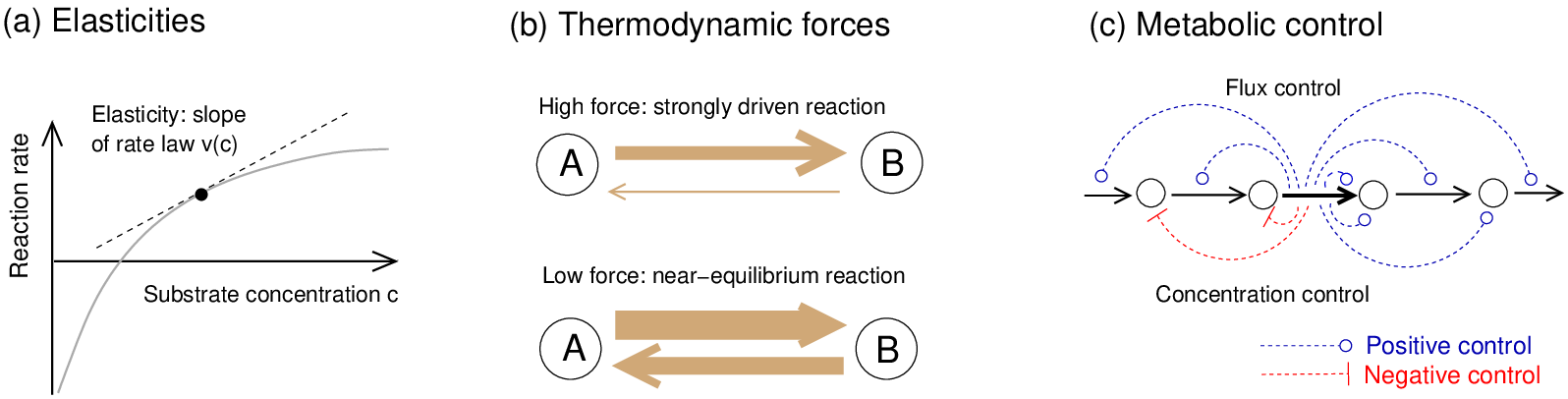}
\end{center}
\caption{\textbf{Metabolic rate laws and dynamics: elasticities,
    thermodynamic forces, and metabolic control.} (a) A reaction rate
  depends on metabolite concentrations as described by a rate law. The
  slope $\partial v/\partial c$ is called reaction elasticity. (b) In
  strongly driven reactions (with a driving force
  $\theta = \ln \frac{v_{+}}{v_{-}}$ much larger than 1), the net rate
  is dominated by the forward rate and the product elasticity is
  almost zero. In contrast, close to chemical equilibrium (with a
  driving force $\theta \approx 0$ and a net rate given by a
  difference between large forward and backward rates), the scaled
  elasticities
  $\Esc^{v_{l}}_{c_{i}} = \frac{c_{i}}{v_{l}}\,\frac{\partial
    v_{l}}{\partial c_{i}}$ are large. (c) Metabolic control
  coefficients describe how perturbations in single reactions shape
  steady-state fluxes and metabolite concentrations across the
  network. If an enzyme is inhibited or repressed, upstream
  metabolites accumulate and downstream metabolites deplete. This
  long-term response depends on network structure, flux distribution,
  and reaction elasticities.  Thermodynamic forces shape metabolic
  control via the elasticities: a strongly driven reaction, with its
  low product elasticity, is insensitive to downstream processes and
  deprives all downstream enzymes of their all flux control.}
 \label{fig:elasticitiesthermoandcontrol}
  \end{adjustwidth}
\end{figure*}

\myparagraph{\mysmallbreak Elasticities and linearised dynamics}
Enzyme kinetics, thermodynamic forces, and metabolic control are
closely related (Figure \ref{fig:elasticitiesthermoandcontrol}).  The
metabolic dynamics close to a reference state depend on reaction
elasticities, defined as derivatives
$\Euns^{v_{l}}_{c_{i}}=\partial v_{l}/\partial c_{i}$ of kinetic rate
laws $\rate_{l}(\cv)$ with respect to metabolite concentrations
$c_{i}$.  At high substrate concentrations, an enzyme becomes
saturated, and additional substrate has little effect, so the
elasticity is low.  Elasticities describe the immediate, local effects
of metabolite perturbations on reaction rates: to define them, we
formally assume that metabolite concentrations are not dynamic
variables, but parameters to be tuned from the outside.  Elasticities
between rates and other variables, e.g.~enzyme levels, are defined
accordingly. Importantly, our (so-called unscaled) elasticities
$ \Eun^{v_l}_{c_{i}}$ can be rewritten as
$ \Eun^{v_l}_{c_{i}} = v_{l}\, \Esc^{v_l}_{c_{i}}\, \frac{1}{c_{i}}$
with unitless scaled elasticities defined as $\Esc^{v_l}_{c_{i}}$
$\Escs^{v_{l}}_{c_{i}} = \frac{c_{i}}{v_{l}}\frac{\partial
  \rate_{l}}{\partial c_{i}}$ (or
$ \Esc^{v_l}_{c_{i}} = \partial \ln |v_{l}|/\partial \ln
c_{i}$). Scaled elasticities describe effective reaction orders. For
irreversible rate laws, we obtain $\Esc^{v_l}_{c_{i}}\approx 0$ in the
linear range and $\Esc^{v_l}_{c_{i}}\approx 1$ near saturation. In
general, $\Esc^{v_l}_{c_{i}}> 1$ indicates that two molecules of the
same substrate are used.  Formulae for second-order elasticities
(defined via second derivatives) are given in Supplementary
Materials. 
To linearise a metabolic model around a reference state, we just need
the stoichiometric matrix and the unscaled elasticity matrix
$\Eunc$. Both matrices can easily be obtained if the reference state
(metabolite levels and fluxes) is known.  A perturbation (e.g.~by an
initial perturbation vector $\Delta \cv(t)$) will \todo{lead to a
  linearised dynamics}; by solving the differential equations
$\frac{\md}{\md t} \Delta \cv(t)= \Amat\,\Delta \cv$ with the Jacobian
matrix $\Amat=\Nint\,\Eunc$ we can
simulate the propagation of dynamic perturbations across the network
and determine their long-term effects \cite{libk:05}.  In models with
conserved moieties, a complication arises: if the internal
stoichiometric matrix has non-full row rank, and we need to focus on
independent metabolite concentrations. Their concentrations follow a
dynamic equation
$\frac{\md}{\md t} \Delta \cv^{\rm ind}(t)= \Amat\,\Delta \cv^{\rm
  ind}$ with Jacobian matrix $\Amat=\NR\,\Eunc\,\Lmat$, and
$\cv(t)=\Lmat\,(\cv^{\rm ind}(t)-\cv^{\rm ind}(0))+\cv(0)$, where
$\Nint=\Lmat\,\NR$ and $\Nint$ has full row rank.

\coout{note that jacobian is called M in
  other papers; here A is better bcs of the network matrices M}

\myparagraph{Elasticities and enzyme efficiency} Enzyme elasticities
are subject to multiple functional requirements.  At low saturation
 (i.e.~high elasticities), the enzyme efficiency (flux per enzyme level)
is low, and the enzyme demand per flux (i.e.~the inverse enzyme
efficiency) is high.  At high saturation, a different  problem
arises: the reaction rate hardly changes with  the substrate level, so the
reaction is  ``stiff'':  fluctuations in inflowing
substrate cannot be buffered and lead to large fluctuations in substrate
levels. Hence, the optimal choice of elasticities (and thus, of 
saturation values) reflects trade-offs between enzyme demand and
favourable control properties. \co{saturation value or saturation level?}

\myparagraph{Elasticities of the modular rate laws} Modular rate laws
(see Supplementary
Materials) 
are generic reversible rate laws based on simple enzyme mechanisms,
with random-order binding. The formulae contain terms of the form
$\beta_{li}^{\rm X} = c_{i}/(c_{i}+k_{li}^{X})$, called saturation
values. A saturation value describes the saturation of an enzyme with
a reactant or small-molecule regulator, where  $c_{i}$ is a
metabolite concentration, and $k^{\rm X}_{li}$ is the dissociation
constant between enzyme and metabolite (X  stands for M
(reactants), A (activators), or I (inhibitors)).  In modular rate laws, which
are based on a quasi-equilibrium approximation \cite{liuk:10}, the
$k^{\rm X}$ are dissociation constants and half-saturation
concentrations at the same time.  Saturation values range between 0
and 1, and from saturation values and metabolite concentrations,
we can reconstruct the dissociation constants $k^{X}_{li}$. The
scaled elasticities of modular rate laws consist of two terms
\cite{liuk:10},
 \begin{eqnarray}
 \label{eq:modelasticities}
 \Esc^{v_l}_{c_{i}} &=& \Esc^{\rm rev}_{li} + \Esc^{\rm kin}_{li}.
\end{eqnarray}
The thermodynamic ``reversibility'' term $\Esc^{\rm rev}_{li}$ depends
directly on the thermodynamic force $\theta_{l}$. The kinetic term
$\Esc^{\rm kin}_{li}$ arises from kinetics and depends on the
saturation values.  Since forces are coupled through Wegscheider
conditions, the elasticities must be interdependent. However, given
the forces and metabolite concentrations, the saturation values can be
freely varied without violating any physical laws (proof see
Supplementary Materials). 
Details about modular rate laws and their (first- and second-order)
elasticities can be found in \cite{liuk:10}.

\myparagraph{Thermodynamic elasticity term} All (thermodynamically
consistent) reversible rate laws share the same numerator, which has
the shape of mass-action rate law \cite{liuk:10}. The numerator leads
to the thermodynamic elasticity term $\Escs^{\rm rev}_{li}$
\cite{liuk:10}
\begin{eqnarray}
\label{eq:thermoterm}
 \Escs^{\rm rev}_{li} = 
 \frac{v_{+l}\,m^{\rm S}_{li} - v_{-l}\,m^{\rm P}_{li}}{v_{l}}
 = \frac{\zeta_{l}\,m^{\rm S}_{li} - m^{\rm P}_{li}}{\zeta_{l}-1}
\end{eqnarray}
with substrate and product molecularities $m_{li}^{\rm S}$ and
$m_{li}^{\rm P}$ and a flux ratio
$\zeta_{l} = \frac{v_{+l}}{v_{-l}}=\e^{\theta_{l}}$.  Coming back to Eq.~(\ref{eq:thermoterm}), the flux ratio,
and therefore the elasticity term $ \Escs^{\rm rev}_{li}$ depends on
the thermodynamic force $\theta_{l}=-\Deltar G_{l}/RT$, with
Boltzmann's gas constant $R$ and absolute temperature $T$.  For
 forward reactions near equilibrium (i.e.~small positive thermodynamic
forces $\theta_{l} \approx 0$), the substrate terms (where
$m^{\rm S}_{li} > 0,\, m^{\rm P}_{li}= 0$) are close to infinity and
the product terms (where $m^{\rm S}_{li} = 0,\, m^{\rm P}_{li}> 0$)
are close to $-\infty$. For completely forward-driven reactions (with
thermodynamic force $\theta_{l}\rightarrow \infty$), we obtain
elasticities $m^{\rm S}_{li}$ (for substrates) and $0$ (for products).
Between these extremes, the elasticities vary with the thermodynamic
force. With a small force of 1 kJ/mol $\approx$ 0.4 RT  (and molecularities and
stoichometric coefficients equal to 1),
Eq.~(\ref{eq:thermoterm}) yields substrate elasticities of $\approx 3$ 
 and  product elasticities of $\approx -2$. In contrast, for a force of 10 kJ/mol we
obtain values of 1.02 (substrate) and -0.02 (product).

\myparagraph{Kinetic elasticity term} The form of the kinetic
elasticity term $\Esc^{\rm kin}_{li}$ depends on the type of modular
rate law. With  mass-action or power-law rate laws without
regulation, the kinetic term $\Escs^{\rm kin}_{li}$ vanishes and the
elasticities   follow directly from  network structure and
driving  forces:
\begin{eqnarray}
\label{eq:Emassaction}
\Ematsc = \Ematsc^{\rm rev} = \diag(\vv)\inv
[\diag(\vv_{+}) \Mmat^{\rm S} - \diag(\vv_{-}) \Mmat^{\rm P}] = \diag(\zeta -
\hat 1)\inv [\diag(\zeta) \Mmat^{\rm S} - \Mmat^{\rm P}].
\end{eqnarray}
With other rate laws, a kinetic elasticity term $ \Escs^{v_l}_{c_i}$
needs to be added. For example, for the simultaneous-binding modular
(SM) rate law with non-competitive activation and inhibition
\cite{liuk:10}, the kinetic term consists of four terms (related to
substrates S, products P, activators A, and inhibitors I):
 \begin{eqnarray}
 \label{eq:smelasticities}
 \Escs^{v_l}_{c_i} &=& 
 \Escs^{\rm rev}_{li}
 - \underbrace{\left[ m^{\rm S}_{li}\, \beta^{\rm M}_{li} 
 + m^{\rm P}_{li}\, \beta^{\rm M}_{li}\right]}_{ \Escs^{\rm den}_{li}}
 + \underbrace{\left[m^{\rm A}_{li} \, (1-\beta^{\rm A}_{li}) - m^{\rm I}_{li}\, \beta^{\rm I}_{li}
\right]}_{ \Escs^{\rm reg}_{li}}.
\end{eqnarray}
The two terms $ \Esc^{\rm reg}_{li} - \Esc^{\rm den}_{li}$ reflect 
two parts of the rate law: a prefactor for regulation and the rate law
denominator.  Eq.~(\ref{eq:smelasticities}) generalises
Structural Kinetic Modelling: the elasticity formula in SKM,
 \begin{eqnarray}
 \label{eq:smelasticitiesa}
 \Escs^{v_l}_{c_i} &=& 
 m^{\rm S}_{li}\, (1-\beta^{\rm M}_{li})
 + m^{\rm A}_{li} \, (1-\beta^{\rm A}_{li}) - m^{\rm I}_{li}\, \beta^{\rm I}_{li},
\end{eqnarray}
is a limiting case of Eq.~(\ref{eq:smelasticities}) for completely
forward-driven reactions, where $\beta^{\rm M}_{li}=0$ and
$\Escs^{\rm rev}_{li} = m^{\rm S}_{li}$. For reversible reactions, a
difference remains: in contrast to SKM, the thermodynamic term
$\Esc^{\rm rev}_{li}$ in STM makes elasticities thermodynamically
consistent and interdependent.  With the common modular rate law
\cite{liuk:10} (or ``convenience kinetics'' \cite{likl:06a}), the
formula for scaled elasticities is more complicated:
 \begin{eqnarray}
 \Escs^{v_{l}}_{c_{j}} = 
 \beta_{lj} 
 \frac { \zeta_{l}\, m^{\rm S}_{lj} - m^{\rm P}_{lj} } { \zeta_{l} - 1 }
 - \beta_{lj} \frac
 { m^{\rm S}_{lj} \psi^{+}_{l} + m^{\rm P}_{lj} \psi^{-}_{l} }
 { \psi^{+}_{l} + \psi^{-}_{l} -1}
 + m^{\rm A}_{li} \, \alpha^{\rm A}_{li} - m^{\rm I}_{li}\, \beta^{\rm I}_{li},
\end{eqnarray} 
where $\psi^{\pm}_{l}=\prod_{l}(1+c_{i}/\kM_{li})^{m^{\pm}_{li}}$.
Formulae for other rate laws and second-order
elasticities can be found in \cite{liuk:10} and Supplementary Materials. 

\myparagraph{Factorised rate laws} There is another useful elasticiy
formula. If the flux directions are known, reversible rate laws can be
written in the factorised form \cite{nflb:13}
\begin{eqnarray}
v= e\,\kcat\,\eta^{\rm rev}(\theta)\,\eta^{\rm kin}(\cv),
\end{eqnarray}
where the efficiency terms are unitless numbers between 0 and 1, and
the thermodynamic efficiency is given by $\eta^{\rm rev}(\theta)=
1-\exp(-\theta)$.  Here, forward $\kcat$ values and equilibrium
constants (or other variables from which they can be derived) serve as
basic parameters, where  $\kcat$, $\keq$, and $\theta$ correspond to  the flux
direction. Like for modular rate laws, there are simple formulae
for the scaled elasticities $\Escs^{v_{l}}_{c_{j}} =
\frac{c_{i}}{v_{l}}\frac{\partial v_{l}}{\partial c_{i}}$: 
we obtain (proof see Supplementary Materials)
\begin{eqnarray}
\label{eq:FacRateLawElast}
\Escs^{v_{l}}_{c_{j}} &=&  \frac{-1}{\e^{-\theta_{l}}-1}n_{il} + \frac{\partial \ln \eta^{\rm kin}_l}{\partial \ln c_{i}},
\end{eqnarray}
where the second term depends on the choice of rate law. Without the
second term (e.g.~assuming full saturation), we obtain an elasticity
formula that depends only on thermodynamic forces. Since we're
interested in the interplay between thermodynamics and saturation,
this formula will not be further considered below.

\co{modular rate laws and factorised rate laws als gleichbereichtigt
  darstellen, kurz vor-und nachteile praesentieren?}  \co{correspond
  to a different schema, with kcatforward (with repsect to the flux
  direction) as a basic variable, and no KV. mod rate laws use kv for
  symmetry reasons (to deal with flux reversals), but separable is
  more convenient whnever flux directions remain fixed} 

\begin{figure*}[t!]
  \begin{adjustwidth}{-\extralength}{0cm}
\begin{center}
\begin{tabular}{lll}
(a) Dependencies between model variables & & (b) Model construction \\[3mm]
 \includegraphics[height=7.4cm]{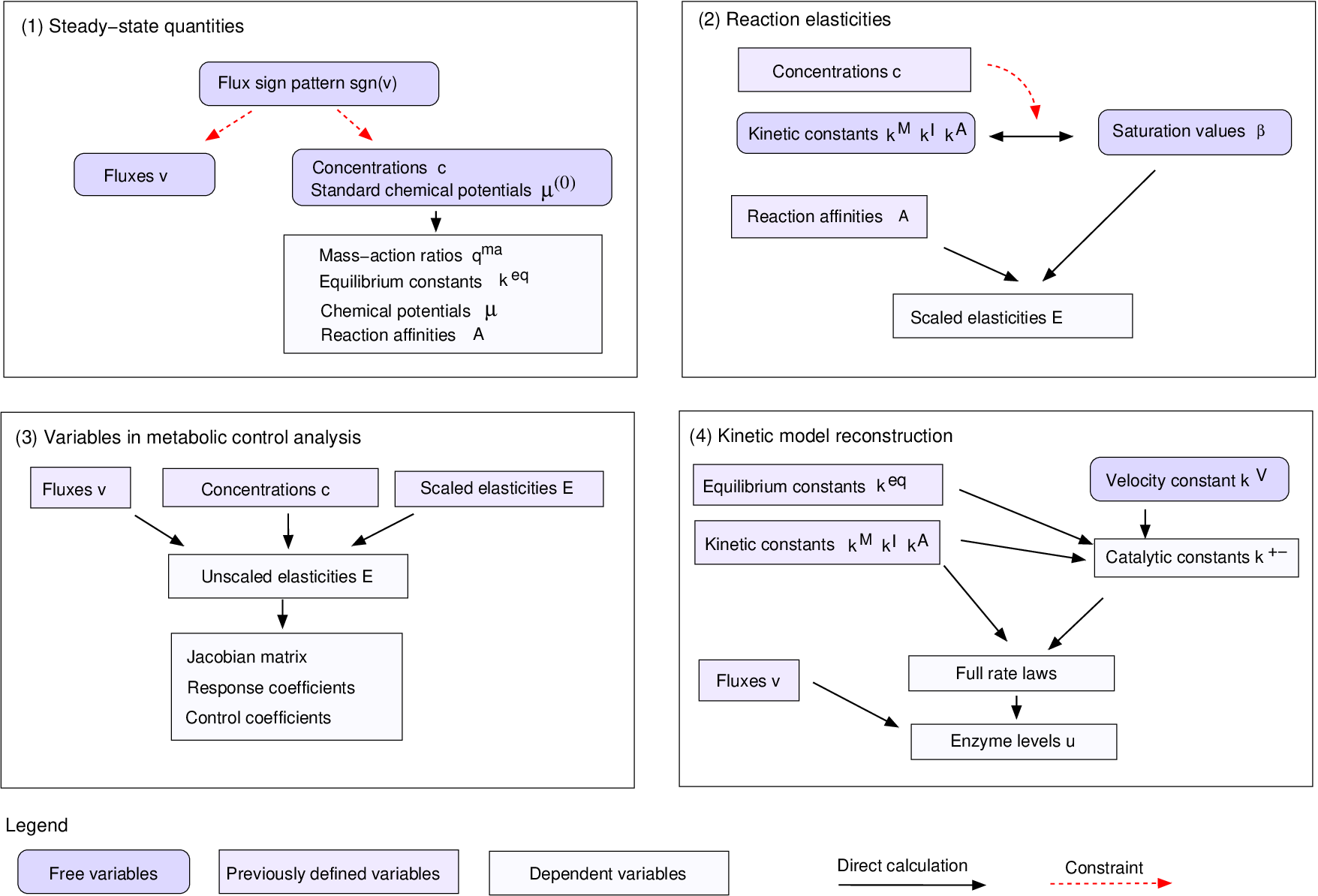}
& \ \hspace{5mm}& 
 \includegraphics[height=7.4cm]{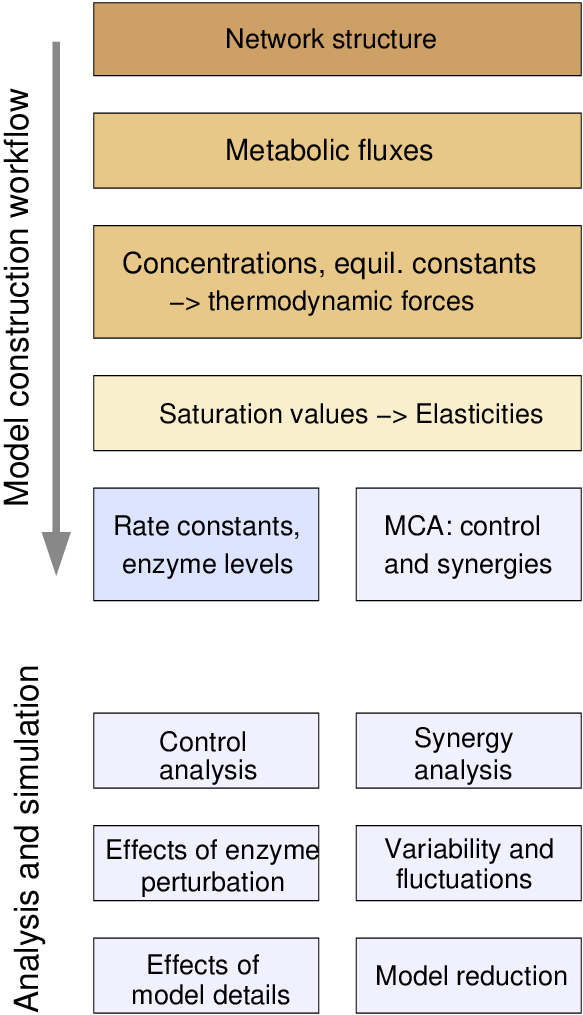} 
\end{tabular}
\end{center}
\caption{\textbf{Dependency schema and systematic model construction}
  (a) Dependencies between model variables (kinetic constants and
  state variables) in kinetic metabolic models. A dependency schema
  describes physical or logical dependencies between variables and can
  serve as a blueprint for model construction. For simplicity it is
  shown in four parts, corresponding to four steps of model
  construction. (1) Metabolic state phase. For physical reasons,
  fluxes and thermodynamic forces must have the same signs.
  Predefined flux directions define sign constraints on fluxes and
  thermodynamic forces, and fluxes and chemical potentials can be
  sampled under these constraints. (2) Kinetics phase. Saturation
  values $\beta^{\rm M}_{li}$, $\beta^{\rm A}_{li}$, and
  $\beta^{\rm I}_{li}$ can be chosen independently between 0 and
  1. Together with the metabolite concentrations and thermodynamic
  forces, they determine kinetic constants ($\kM_{li}$, $\kA_{li}$,
  and $\kI_{li}$) and reaction elasticities. The elasticities further
  determine control properties (3) as well as kinetic constants and
  enzyme concentrations (4), allowing us to reconstruct the entire
  kinetic model. In the graphics, some of the variables stem from
  previous steps (types of variables marked by colours).  (b) Model
  construction around a metabolic reference state. Based on a
  dependency schema, basic model variables can be freely chosen or
  sampled, while derived variables are computed from them.}
 \label{fig:samplingscheme}
  \end{adjustwidth}
\end{figure*}

\subsection{Model construction}

\myparagraph{\mysmallbreak Systematic model construction} STM relies
on a dependency schema (Figure \ref{fig:samplingscheme} (a)) in which
independent ``basic'' variables determine the remaining ``derived''
variables (where for brevity, I use the term ``variables'' not only
for fluxes, metabolite concentrations and enzyme concentrations, but
also for kinetic and thermodynamic constants.).  Following the schema,
we can build models systematically step by step (Figure
\ref{fig:samplingscheme} (b)): we start from a network (reaction
stoichiometries and regulation arrows), determine the state variables
(concentrations, fluxes, equilibrium constants), choose the saturation
values, and compute consistent (first and second order) elasticities
for our rate law by using Eq.~(\ref{eq:modelasticities}).  The kinetic
constants can be reconstructed easily.  From the second-order
elasticities, we obtain second-order control and response
coefficients, describing enzyme synergies for static \cite{hohe:93} or
periodic perturbations \cite{lieb:2005}.  Any feasible model with this
type of rate laws can be obtained in this way (for details, see
Supplementary Materials). 
 
\coout{(So far I understood you generate only one model. Make it clear
 that the method generates multiple models with the
 advantages/disadvantages of that.} \coout{(see an SI table to be
 made); possibilities to choose from; usage of known information,
 including kin const. gute dokumentation der ganzen
 verarbeitungsschritte, alternative moeglichkeiten!!}

\co{in next parag: assume that states can be found by modellers' favourite
  methods, possibly with the help of equilibrator, thermo FBA,
  Parameter Balancing etc. / then standard choice of kinetics
  (saturation 1/2), usage of kinetic data, or sampling} 

\myparagraph{Turning metabolic networks into kinetic models} During
model construction, the model variables can be chosen in various ways:
they can be replaced by data values, sampled, fitted, or
optimised. Below, we
assume that fluxes are fitted to data or determined by FBA, metabolite
concentrations are chosen within feasible ranges, and saturation
values are sampled. In the case, the hardest step in the calculation
is the choice of fluxes, which need to be thermo-physiologically
feasible: they need to be thermodynamically realisable with metabolite
concentrations within physiological ranges, which implies that the
flux distribution must be loopless \cite{prtp:06}.  Such fluxes can be
obtained by EBA or thermodynamic FBA (based on mixed-integer linear
problems) or by starting from given fluxes and removing infeasible
loops.  On the contrary, STM can be applied to non-steady reference
states, for example to flux data that look non-stationary when mapped
to a pathway model (because some in- and outfluxes are ignored in the
model). However, the following step -- Metabolic Control Theory --
requires a steady reference state. To construct such a steady state
(from realistic flux data, and in a small pathway model), it may be
necessary to add some extra reactions to account for incoming and
outgoing fluxes.  Given the fluxes, feasible metabolite concentrations
can be determined, e.g., by the Max-Min Driving Force (MDF) method
\cite{nbfr:14}. If the chosen flux distribution is thermodynamically
infeasible, some of the thermodynamic forces will be opposite to the
flux direction. To apply STM in this case, we may modify the forces by
extra terms (which may be attributed to additional, hypothetical
metabolites; see Supplementary
Materials). 
To construct a model ensemble with stable states, we may follow the
procedure of SKM: we generate multiple model instances, check that the
model has a stable reference state (the eigenvalues of the Jacobian
matrix must have negative real parts), and select the stable
models. It is not generally known how flux patterns, metabolite
concentrations, thermodynamic forces, and enzyme saturation contribute
to stable states, and how the chances of finding stable states depend
on network size. This would be an interesting question to be studied
by STM.

\co{next parag: note WD
  with beginning of previous paragraph} 

\myparagraph{Model construction based on kinetic data} In model construction by STM,
the basic model variables can be sampled, freely chosen, fitted to
data, or optimised for a fitness objective. We can use these different
options to incorporate extra knowledge or data. A complete set of
kinetic constants (as input data for STM) can be obtained by parameter
balancing, a method for translating incomplete, uncertain, and
contradictory data (for kinetic constants and state variables) into
complete, consistent sets of model variables
\cite{lskl:10,luli:19}. Given such data, a known $\kM$ value can be
converted into a saturation value. This value can then be inserted
directly, or saturation values can be sampled around it (in a range or
with a beta distribution) to account for uncertainties. Known $\kcat$
values can be used similarly: in the dependency schema, the $\kV$
values, together with equilibrium constants and fluxes, determine the
turnover rates $\kcatp$ and $\kcatm$ and the enzyme concentrations $e$
for the same reaction. To match them to data, we can first choose
$\kV$ values at random and then modify them to obtain a good fit of
forward $\kcat$ value, enzyme concentration, and flux; or we use a
different schema with forward $\kcat$ values (instead of $\kV$) as
basic variables.  Finally, STM may also be used for Bayesian model
fitting and to predict biologically optimal states. In these cases,
model variables are not sampled but treated as choice variables, which
can be fitted or optimised. Unfortunately, the resulting optimality
problems are typically non-convex and hard to solve. Model balancing,
which defines simplified convex optimality problems, may be used
instead \cite{lino:21}.

\myparagraph{Model ensembles} By repeatedly sampling model parameters,
we obtain an ensemble of models that all share the same structure but show
different parameter values \cite{wabh:04,likl:05,trrl:08}. More generally, a model
ensemble can reflect the network structure and some fixed choices made by the
modeller (e.g.~given  fluxes and metabolite concentrations), but
variability in all other variables (representing either  objective
variability or subjective uncertainty). Different model assumptions or
model variants will lead to different model ensembles. To study how
choices of model structure, metabolic state, or kinetics
influence metabolic dynamics, we generate  model variants, translate them into 
model ensembles, and search for significant differences in their
 behaviour. Significance tests are described in Supplementary Materials.
 Some models may have unstable states, as shown by positive
 eigenvalues of the Jacobian matrix.  To sample stable models only, we
 can proceed like in SKM: when generating a model ensemble, we discard
 all models with unstable states. In the resulting ensemble, the model
 variables may show different distributions and correlations:
 e.g.~saturation values may not follow the original distributions and
 become correlated.

\myparagraph{Variants of the algorithm: different rate laws, growing
  cells, and multiple steady states} The STM algorithm can be modified in
various ways.
\begin{enumerate}
\item To model metabolism in growing cells, the formulae must be
  adapted.  For balanced growth at a growth rate $\lambda$, all
  compounds have to be reproduced continuously, so the fluxes follow a
  modified mass balance condition $\Nint\,\vv-\lambda \,\cv=0$ with an
  extra dilution term.  Metabolite concentrations and fluxes are
  tightly coupled by this equation, and in model construction they
  must now be chosen together. The Jacobian matrix (which also appears
  in formulae for the control matrices) contains an extra term
  $-\lambda\, \Imat$. In the kinetic model, a dilution term
  $-\lambda \,c_{i}$ must be added to the ODE of each metabolite.
  Moreover, conserved moieties in such models must vanish because
  otherwise they would be diluted, making a steady state impossible.
\item Instead of choosing saturation values and concentrations and
  computing the constants $\kX_{li}$, we may follow a different
  dependency schema in which the constants $\kX_{li}$ are basic
  variables and the saturation values are computed.  While case
  saturation values may be sampled, for example, from uniform or beta
  distributions, kinetic constants may be sampled from log-normal or
  gamma distributions.  Distributions for ratios $c/\kM$ and for the
  corresponding saturation values
  $\beta = \frac{c/\kM}{1+c/\kM}=\frac{c}{\kM+c}$ correspond to each
  other. If $\beta$ is uniformly distributed in $]0,1[$, the ratio
  $c/\kM$ shows a probability density function
  $\mbox{prob}(c/\kM)=\frac{1}{(1+c/\kM)^{2}}$, i.e.~$\ln (c/\kM)$
  follows a logistic distribution with location parameter 0 and scale
  parameter 1 (see Supplementary
  Materials). 
  By sampling saturation values not within $]0,1[$ but in a smaller
  range, one may avoid highly saturated enzymes, and one may sample
  $\kM$, $\kA$, and $\kI$ values around known experimental values.
\item To build a model with multiple steady states, in the ``metabolic
  state phase'', we choose one set of equilibrium constants, but
  several sets of concentrations and fluxes (for the different steady
  states); in the ``kinetics phase'', the constants
  $\kM_{li}, \kA_{li}$, and $\kI_{li}$ and the velocity constants
  $\kV_{l}$ (geometric means of forward and backward catalytic
  constants) are chosen or sampled, for example, by parameter
  balancing \cite{lskl:10,slsk:13,luli:19}, and elasticities are
  computed from them. Finally, enzyme concentrations for each state
  are computed by matching reaction rates (from the rate laws) to
  predefined fluxes.
\end{enumerate}
Eventually, to make models more realistic, the generic rate laws of
some reactions may be replaced by manually chosen laws obtained from
enzyme assays \cite{bghs:09}. For other extensions and modifications
of STM, see Supplementary Materials. \co{REF}

\subsection{Metabolic control and synergy effects}

\co{allgemein woerter
  Synergy coefficient / synergy / synergy effect klarkriegen}

\myparagraph{\mysmallbreak Metabolic control} If
enzymes are perturbed by inhibition or transcriptionally, what will be
the metabolic effects?  Inhibiting an enzyme makes the substrate
accumulate and the product deplete. These changes have further
effects, which counteract the original effect and eventually lead to
shifts of the steady-state in the entire network. To understand such
indirect effects, we can start with a simple example: a reaction
perturbed by a single parameter (e.g.~an enzyme or external substrate
concentration).  In (first-order) MCT, the direct effect (on the
reaction rate) is described by a parameter elasticity, and the further
effects (on steady-state metabolite concentrations and fluxes) are
described by control coefficients. By multiplying parameter elasticity
and control coefficient, we obtain the response coefficient (the
sensitivity between parameter perturbation and metabolite
concentrations or fluxes). If perturbations are small, we can treat
them as additive: the effects of simultaneous perturbations can simply
be summed over.  A variation $\delta \esymbolv$ of enzyme activities,
at fixed external metabolite concentrations
($\delta \cv^{\rm ext}=0$), will thus lead to a flux variation
$\delta \vv = \Cvmat\,\Eunc\,\delta \esymbolv$, with the unscaled
control coefficient matrix
$\Cvmat = \Imat - \Eunc\,\Lmat\,(\Nr\,\Eunc\,\Lmat)\inv \,\NR$.  Each
matrix column describes the (positive or negative) effect of one
enzyme on all fluxes.  Accordingly, in general increasing an enzyme
activity can increase or decrease fluxes across the network. The
control coefficients (matrix elements) are usually unknown, but STM
estimates them based on (partial) information about network structure,
reference fluxes, metabolite concentrations, thermodynamic forces, and
enzyme saturation.  \co{also effect of external metabolites?}  \co{for
  2nd order see SI} \co{mention dynamic simulation - and use it for
  something below? at least one example simulation?}

\myparagraph{Synergy effects} Enzyme
perturbations (e.g.~by transcriptional repression or enzyme inhibition)
will change fluxes and other target variables. For small
perturbations, this can be described by a linear approximation: a variation
$\delta e_j$ of the enzyme level
 leads to flux changes
$\delta v_{l} = R_{e_{j}}^{v_{l}}\,\delta e_{j}$. For larger
perturbations, however, the results may be different, and we describe
this as synergy effects: an enzyme perturbation changes not only the
target variable itself, but also the very effects of enzymes on this
variable.  The extra effect is called synergism (or ``antagonism'' for
negative synergisms).  Antagonisms arise, for example, if
two enzymes share the same substrate: as one enzyme concentration goes
down, the substate concentration increases and the flux catalysed by
the other enzyme goes up, increasing its flux control. Synergisms do
not only exist between enzymes, but  between any variables (or
discrete network features) that affect the reaction rates. Besides
enzyme concentrations, this may include enzyme inhibition, knock-outs,
differential expression, or perturbations of external metabolite
concentrations. Generally, to quantify synergies, we consider two
perturbations (e.g.~of enzyme activities) and a target variable
(e.g.~the biomass production rate). If single perturbations a and  b change the
target value by factors $w_{\rm a}$ or $w_{\rm b}$ (typically smaller
than 1, if the  target is a cell objective to be
maximised), we may expect, as a guess, that a double perturbation will
lead to a change $w_{\rm a}\cdot w_{\rm b}$. We now compare this guess
to the actual change $w_{ab}$.  If the two values differ, we describe
this by a synergy effect
$\etasc^{\rm z}_{\rm ab} = \ln \frac{w_{\rm ab}}{w_{\rm a}\,w_{\rm
    b}}$. Otherwise, if we expect additive (instead of
multiplicative) changes, the difference
$\etaun^{\rm z}_{\rm ab} = w_{\rm ab} - (w_{\rm a} + w_{\rm b})$ can
be used as a synergy measure.

\myparagraph{Enzyme synergies described by metabolic control theory}
How can synergies be predicted from
models?  All we need is a model that predicts changes in a metabolic
target variable (e.g.~a flux or metabolite concentration) after single
or double perturbations of enzyme concentrations (in kinetic models)
or of fluxes (in flux analysis).
Synergies in kinetic models  can be computed numerically: we just
vary two enzyme levels and simulate the effect on the 
steady-state fluxes. A first inhibition will change the
flux control coefficients of all enzymes, and therefore the effect of
a second inhibition. For small perturbations, the synergistic effect can be
approximated by a second-order approximation.  While the usual
(first-order) response coefficients capture linear effects of single
enzymes, the second-order response coefficients (or ``synergy
coefficients'') capture synergies of enzyme pairs (and second-order
effects of single enzymes).  If variables  cannot be negative, such as concentrations,
it can make sense to describe their changes on log scale
(while the usage of log scale, applied to fluxes,  would make it
impossible for fluxes to change their direction). With logarithmic
enzyme changes $\Delta \ln e_{\rm a}$ and $\Delta \ln e_{\rm b}$
(where
$\Delta \ln e_{\rm a} \approx \frac{\Delta e_{\rm a}}{e_{\rm a}}$),
the synergy is given by
$\etasc^{\rm z}_{\rm ab} \approx \Rsc^{z}_{e_{a} e_{b}} \Delta \ln
e_{\rm a}\, \Delta \ln e_{\rm b}$, where $\Rsc^{z}_{e_{a} e_{b}}$ is
the scaled synergy coefficient (another word for second-order response
coefficients). If two enzymes are inhibited, negative
synergies ($\Rsc^{z}_{e_{a} e_{b}}<0$) are called aggravating while
positive synergies ($\Rsc^{z}_{e_{a} e_{b}}>0$) are called buffering.
The second-order effects of single-enzyme perturbations are described
by self-synergies: they are usually aggravating because 
inhibition tends to  increase an enzyme's control, which increases the
effect of the inhibition on the target variable.

\co{FBA with molecular crowding
  simulieren?}  \co{note that objective or target need not be the
  same! (except for in "growth sensitivities" or in epistasis!)}

\myparagraph{Enzyme synergies in constraint-based models} Different
modelling frameworks rely on different assumptions. FBA tries to
maximise the metabolic objective even after a perturbation, MoMA
assumes that fluxes show minimal changes despite a perturbation
(attempted homeostasis), while MCT predicts flux responses from
metabolic dynamics, assuming that enzyme concentrations remain
constant (as considered here) or are optimally adapted (see
\cite{lksh:04}).  How does this work in practice?  In FBA or kinetic
model simulations, synergies are simulated one by one for each enzyme
pair.  To model flux perturbations by FBA, we first solve an FBA
problem with a given objective function (e.g.~biomass production flux)
and determine a flux distribution as well as our target variable
(typically, the objective itself). To
mimic an enzyme inhibition, we put a bound on the catalysed flux,
constraining it to a certain percentage of the original flux
(e.g.~90\%~for a small perturbation, 50\%~for a large perturbation or
knock-down, 0\%~for complete inhibition or knockout). When solving the
FBA problem again, we obtain a new target value. By repeating this
procedure for single and double perturbations we can compute all
synergy values. Synergies in MoMA are computed similarly: again, the inhibited reactions are constrained, but now the new
fluxes are determined by requiring 
a minimal flux change compared to the unperturbed flux. With  synergies   computed like this,
a double inhibition in a linear pathway has a simple effect: the
new pathway flux is the minimum of the two inhibited fluxes. Thus,
after a first inhibition, the second inhibition has either its full
effect or no effect at all. In both cases we obtain buffering
synergies.  In MCT, synergies can be  obtained directly from 
second-order elasticities, which  can be sampled by STM (see
Supplementary Materials).
Epistatis -- synergistic effects of gene deletions on cell fitness --
can be predicted similarly: we consider synergies referring to gene
knockouts as (large) perturbations and cell growth as the target
variable.  To obtain a good epistasis measure,
with comparable ranges for positive and negative epistatis from FBA
calculations, Segr\`e et al.~\cite{sdck:05} introduced a correction
for buffering interactions (see Supplementary
Materials).
With MCT, this is not necessary because positive and negative
synergies are already in similar ranges.

\co{target uea statt output?}

\section{Results}

\subsection{Structural thermokinetic modelling}

\myparagraph{\mysmallbreak Structural thermokinetic modelling and
  dependency schemas} Structural Thermokinetic Modelling (STM) is a
framework for building kinetic metabolic models with reversible rate
laws. It is simple, can flexibly integrate available data, resulting
in consistent models, and can be used for semi-automatic model
construction. In contrast to SKM, elasticities are not given by
saturation values, but they also depend on thermodynamic forces.
Elasticities can be computed from thermodynamic forces and saturation
values for a number of rate laws \cite{liuk:10}. STM is based on a
dependency schema, describing the dependencies between variables by
(linear or nonlinear) functions. Using Eq.~(\ref{eq:modelasticities}),
elasticities can be computed for a number of reversible rate laws. The
resulting elasticity matrices reflect network structure, fluxes,
thermodynamics, and enzyme saturation with reactants and regulators,
and full kinetic models can be reconstructed. Any choice of the basic
variables leads to consistent models, and any consistent model can be
obtained by a choice of the basic variables. The same schema can be
used for error propagation or for tracing small perturbations: in this
case, arrows in the schema correspond to ``connection'' matrices
(containing derivatives between model variables). It can also be used
to define probability distributions: by defining an probability
distribution of the independent variables, we obtain distributions and
correlations of all variables. In turn, any feasible distributions of
model variables can be defined by distributions of the basic
variables.

\myparagraph{Model construction} To construct metabolic states and
kinetic models, STM follows the dependency schema (Figure
\ref{fig:samplingscheme}): basic variables are freely chosen, while
derived variables are computed from them. In practice, choosing can
mean that variables are sampled, chosen manually, fitted to data, or
optimised (assuming a given metabolic optimality problem).  During
this stepwise model construction, various pieces of data can be
included. In the ``metabolic state'' phase, we choose
thermodynamically feasible fluxes, metabolite concentrations, and
equilibrium constants (where Gibbs free energies of formation or
independent $\keq$ values may serve as basic variables which determine
chemical potentials and thermodynamic forces).  The flux distribution
must be thermodynamically feasible: flux directions must follow the
thermodynamic forces, which depend on metabolite concentrations and
equilibrium constants.  In practice, equilibrium constants can be
obtained by eQuilibrator \cite{fnbm:12} and feasible metabolite
concentrations can be obtained by MDF \cite{nbfr:14}. All these
variables can be obtained by existing methods for thermodynamic flux
modelling. In the ``kinetics'' phase, saturation values are chosen in
the range $]0,1[$ (or possibly in a smaller range, or using a beta
distribution), and the elasticity matrix is computed using
Eq.~(\ref{eq:modelasticities}). The resulting elasticity matrix
corresponds to a consistent kinetic model with reversible rate laws,
which can be easily reconstructed.  From the elasticity matrix, we
also obtain the unscaled elasticities $\Eun^{v_{l}}_{c_{i}}$, the
Jacobian matrix and response or control matrices used in Metabolic
Control Theory. Using these matrices, we can study model properties
such dynamic stability, oscillations, linearised temporal dynamics, or
propagation of noise. Second-order elasticities and response or
control coefficients, describing synergies, can also be obtained.

\begin{figure*}[t!]
  \begin{adjustwidth}{-\extralength}{0cm}
 \begin{center}
{\footnotesize
  \begin{tabular}{l}
  Network structure \\[4mm]
  \includegraphics[height=8.8cm]{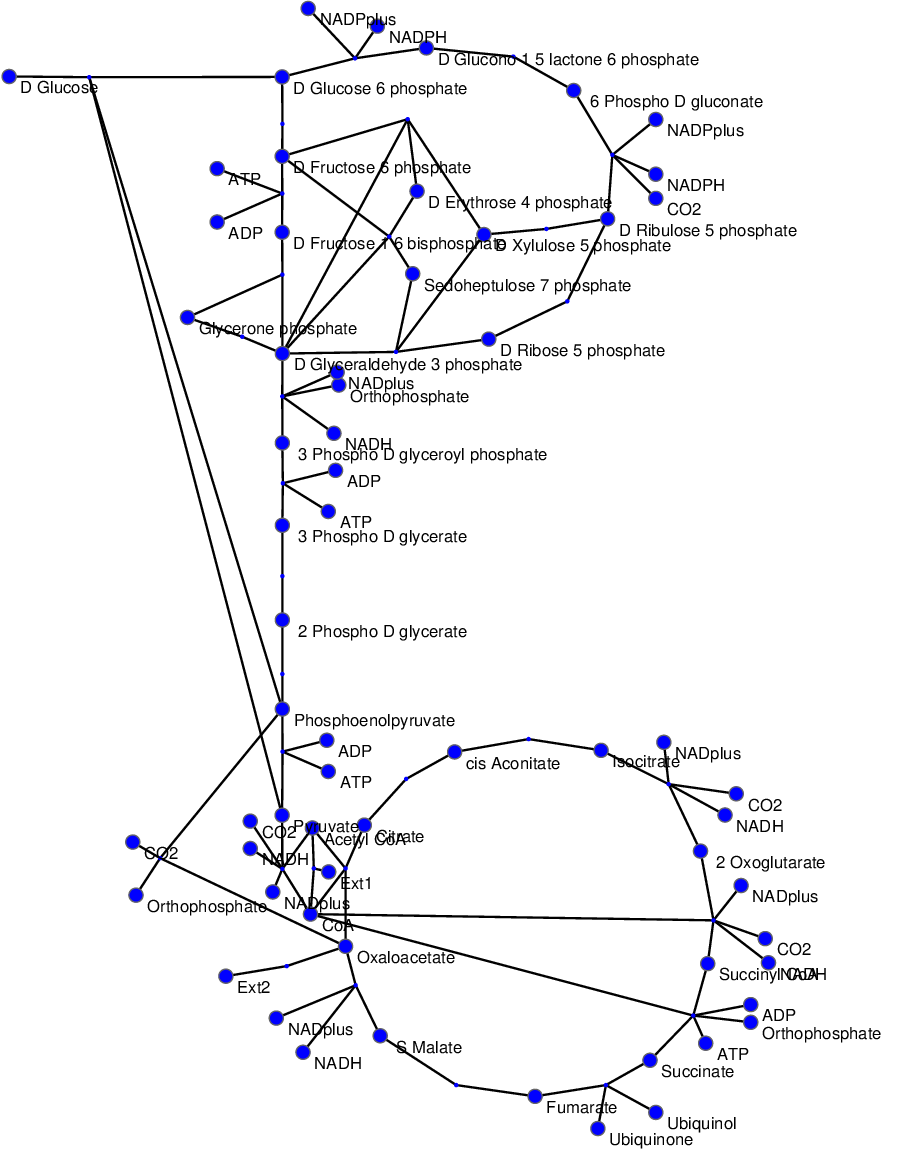}\\[4mm]
  \end{tabular}
  \begin{tabular}{ll}
 (a) Metabolic fluxes (mM/s) & 
 (b) log$_{10}$ Concentrations (mM)\\
 \includegraphics[height=5.0cm]{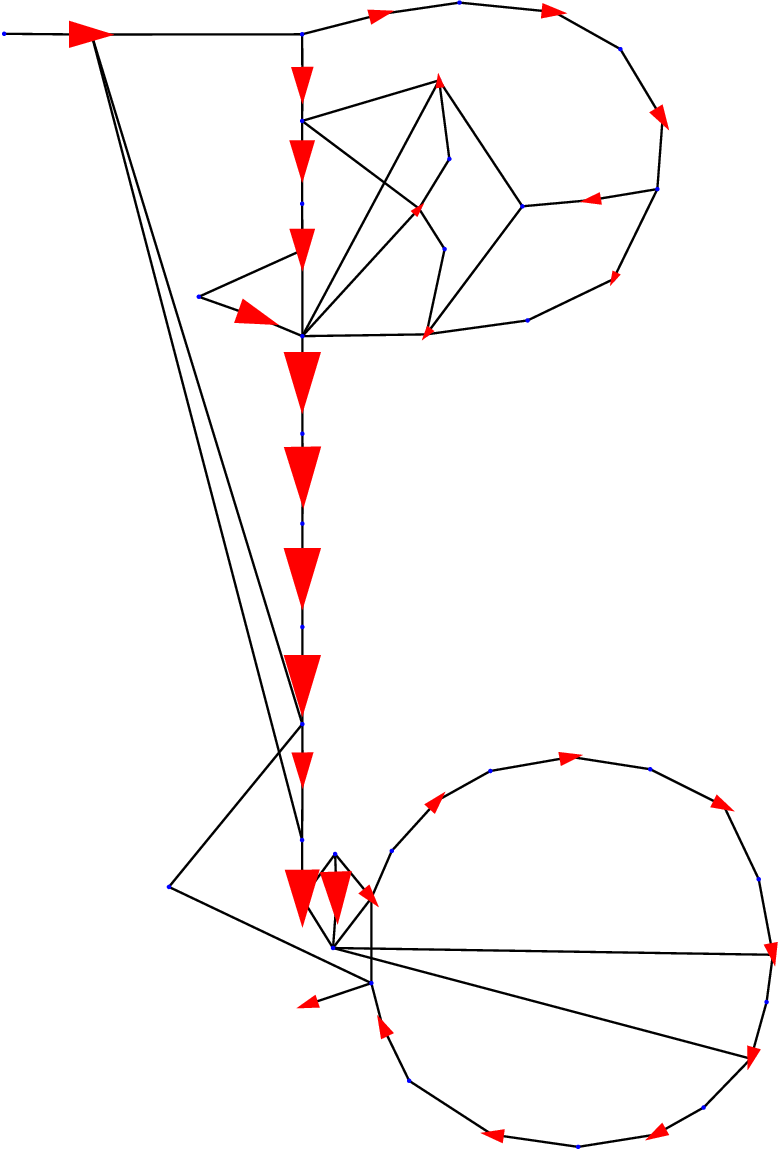}&
 \includegraphics[height=5.0cm]{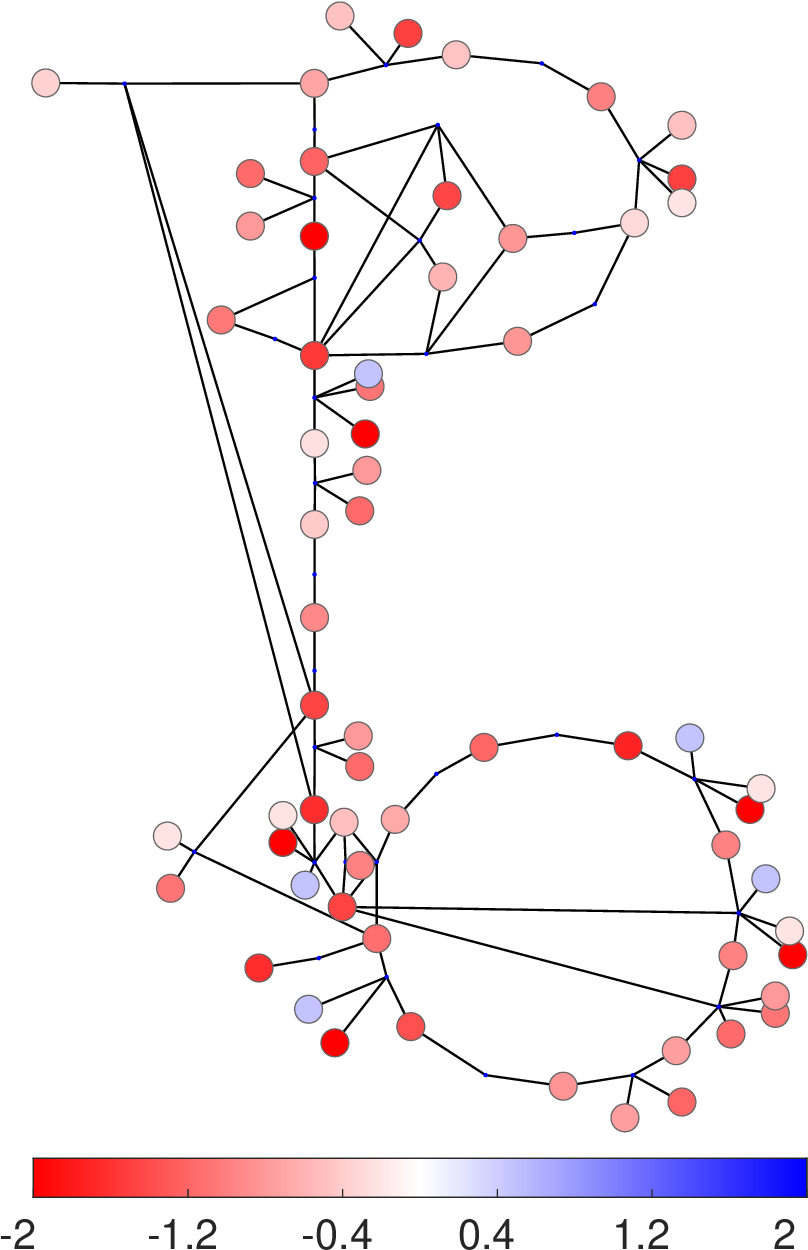}\\[2mm]
 (c) Chem.~pot.~[kJ/mol] & 
 (d) Thermodynamic forces [RT]\\
 \includegraphics[height=5.0cm]{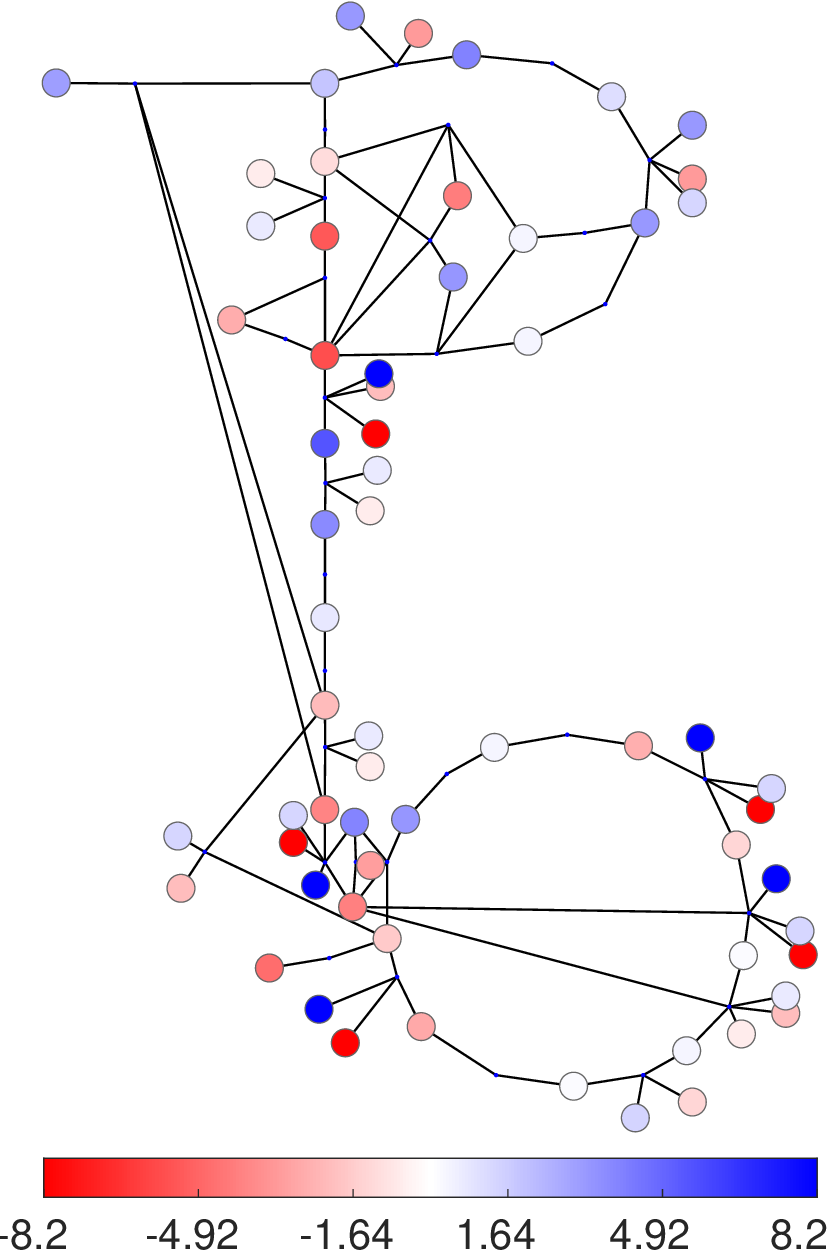}&
 \includegraphics[height=5.0cm]{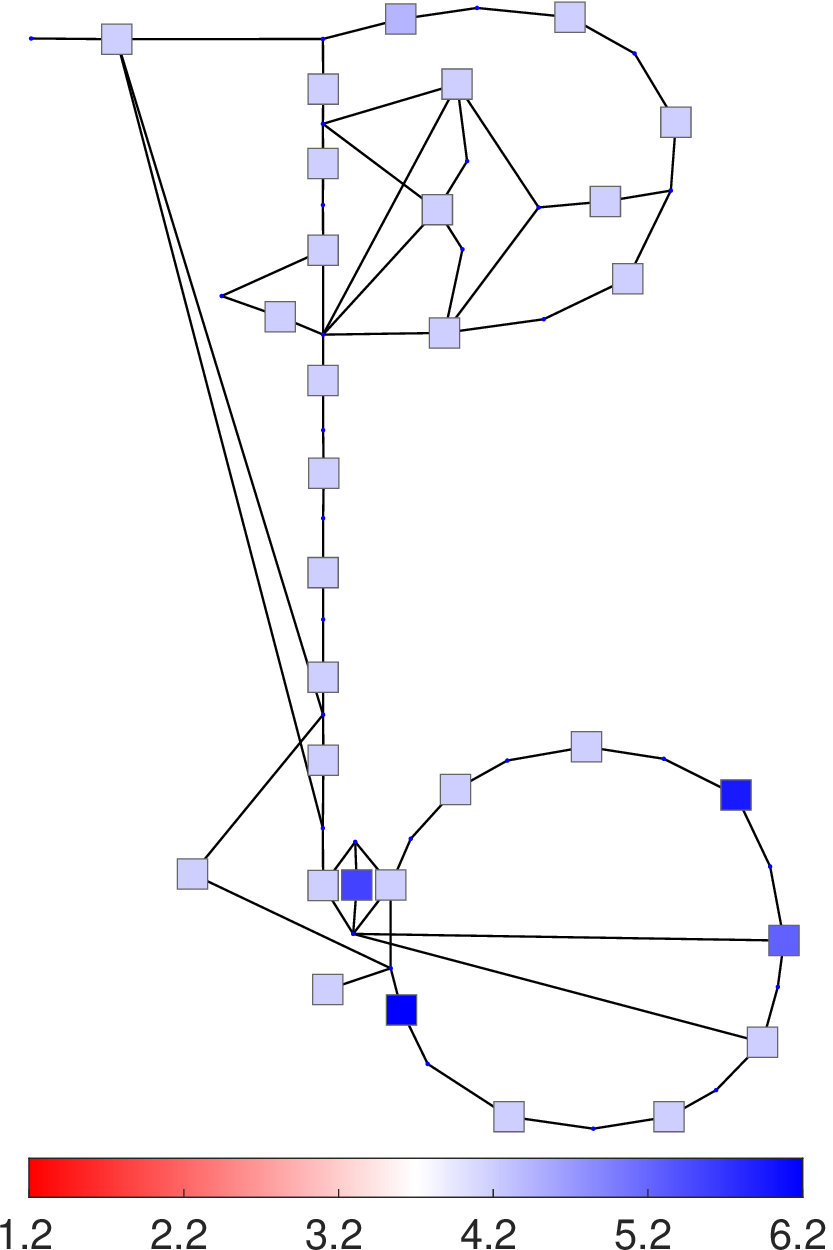}
\end{tabular}\ \\[2mm]
\begin{tabular}{llll}
 (e) Enzyme saturation & 
 (f) Reaction elasticities &
 (g) Flux control (on PFK)& 
 (h) Enzyme synergies (on PFK)\\[2mm]
 \includegraphics[height=5.0cm]{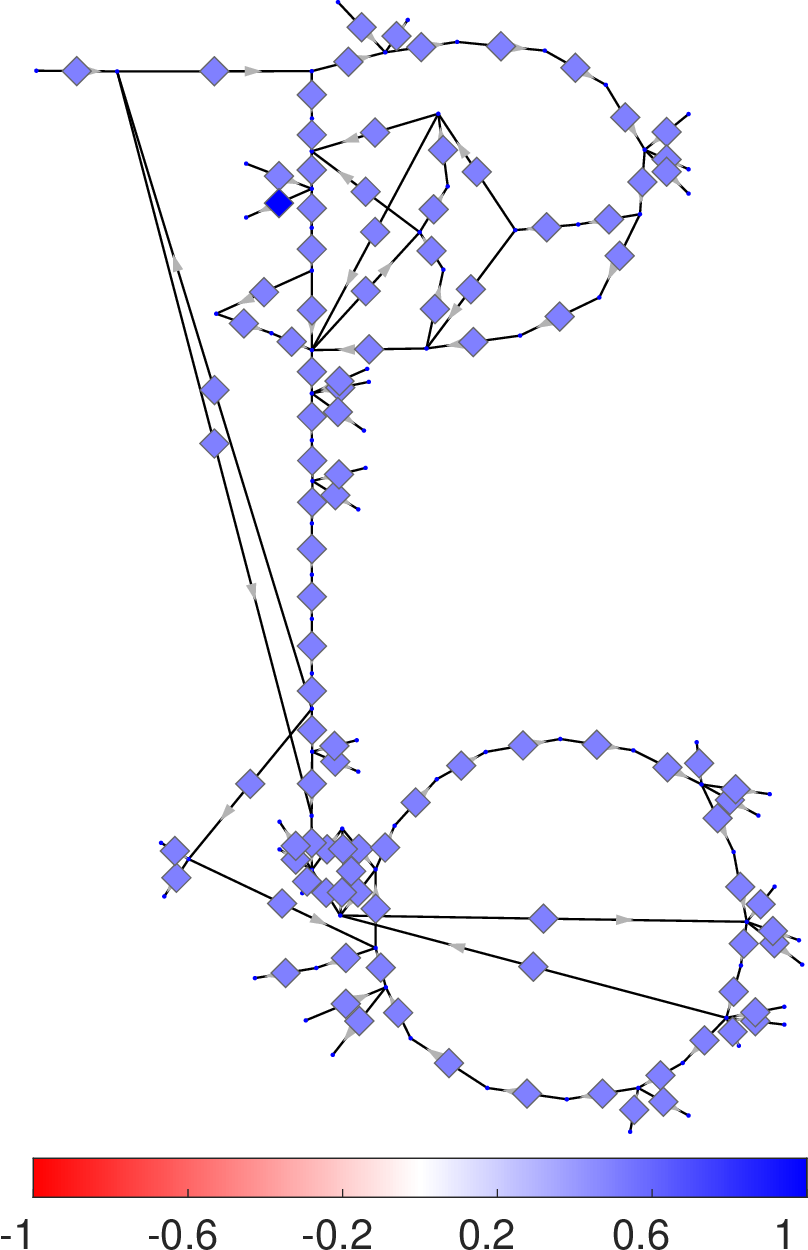}&
 \includegraphics[height=5.0cm]{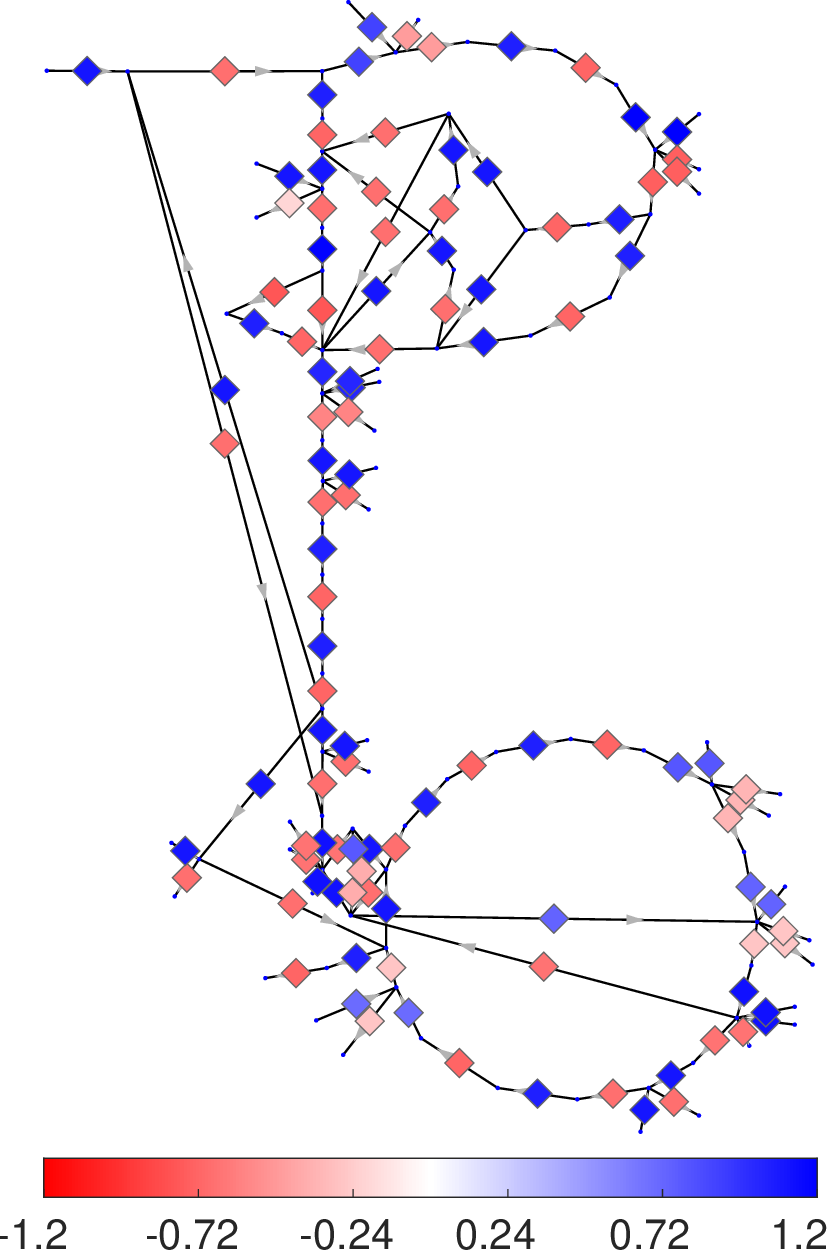}& 
 \includegraphics[height=5.0cm]{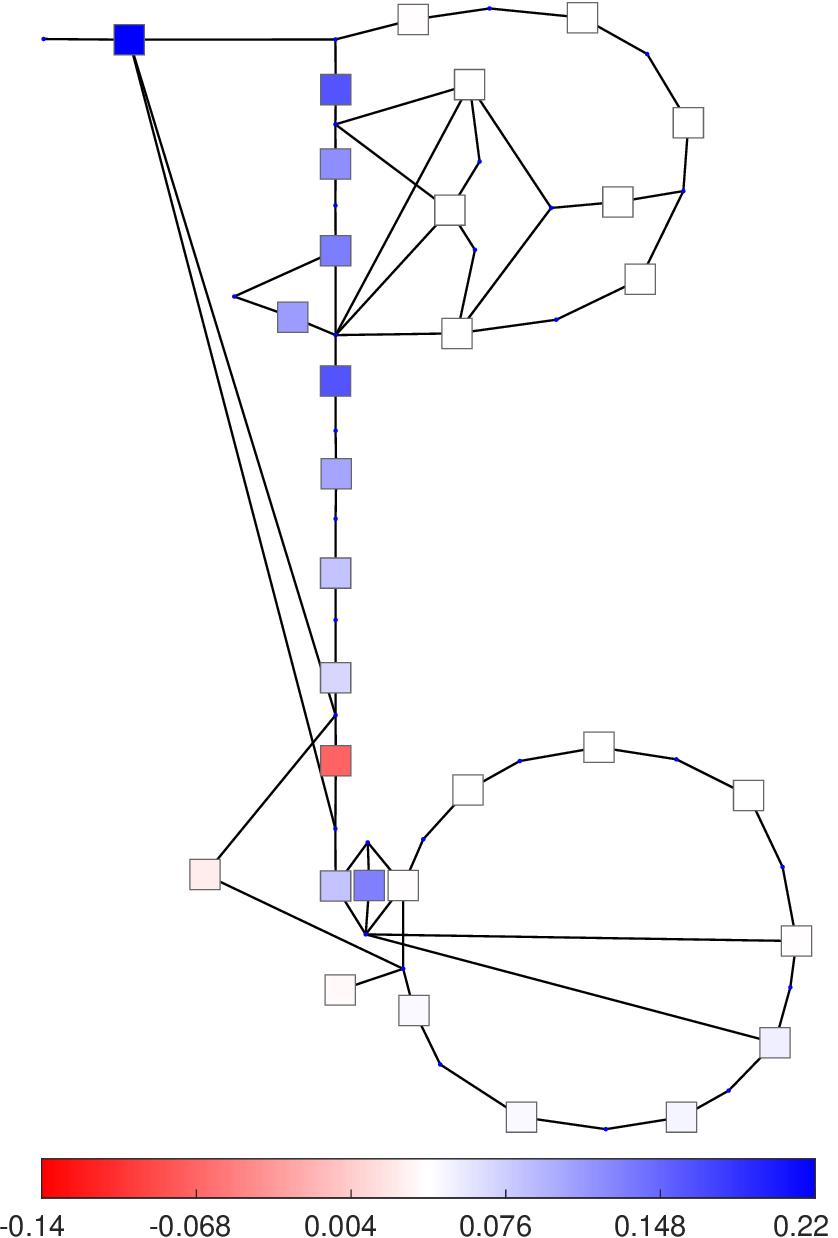}& 
 \includegraphics[height=5.0cm]{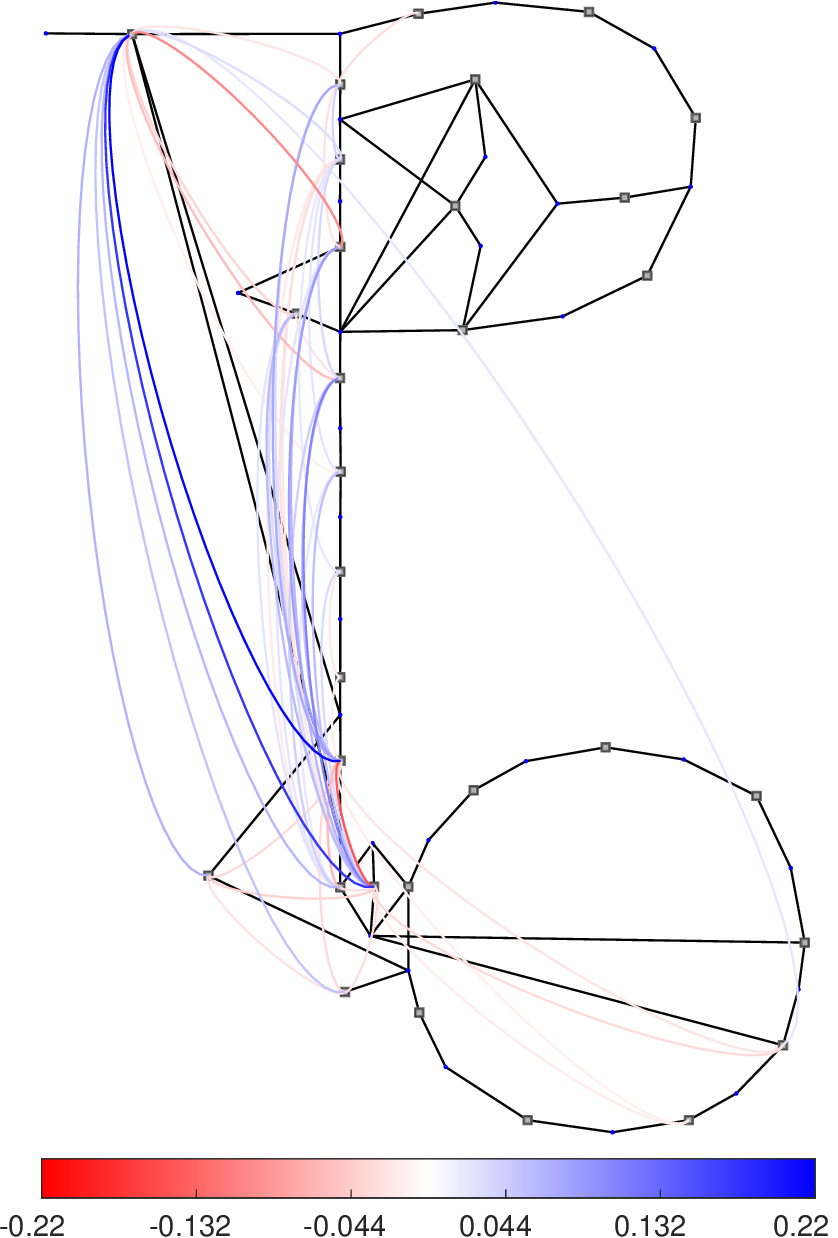}
\end{tabular}
}
 \end{center}
 \caption{\textbf{Systematic model construction by STM.}  Model of
   central carbon metabolism in \emph{Escherichia coli} (for model
   details, see \url{github.com/liebermeister/stm}). Network structure
   and fluxes (respiration on glucose) were taken from
   \cite{nfbd:16}. The panels show different types of variables
   obtained step by step during model construction following the
   schema in Fig.~\todo{2} (circles: metabolites; arrows and squares:
   reactions). By sampling the basic variables repeatedly and
   computing the others, a model ensemble can be constructed. (a)
   Thermodynamically feasible fluxes (grey arrows) obtained by flux
   minimisation (data from \cite{gbhb:10}).  (b) Metabolite
   log-concentrations. Metabolite concentrations and thermodynamic
   forces were determined by thermodynamic balancing \cite{luli:19} of
   measured metabolite and reaction Gibbs free energy data.  (c)
   Chemical potentials. (d) Thermodynamic forces in units of RT.  (e)
   Saturation values were set to the standard value of
   1/2. Alternatively, the saturation values could be determined from
   data or be sampled at random between 0 and 1. (f) Scaled reaction
   elasticities $\Esc_{c_{i}}^{v_{l}}$. (g) Scaled control
   coefficients $\hat(C)_{v_{l}}^{r_{\rm PFK}}$ for the flux in upper
   glycolysis (PFK reaction) as the target variable. (h) Enzyme
   synergies (scaled second-order control coefficients) for the
   glycolytic flux. Positive values are shown in blue, negative values
   in red, zero values in white. For clarity, only enzyme synergies in
   the outer 5 percent quantiles are shown.}
 \label{fig:ATPrephosphorylation}
  \end{adjustwidth}
\end{figure*}

\co{Infos zur grafik: modell: ecoli\_ccm; skript:
  ecoli\_ccm\_reference\_state; target reaction: PFK flux} \co{hier
  sind standard saturation value =1/2, oder? noch was zu gesampleten
  machen?} \co{neue extra-abbildung mit differential expression
  (protein level from data) + following metabolite and flux changes;
  sowie ext glucose change (and resulting metabolite and flux
  changes)} \co{note that energy dissipation could also be plotted!}
   
\myparagraph{Example: \emph{E.~coli} model} Figure
\ref{fig:ATPrephosphorylation} shows how a model of central metabolism
in \emph{Escherichia coli} can be constructed step by step.  In model,
a modified version of the model from \cite{nfbd:16}, exchange
reactions were added to some biomass precursors, and a stationary flux
distribution was obtained from measured metabolic fluxes by projecting
flux data onto the space of stationary fluxes while constraining the
flux directions. Metabolite concentrations and thermodynamic forces
for the reference state were determined by balancing \cite{lskl:10} of
metabolite and reaction Gibbs free energy data.  Data as in
\cite{nfbd:16} were used in all cases.  Model structure and data were
taken from \cite{nfbd:16} (for details, see
\url{github.com/liebermeister/stm}), and metabolite concentrations,
consistent with the flux direction, were determined by thermodynamic
balancing. The reference state satisfies all relevant constraints:
stationary fluxes, Wegscheider conditions for equilibrium constants
and thermodynamic forces, \todo{Haldane relationships for kinetic
  constants}, and consistent directions of fluxes and driving forces.
The figure shows saturation values assuming a half-saturation of
enzymes, i.e.~setting all saturation values to standard values of 1/2
(in an irreversible Michaelis-Menten kinetics, an enzyme works at its
half-maximal rate at a substrate concentration $c=\kM$, that is, at a
scaled substrate elasticity of 1/2). Alternatively, we can have inserted known
saturation values for model fitting (obtained from known $\kM$ values
(note that superscripts are used for convenience, as in $\kM$ instead
of $K_{\rm M}$) and metabolite concentrations), or we could have
sampled them in the range $[0,1[$, either uniformly or around
plausible values) for ensemble modelling.  \co{say: no small-molecule
  regulation was considered, but it could be included easily.}

 \co{parag davor: Proof:
  $\frac{\md}{\md x} \frac{1}{1+x} = \frac{1}{(1+x)^2}$. In the point
  $x=1$, the function value is 1/2 and the value of the derivative is
  1/4, so the scaled elasticity in this point is given by
  $\frac{1/4}{1/2} = 1/2$.} \co{clarify distinction between
  dissociation constant (thermodynamic, half-binding) and KM value
  (kinetic, half-maximal rate), which may also be affected by the
  speed of the conversion step (in the quasi-steady-state derivation
  of MM kinetics).} 

\myparagraph{Quality of a reconstructed kinetic model} Will models
constructed by STM be biologically plausible? To
test this, I started from a model of the threonine pathway in
\emph{E.~coli} \cite{crqf:01} (from BioModels Database
\cite{LeNov`ere2006a}) and constructed a twin version with the same
reference state and with elasticities set to 1/2.  To compare the two
models, I simulated a sudden increase of the external substrate
concentration (aspartate) (see Supplementary Figure
\ref{fig:dynamicsimulations}).  Despite the different kinetic constants,
the simulations show similar qualitative dynamics and even similar  time
scales. Hence, realistic fluxes and metabolite
concentrations, together with plausible assumptions about enzyme
saturation can  suffice to obtain realistic (unscaled) elasticities, and
hence a  plausible dynamic behaviour. \co{JA!  woher kommen die
  gleichgewichtskonstanten? note whether original kinetics were
  reversible.} \co{lieber noch samplen und baender plotten?}

\myparagraph{Model ensembles and usage of STM} Aside from constructing
a single model, we can build model ensembles with broad parameter
distributions. Model ensembles can be used to study the effects of
model structure, flux distribution, thermodynamics, or enzyme
saturation on metabolic behaviour, and to see which details have
significant effects on model outputs. For example, to study the role
of thermodynamic forces in flux control, we may generate model
ensembles with the same fluxes, but different forces, determine the
flux control coefficients, and check which control coefficients differ
significantly between the ensembles. To assess significance for a
single control coefficient, its distributions (arising from sampled
saturation values) are compared between the different ensembles
(representing different choices of forces). Of course, a model
contains many control coefficients (one between each reaction and each
target variable), and so the problem of multiple testing must be
addressed (see Methods and Supplementary Materials).
Ensemble models can answer a wide range of questions,
e.g.~how metabolic dynamics, homeostasis, and control depend on
network structure, thermodynamics, enzyme saturation, or
regulation. Even with little data, we can study control or synergy
coefficients, compute their distributions and correlations, and see
which control coefficients differ significantly from zero. All these
predictions are probabilistic, reflecting the uncertainties arising
due to missing or imprecise data. While model ensembles can be built
without any kinetic data at all, inserting data (e.g.~kinetic
constants) will decrease variability and make model results more
precise.  Hence, in STM, data that would usually not suffice for model
fitting can still be used for model predictions and to assess their
uncertainty ranges. The basic STM approach can be extended in various
ways: SI section \ref{Sec:SIextensions} describes ``biological''
extensions taking into account cell compartments, metabolite dilution
by cell growth, the treatment of thermodynamically infeasible fluxes,
avoiding divergencies close to chemical equilibrium, enzyme reactions
composed of elementary steps, multiple steady states, and an assumed
adaptation of enzyme levels, as well as ``statistical'' extensions
regarding prior distributions for saturation values, the analysis of
sampled target variables, significant differences between model
variants, and ways to choose the distributions of target variables.

\coout{Signifikanz erklaeren: gutes beispiel fuer signifikante
  unterschiede: branch point-beispiel fuer synergien. einfacher zu
  verstehen? // daran erklaeren: signifikante unterschiede zwischen
  modellvarianten // beweis von einfluessen einer bestimmten
  groesse. hier nur andeuten, genauer in methods/SI: signifikante
  unterschiede zwischen (i) strukturellen varianten (ii) varianten mit
  verschiedenen festen zahlen (iii) varianten mit verschiedenen
  verteilungen: Aufteilen von varianzen, erklaerung durch verschiedene
  teileffekte. // Beispiel: analysis of three glycolysis models -
  statistics of comparison: problem: zahlen sind evtl nicht sehr
  realistisch}

\myparagraph{Usage of STM} STM helps us build realistic metabolic
models, study their control properties, and assess how they vary. The
underlying dependency schema shows how model variables depend on each
other. Equipped with these methods, we can now study a general
question: the role of thermodynamic forces in metabolic dynamics and
control. In the following sections I show how different aspects of
metabolic dynamics -- including flux control, linearised metabolic
dynamics, enzyme synergies, and metabolic fluctuations -- are shaped
by the pattern of thermodynamic forces in the network.

\subsection{Metabolic effects of  gene expression changes}

\myparagraph{\mysmallbreak Enzyme changes and their effect on
  metabolite levels and fluxes} Expression changes of a single enzyme
can change the network-wide fluxes. In Flux Balance
Analysis (FBA), these effects have been modelled by changing the flux
bounds (as a proxy for changing enzyme activities) and re-optimising
the fluxes.  In MCT, in contrast, the effects of enzyme changes are
described by control and response coefficients, while synergy
coefficients can be used in addition for more precise approximations
\cite{likl:05} or to account for the adaptation of other enzymes
\cite{lksh:04}.
\co{FN: this is just one example: can also be inhibition or substrate
  changes} 

\co{JA!!} \co{ankuendigen: nur theoretische vorhersagen, aber mit 2,
 ordnung} \co{hier pictures with sampled CC in SI, ref to murabito,
 say that this works also for second order. dann erst das beispiel
 hier (neg or pos?) einfuehren} \co{sagen: cc berechnet (siehe abb;
 siehe abb mit verteilungen in SI; cite murabito // moegliche
 anwendung: vorhersage von flussaenderungen aus expressionsdaten in
 b-subtilis-modell: schauen, ob bscm-flussvorhersage relevant ist
 mehr daten? e-coli, heinemann + dans fba-daten? However, we can now
 study the effects of thermodynamic in more detail, and this can
 yield general insights}

\myparagraph{Control over metabolic objectives} If a target variable
(such as ATP or biomass production) contributes to cell fitness,
enzymes should have a positive influence on this variable: if an
enzyme had a negative influence, the cell would benefit from
downregulating this enzyme \cite{lieb:18theory}, and may have done
this already!  In enzyme-optimal states \cite{klhe:99,lieb:14a}, the
marginal cost and benefit of each enzyme must be balanced, so each
enzyme must have a positive control over the metabolic objective,
i.e.~a positive marginal benefit \cite{lieb:14a}.  Likewise, in FBA
models (and assuming flux bounds proportional to enzyme levels), an
enzyme knock-down can decrease the metabolic benefit, but can never
increase it (otherwise FBA would have chosen a smaller flux from the
start). But this holds only if we assume an optimal state. Generally,
without optimality assumption, enzymes may have positive or negative
control over different fluxes: for example, inducing some enzymes may
reduce biomass production.  To test whether enzymes are likely to have
a negative flux control on a flux objective, I built a series of
models describing the flux distributions in central metabolism of
human hepatocytes. Starting from the large Hepatonet1 network, sparse
flux distributions for specific objectives (for example, ATP
regeneration during aerobic growth on glucose) were determined by FBA
with flux minimisation. In the original paper \cite{holz:04}, for each
of these flux distributions, a network model was build by omitting
reactions with inactive fluxes, and applying STM like with the
\emph{E.~coli} model above. Metabolite concentrations were determined
by thermodynamic balancing of measured concentrations in
\emph{E.~coli} (as a substitute for human hepatocyte data). using these flux distributions, I obtained models referring to
a large number of different flux objectives.  For example, with ATP
production as the flux objective (details in SI section
\ref{sec:SIhepatocyteSI}), all active enzymes have a positive control
over ATP production. This was a typical case: usually, most of the
active enzymes had a positive control on the flux objective, even if
the models were not constructed to be in enzyme-optimal
states. Apparently, suppressing unnecessary fluxes (in the FBA step,
assuming a flux objective) already led to a state in which -- once
kinetics are considered -- most enzymes have a positive control on the
flux target. While this makes sense intuitively, it provide strong
support for FBA: it shows that even if FBA ignores kinetics, it
provides a good starting point for kinetic models, providing fluxes
that are likely to support enzyme-optimal states.

 \co{e
  coli data were used for concentrations // lieber konzentrationen aus
  krebszellen?  // neues rabinovitz-paper als argument fuer metabkonz
  aus anderen organismen}

\begin{figure}[t!]
\begin{center}
    \begin{tabular}{lll}
      (a) Flux after enzyme upshift & (b) Enzyme-flux response  & (c) Scaling of forces   \\[2mm]
      \includegraphics[height=3.3cm]{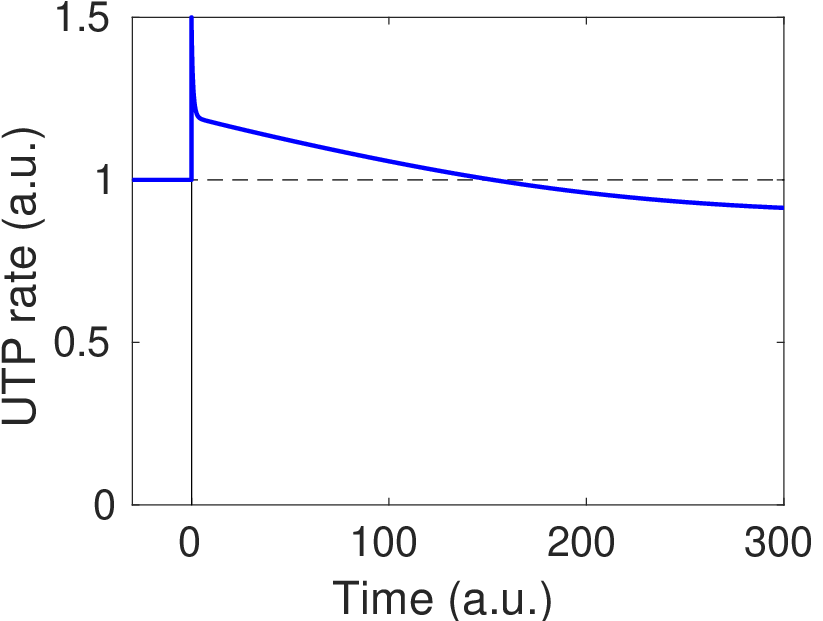}&
      \includegraphics[height=3.3cm]{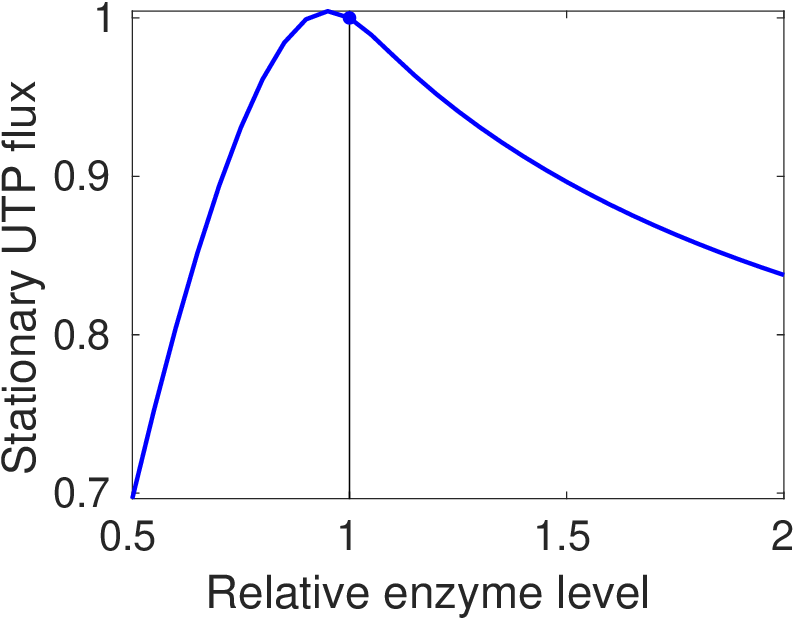}&
      \includegraphics[height=3.3cm]{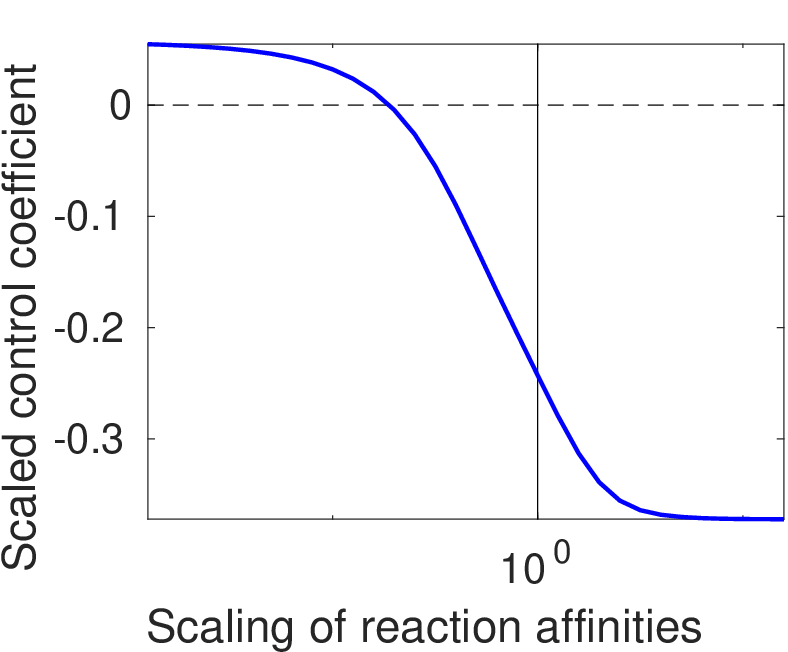}
    \end{tabular}
  \end{center}
  \caption{\textbf{Paradoxical effect of flux control by enzymes: a
      model of UTP rephosphorylation in human hepatocytes.}  Fluxes in
    central metabolism were obtained in \cite{gbhb:10} by FBA with UTP
    production as the objective.  Flux control coefficients (blue:
    positive; pink: negative) obtained by STM with standard
    elasticities, i.e.~saturation values set to 0.5.  Paradoxically,
    the UTP-regenerating enzyme NDK has a negative influence on its
    own flux. (a) Metabolic dynamics after an NDK upshift. A
    higher NDK activity first speeds up the reaction, but later the
    rate drops below its initial value. (b) Dose-response curve
    between NDK amount and steady-state UTP production.  In the
    reference state (vertical line), the curve slope (response
    coefficient) is negative: an increase of the enzyme decreases the
    flux.  (c) The response coefficient depends on thermodynamic
    forces. To see this, all thermodynamic forces were increased by a
    common scaling factor (x-axis): at higher thermodynamic forces,
    the response (y-axis) becomes more negative.}
 \label{fig:counterproductive}
\end{figure}

 \coout{abbildung a (fluesse) geloescht, weil bei
      nicht-doppelten flusspfeilen fehler in der grafik auftreten; es
      ist ein problem in netgraph\_draw mit option single\_arrow bei
      diesem modell (andere gehen); ich hab lange gesucht und das
      problem nicht gefunden (a) UTP-generating fluxes in human
      hepatocytes.}
    
 \co{parag below: FN: remember that CJ = I-E (NEL)inv N; I in the diagonal is
  expected to be bigger than the second term; or the diagonal in the
  second term should be positive. why? // what's the intuition? what
  does this mean for their value?}
    
\myparagraph{Paradoxical self-inhibition effect in UTP regeneration}
While enzymes can have a negative flux control (e.g.~a control on
reactions that compete with the enzyme for substrate), it is hard to
imagine that they exert a negative control on their own catalysed
flux.  But there are examples of this.  Figure
\ref{fig:counterproductive} (a) shows results from a variant of our
hepatocyte model, but with UTP production during anaerobic growth on
glucose as the objective: UTP is regenerated from UDP via the
phosphotransferase reaction UDP + ATP $\leftrightarrow$ UTP + ADP,
catalysed by nucleoside diphosphate kinase (NDK).  In a kinetic model
(constructed by STM and assuming half-saturated enzymes), the
influence of NDK on its own steady-state flux is described by the flux
control coefficient $\Cun^{\rm v_{UTP}}_{\rm e_{UTP}}$. Surprisingly,
this control is negative: a higher enzyme level decreases the flux!
This paradoxical effect also shows up in dynamic simulations: when the
enzyme level increases, the flux increases as well, but then drops
below the original flux. If we plot the steady-state flux against the
enzyme concentration, the slope of the curve -- i.e.~the response
coefficient $\Run^{\rm v_{UTP}}_{\rm e_{UTP}}$ -- is negative in the
reference state. Probably, the reason is that UTP production consumes
ATP; due to the turbo design of glycolysis \cite{twdw:98}, a high ATP
consumption reduces the ATP level drastically, entailing a decrease in
UTP rephosphorylation. In contrast, a lower NDK level allows ATP to
recover, and UTP rephosphorylation increases. Is this pardoxical
behaviour typical or a rare exception, maybe caused by our precisely
half-saturated enzymes? To see this, we can use STM: with saturation
values random drawn between 0 and 1, about 90 percent of the models
show the paradoxical self-repression effect. Hence, this effect does
not depend on fine-tuned parameters, but is made very likely by
network structure, flux distribution, and metabolite
concentrations. Next, to assess the role of the thermodynamic forces,
I varied them proportionally (see Supplementary
Information): 
\co{by adjusting metabolite levels, see SI (kam schon ein stueck
  weiter oben?)}\co{sagen: hypothetical. groessere auessere pot
  differenz wird einfach gleichmaessig verteilt.} 
when bringing all
reactions closer to equilibrium, the effective self-inhibition
stopped, while increasing the driving force increased
self-repression. \co{erklaerung?}  \co{betonen (hier oder auch am
  anfange des abschnitts, oder in diskussion?): STM kann uns helfen,
  die URSACHE eines verhaltens zu verstehen und zu sehen, ob es
  (gegeben netzwerk und fluesse) ein seltener oder ein
  wahrscheinlicher fall ist, und von welchen kenngroesssen (netzwerk,
  fluesse, konz, kraefte, elast) es bestimmt wird, und in welchem
  mass} \co{what would the self-inhibition mean for a meaningful
  (feedback) regulation of this enzyme?}

\subsection{Enzyme synergies and epistasis}

 \co{FN: In FBA, the target must be a flux, for example the
  flux that is also optimised}  \co{also use
   FBA with molecular crowding!}
 
\begin{figure}[t!]
\begin{center}
\parbox{13cm}{(a) Network model \\[2mm]
 \includegraphics[width=6cm]{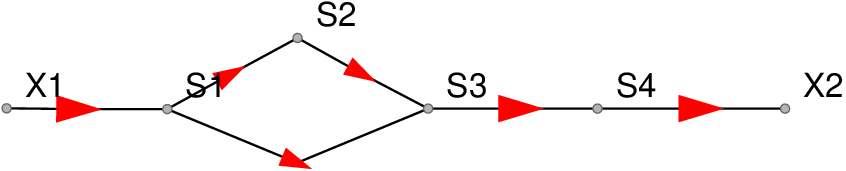}\\[4mm]
 \begin{tabular}{lll}
 (b) FBA & 
 (c) MoMA &
 (d) Metabolic control theory\\[2mm]
 \includegraphics[width=4cm]{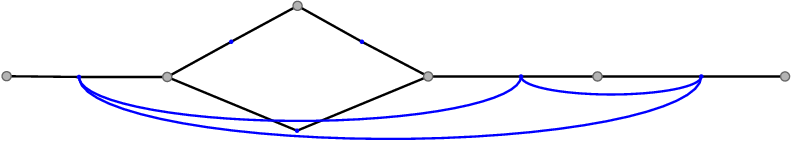}&
 \includegraphics[width=4cm]{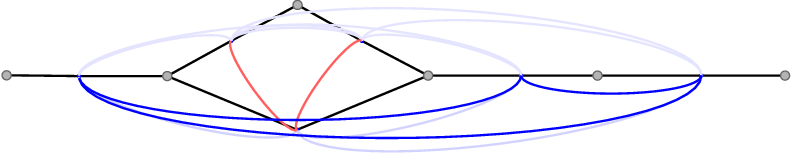}&
 \includegraphics[width=4cm]{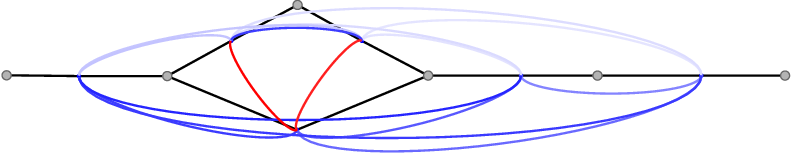}\\
 \includegraphics[width=4cm]{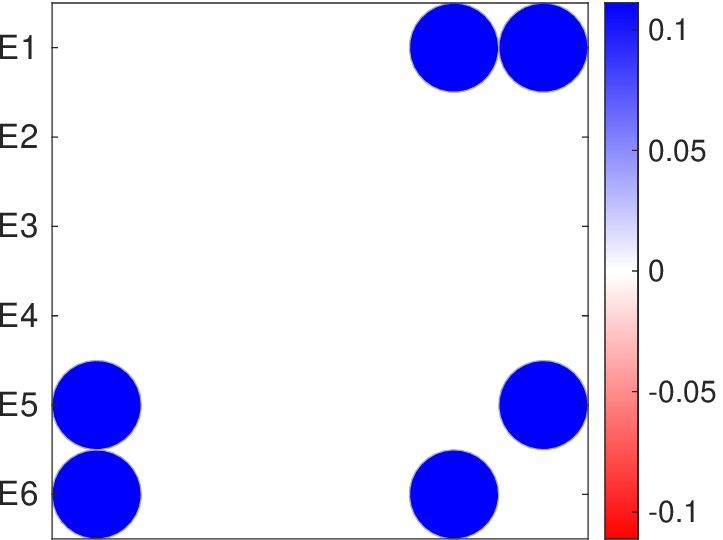}&
 \includegraphics[width=4cm]{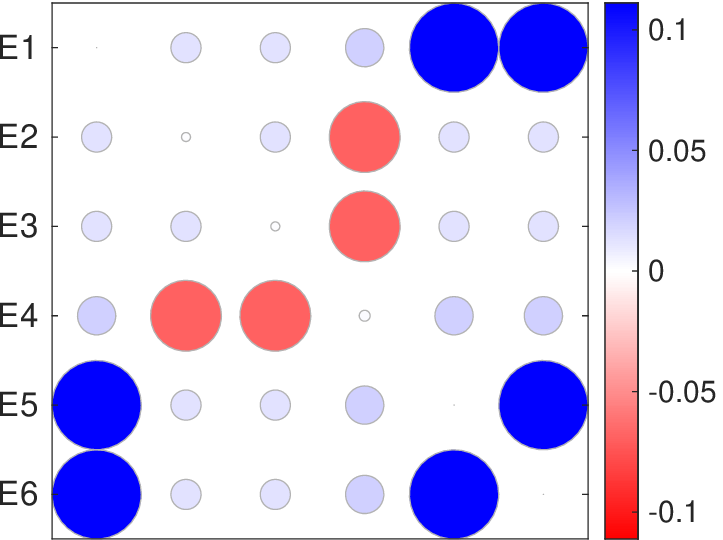}& 
 \includegraphics[width=4cm]{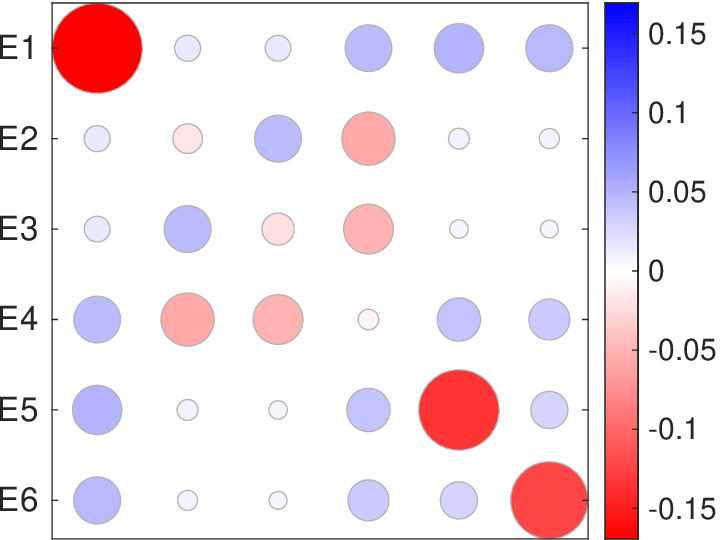}\\
 \end{tabular}
 }
\end{center}
\caption{\co{warum keine negativen synergien FBA? weil bei dem
    gewaehlten szenario (vorgeben einer flussverteilung, dann
    beschraenken einzelner fluesse)die beiden parallelfluesse
    unabhaengig sind!}   \textbf{Enzyme synergies in a simple pathway.} \co{auch
    mcFBA zeigen! damit vier beispiele!}  \co{calculation could also
    be done with EMC2-model alone. here the results: all results are
    roughly the same, ie thermodynamics is most important, saturation
    less so.} Enzyme synergies in a linear example pathway with
  alternative routes. The (scaled) synergies towards a
  fitness-relevant (e.g.~biomass production) objective indicate
  epistasis. (a) Network structure. A conversion from S$_{1}$ to
  S$_{3}$ can occur directly or via the intermediate S$_{2}$. The
  following plots show synergies for double inhibitions predicted by
  FBA (b) or MoMA (c). To simulate enzyme inhibitions, a flux decrease
  to 90 percent of the orginal value was imposed (by setting a bound
  at 0.9 times the previous flux). The double inhibition of an enzyme
  is simulated by applying the inhibition twice, i.e.~leading to a
  relative flux decrease factor of 0.81. Synergies are shown by arc
  colours (red: aggravating, blue: buffering) and as a matrix.  Colour
  ranges span the observed synergies in each panel (red: negative;
  white: zero; blue: positive). Small values (below one percent of the
  maximal absolute value) are not shown. (d) Synergies computed by
  Metabolic Control Theory, assuming common modular rate laws and
  half-saturated enzymes.  With saturation values sampled from a
  uniform distribution, the ensemble mean yields almost the same
  results. A different rate law (the simultaneous binding modular rate
  law) yields very similar results (see Supplementary
  Materials). 
  Compare Figure 3 for synergies in \emph{E.~coli}.}
 \label{fig:FBAsynergiesExample} 
\end{figure}

\myparagraph{\mysmallbreak Constraint-based models and kinetic models
  yield similar synergy patterns} Interactive effects between enzyme
concentrations or other parameters on target variables are called
synergies.  In STM, synergies are obtained from the second-order
response coefficients, also  called synergy coefficients (see Figure 3(h) for
 the \emph{E.~coli} model).  While synergies can be
computed for any perturbation parameters, including concentration
variations in the growth medium, let us   focus on 
enzyme pairs. How do  STM  predictions compare to predictions
by FBA? Since synergy patterns tend to reflect
network structure and flux distribution, \todo{there are similar tendencies}:
Figure \ref{fig:FBAsynergiesExample} shows a comparison between STM
and two types of constraint-based models: classical FBA and MoMA (see Methods). In the
example, synergies computed by  MCT are distributed more uniformly than those
predicted by FBA, while MoMA ranges in between. The synergies
predicted by MCT confirm our expectations (see Supplementary
Materials):
cooperating enzymes (e.g.~in the same metabolic pathway) show
buffering synergies (inhibiting one of them impairs the pathway
function, and inhibiting the second enzyme has less extra effect),
while enzymes in alternative pathways show aggravating synergies
(because an inhibited pathway can still be bypassed, and only the two
inhibitions together take effect).

\co{NOCH ZU ZEIGEN! Woher kommt das ergebnis
 hier? for fba, segre et al showed that .. CITE! (iii) synergies between different pathways tend to be
 monochromatic, i.e.~typically, the synergy between two enzymes does
 not depend on their exact position in the network, but only on the
 pathways in  which they are located: there are pairs of pathways whose
 (inter-pathway) synergies tend to be positive, and others whose
 (inter-pathway) synergies tend to be negative; But these are just tendencies. The
 quantitative synergies depend, of course, on rate laws, enzyme
 saturation, and regulation assumed in the model and both the
 tendencies and the results for specific model instances can be
 studied by STM.}

\coout{fig: Schematische Erklaerung fuer epistase
  zeigen?}

\co{JA! EINFACH SCHREIBEN! Eigenschaften von Synergien: allgemeine ergebnisse .. abb mit
 grossem ecoli-modell: synergy statistics; monochromaticity!!
 enzymes with
 large influences on a target flux also tend to show larger (positive
 or negative) synergies.

 The formulae suggest:
 
 1. Enzyme mit hoeherer (absoluter) kontrolle haben hoehere
 (absolute) interaktionen bzw hoeheren synergy degree; sichtbar in rechnungen?;
 (WAR AUCH MAL sichtbar fuer grosses
 e-coli-netzwerk) insbesondere: synergy vs CC-produkt koennte
 informative sein!

 2. signifikante zyklen in synergien! signifikante verteilung der vorzeichen!

 3. synergien in untersystemen zeigen monochromatisches verhalten
 (WAR SICHTBAR IM grossen e-coli-netz)

 Am Rande: Interaktionen in hefe koennten theoretisch mit brendas
 daten verglichen werden dazu brauechte ich das gene mapping (hab
 ich fuer jol teilweise ..)

 \co{Yeast epistasis}
 \co{etwas zu dreier-epistase sagen?}
}

\co{NAECHSTER Parag: how does STM help us understand these patterns?} \co{mention
  beweis: flux control matrix has symmetric signs (in additional
  material) under what assumptions? noetig wofuer?}

\myparagraph{Mathematical explanations for MCT synergy patterns}
In
constraint-based models, there is no direct formula for synergies:
they need to be computed numerically one by one.  The formulae of MCT, in contrast,
show how synergies reflect network
structure and flux directions (which define ``upstream'' and
``downstream'' enzymes). The synergy coefficients depend on three
factors (formulae in  Supplementary Materials):
on flux control coefficients $\Cun^{j}_{v}$ between the two catalysed
reactions; on control coefficients $\Cun^{y}_{u}$ between these
reactions and the target variables; and on second-order elasticities
in the entire network. The formula suggests that a higher first-order
control between perturbed reactions and target variable, as well as a
large control between the reactions, increase the (positive or
negative) synergies.  Enzyme synergies, computed by MCT, depend on
which enzymes are perturbed, but not on how the perturbation happens:
there is no difference between enzyme knock-outs, enzyme-inhibiting
drugs, or other perturbations that decrease enzyme
activities. Therefore, the synergies for a double knock-down, a double
enzyme inhibition, and a combination of knock-down and enzyme
inhibition.  This is in line with experiments: genes with many
epistasis partners are also more likely to show gene-drug interactions
\cite{cbbk:10}. In reality, cells may compensate enzyme perturbations
by global expression changes, which changes the overall
synergies. This effect was not considered here, but there are ways to
correct our synergies based on predicted optimal enzyme adjustments
\cite{lksh:04}. 
Importantly, MCT defines synergies not only between enzyme levels and
metabolic fluxes, but between any pairs of model parameters (including
external metabolite concentrations) and with respect to any target
variable (including metabolic fluxes and concentrations).  Enzyme
synergies are just a specific example.

 \co{details about optimal enzyme adaptation? say that an adjustment of other enzyme
  levels can be taken into account and will change the synergies (just
  like it changes the effects of a single enzyme inhibition)}

\subsection{Uncertain  states and metabolic fluctuations}
\label{sec: fluctuations}

\co{den abschnitt sortieren:
  
1  static and dynamic trennen

2  white noise (eg chemical) and ``slow'' noise (e.g. enzyme levels) trennen

3  noise generation and noise propagation trennen

ENDE, extra, kurz:  knowledge vs actual distribution 
}

\co{MAIN\\
  o matrix ${\mathcal S}$, diagonal for independent flutuations, frequency-dependent\\
  o uncorrelated white noise:  ${\mathcal S}(\omega)=\Imat$\\
  o propagation   ${\mathcal S}_{c}(\omega)=R_{p}^{c}\,{\mathcal S}(\omega)_{p}\,{R_{p}^{c}}^{\dag}$\\
  spectral density for single concentration $c$ in diagonal element; time correlations between $c$'s in (complex valued) off-diagonal elements\\
  o descrinbing the contribution of a single frequency nois component on an infinite time interval

  SI\\
  But what does this mean for measurable noise and uncertainties?\\
  IN practice, ``noise'' manifests itself as variation on a particular time scale of interest. For instance, if we consider metabolite flutuations on the time scale of second, we assume a measurement device that averages faster noise (below the second range) and whose output will vary on a slower time scale (while some variability is not even noted bcs of the finite duration of time series measured.)
}

  \co{naechster parag: besser statisch und dynamisch von anfang an als zwei
  faelle (zb zwischen zellen, in einer zelle) und dann sagen, wie sie
  zusammenhaengen}

\myparagraph{\mysmallbreak Varying metabolite concentrations} Model
ensembles can help us explore uncertainty in a model (due to missing
or uncertain data), variability between cells in a population, and
fluctuations in time (e.g.~of metabolite concentrations inside a
cell). Notably, cell variables also co-vary, and their variability and
correlations are shaped by network structure and enzyme kinetics.  We
can imagine this easily. If metabolite concentrations and fluxes vary
in time, we can describe this by random fluctuations that propagate in
the network. Variability and uncertainties \co{explain words} in
a cell can emerge from variable enzyme concentrations, external
perturbations or chemical noise.  All these fluctuations percolate
through the network, add up, and cause variability and fluctuations in
metabolite concentrations and fluxes.  If variations are slow and
differ between cells, we can describe the cells by a random ensemble
of steady states.  All sources of noise have their own
dynamics. Fluctuating enzyme concentrations, for example, can be
caused intrinsically (by noise in transcription and translation) or
extrinsically (by variability in other cell variables)
\cite{elss:02,swes:02}. They show typical frequencies on the timescale
of protein dynamics (protein degradation or dilution in growing cells)
or slower. Chemical noise arises from the stochastic rates of single
reactions (because single reaction events occur randomly).  \co{JA in
  paper klar sagen: enzyme fluct + chemical noise sind zwei beispiel
  fuer noise; noise betrifft hier reaktionsrate und kann durch
  verschiedene einfluesse entstehen (ext met, enz act, chemical nois);
  spektrum reicht von schnellen dynamischen bis statischen flukt;
  behandlung mathematisch immer gleich}

\myparagraph{Mathematical description} \co{also refs to SI} How can 
 the variability in cell variables be described, and how are correlations
and temporal fluctuations  shaped by network structure, reference
fluxes, and thermodynamic forces?  To describe a cell population, we
may use the same model for all cells, but with differences in some  input variables that
determine the steady states  (e.g.~variability in the
enzyme concentrations $e_{l}$ with a covariance matrix
$\Sigmamat_{\ev}$). In a
first-order approximation, the covariance matrix of  metabolite
concentrations  is given by
$\Sigmamat_{\cv} =
\Rmat_{\ev}^{\cv}\,\Sigmamat_{\ev}\,{\Rmat_{\ev}^{\cv}}\trans$, with the  response coefficient matrix
$\Rmatun^{\cv}_{\ev}$. A more
precise second-order approximation requires synergy coefficients
\cite{likl:05}. Similar  formulae hold for all perturbation
parameters (e.g.~external metabolite concentrations, kinetic
constants),  fluxes, and even  dynamic
behaviour. \co{ref roberto} The same covariance formula can also be used to describe
subjective uncertainty (e.g.~about predicted steady states, based on
uncertain model parameters) \cite{likl:05}.  Fluctuations in time can
be described similarly \cite{lieb:2005}.  Fluctuating parameters and
variables are characterised by their spectral power density matrices,
which resemble covariance matrices, but are 
complex-valued (Hermitian instead of symmetric) and frequency-dependent. \co{ext conc may
  fluctuate in the environment (freq spectrum?)}  \co{erst staic case,
  cov - cov beschreiben. forml, bsp met in(enz and met ext ); dann
  dynamic} If the spectral densities of fluctuating parameters are
known, the spectral densities of state variables can be computed in a linear approximation, like
for static perturbations, assuming small noise amplitudes (see Supplementary Materials).
The effects of enzyme
fluctuations depend on the frequency: if enzyme fluctuations are
slow, their effects on dynamics can be modelled as quasi-static,
creating permanent differences between cells.

\co{es gibt bilder - nicht zeigen} \co{The random distributions of
  steady-state fluxes is described similarly, with cov ..}

\co{A spectral density matrix ${\mathcal S}(\omega)$, describes the
  variance of Fourier-transformed noise, at frequency $\omega$. For
  variables with independent fluctuations, it is a diagonal matrix,
  and for uncorrelated white noise it is given by
  ${\mathcal S}(\omega)=\Imat$. Chemical noise
  ${\mathcal S}(\omega)\sim\diag(\sqrt(|\vv|))$.

  Spectral density matrix of propagated noise (coming from parameters with spectral density
  ${\mathcal S}_{\pv}$) ${\mathcal S}_{\cv} =
  \Rmat_{\pv}^{\cv}\,{\mathcal S}_{\pv}\,{\Rmat_{\pv}^{\cv}}\trans$
  (Hermitian).

  Propagation: spectral density for each $c_{i}$ (or other output variable of interest) in (real-valued) diagonal element of ${\mathcal S}_{\cv}$, correlations (including phase correlations)
  in off-diagonal elements. The spectral density describes the contribution of a single frequency on an infinite time interval.
}

\begin{figure*}[t!] 
 \begin{adjustwidth}{-\extralength}{0cm}
 \begin{center} 
 \begin{tabular}{lll}
 (a) Control and correlation & 
 (b) Static variability &
 (c) Static covariances \\
 \includegraphics[height=5.8cm]{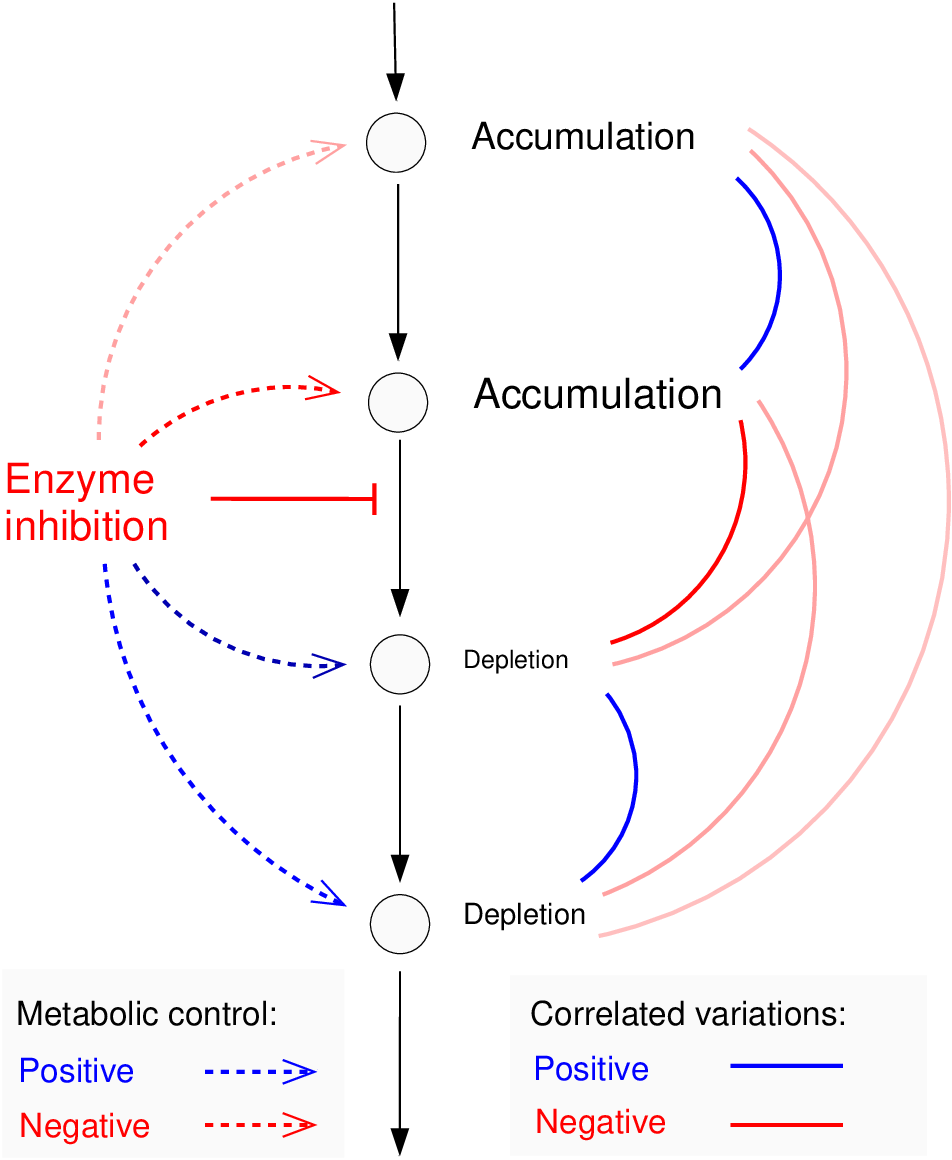}&
 \includegraphics[height=5.8cm]{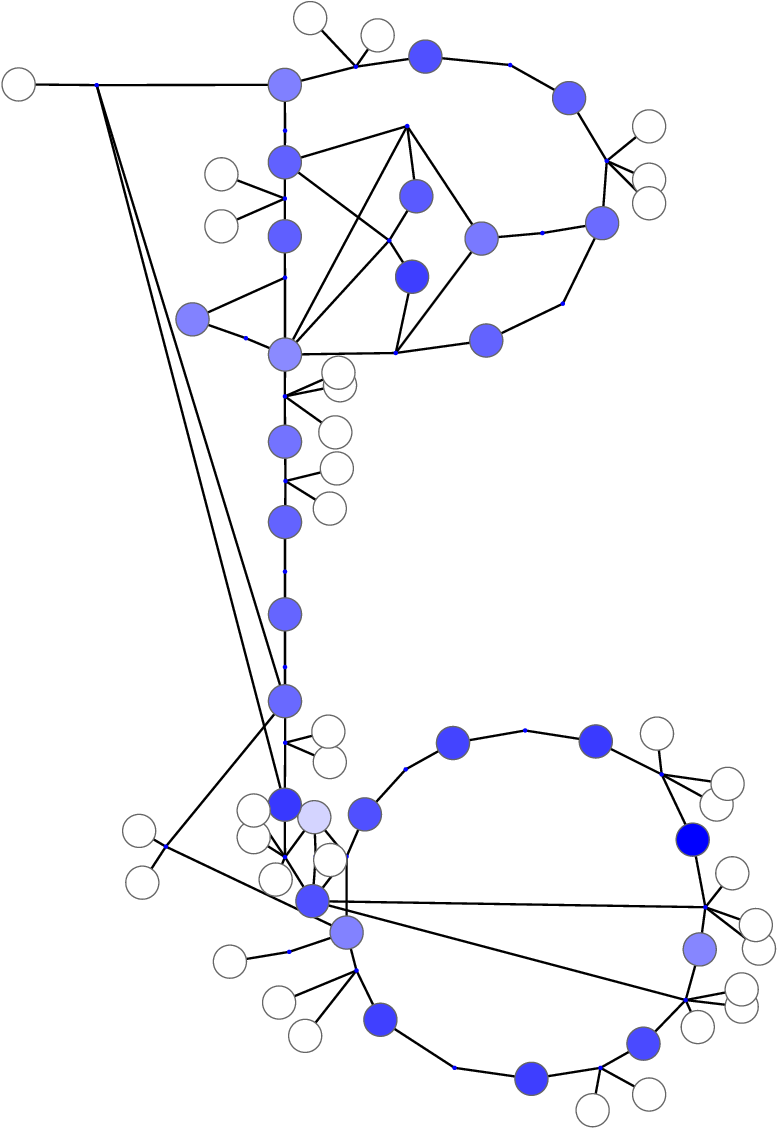}&
 \includegraphics[height=5.8cm]{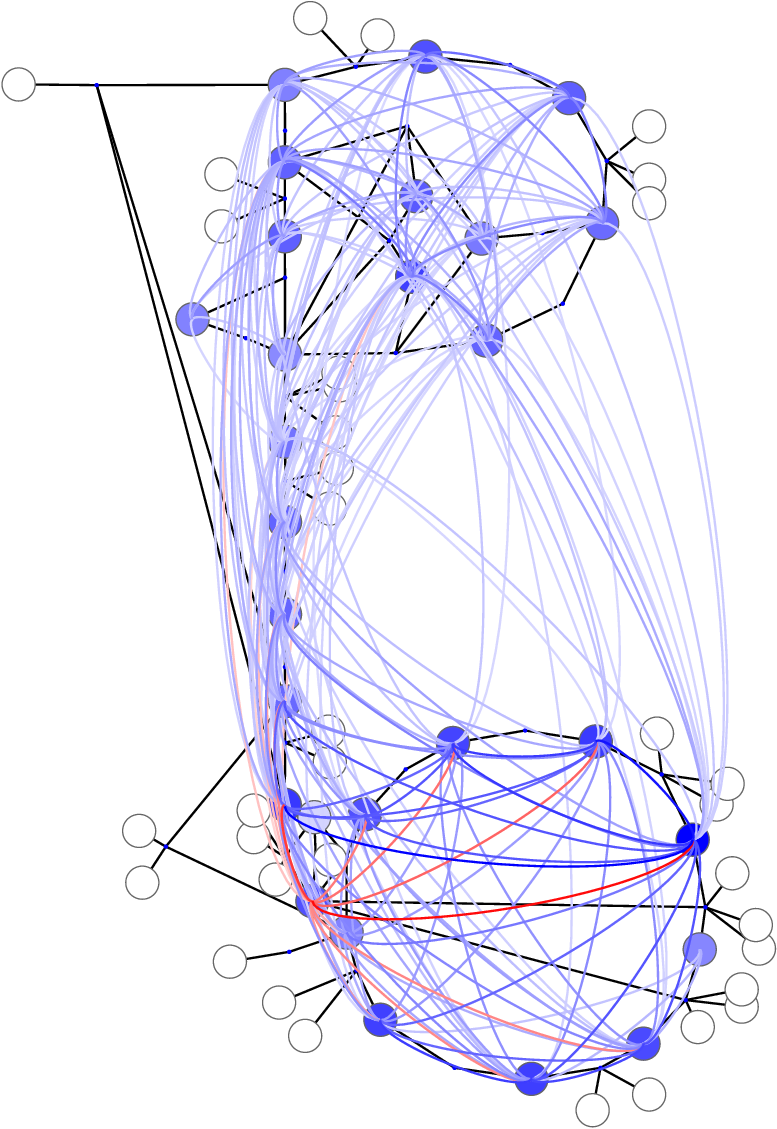}\\[3mm]
 (d) Thermodynamic forces & 
 (e) Noise production & 
 (f) Concentration fluctuations\\[1mm]
 \includegraphics[height=5.8cm]{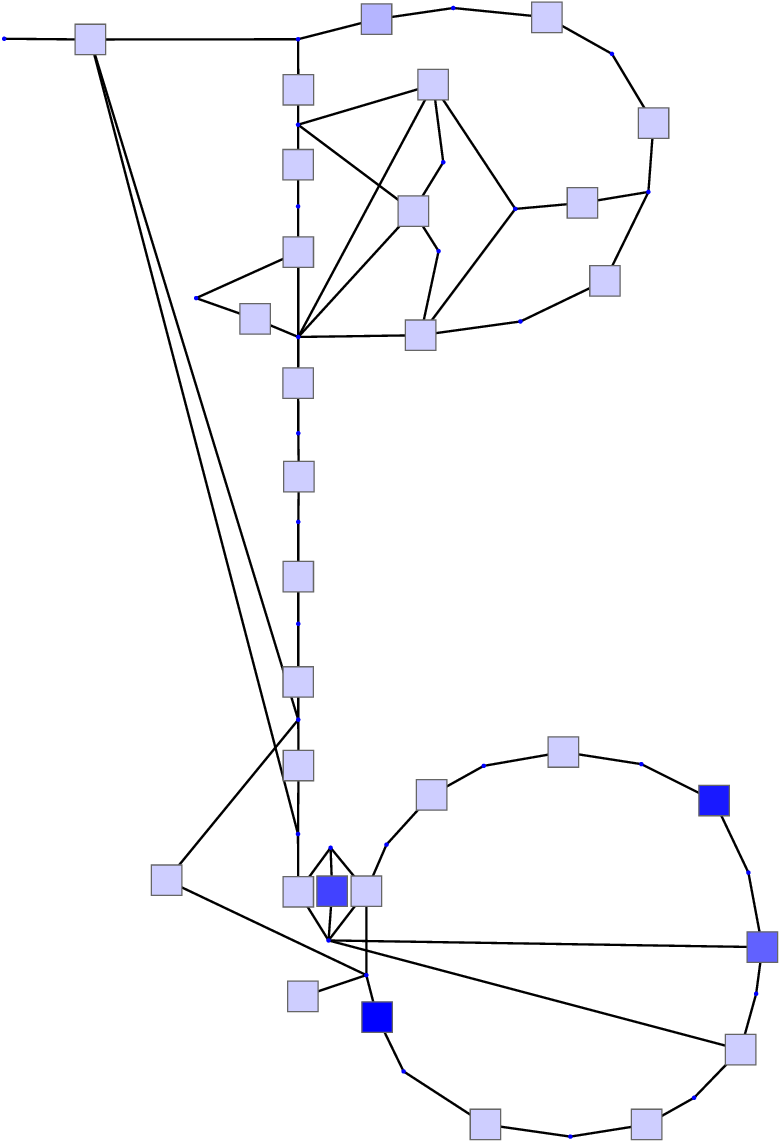}&
 \includegraphics[height=5.8cm]{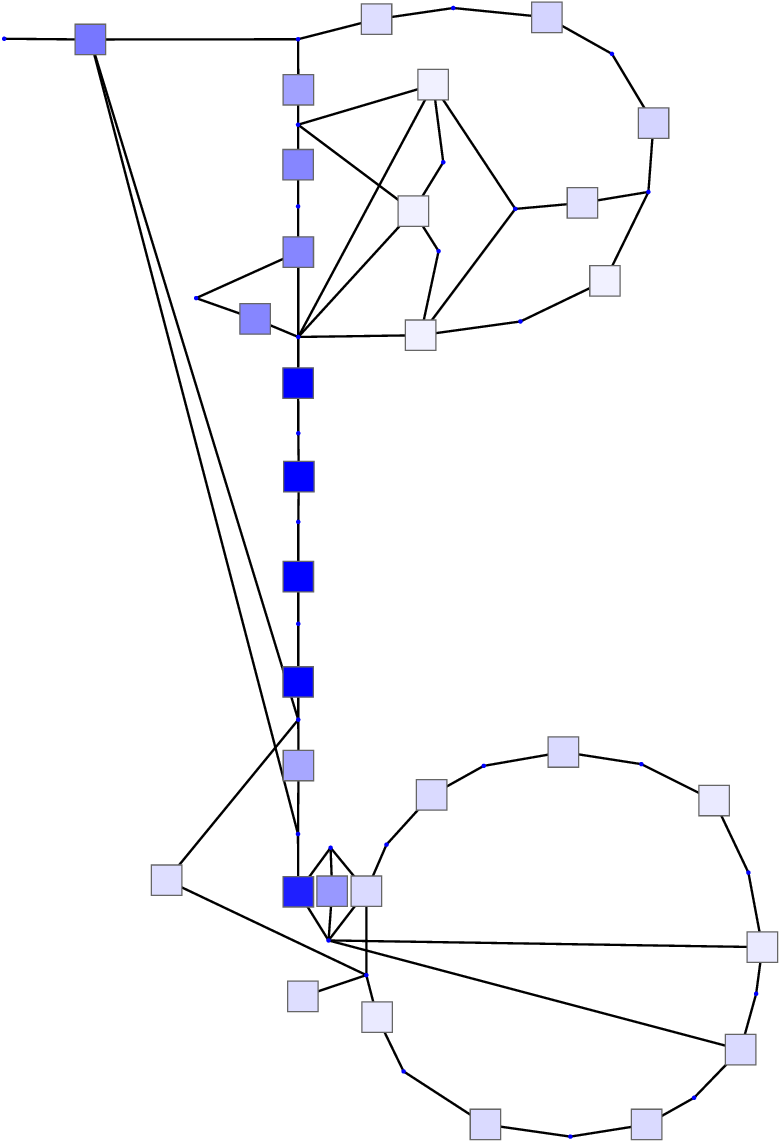}& 
 \includegraphics[height=5.8cm]{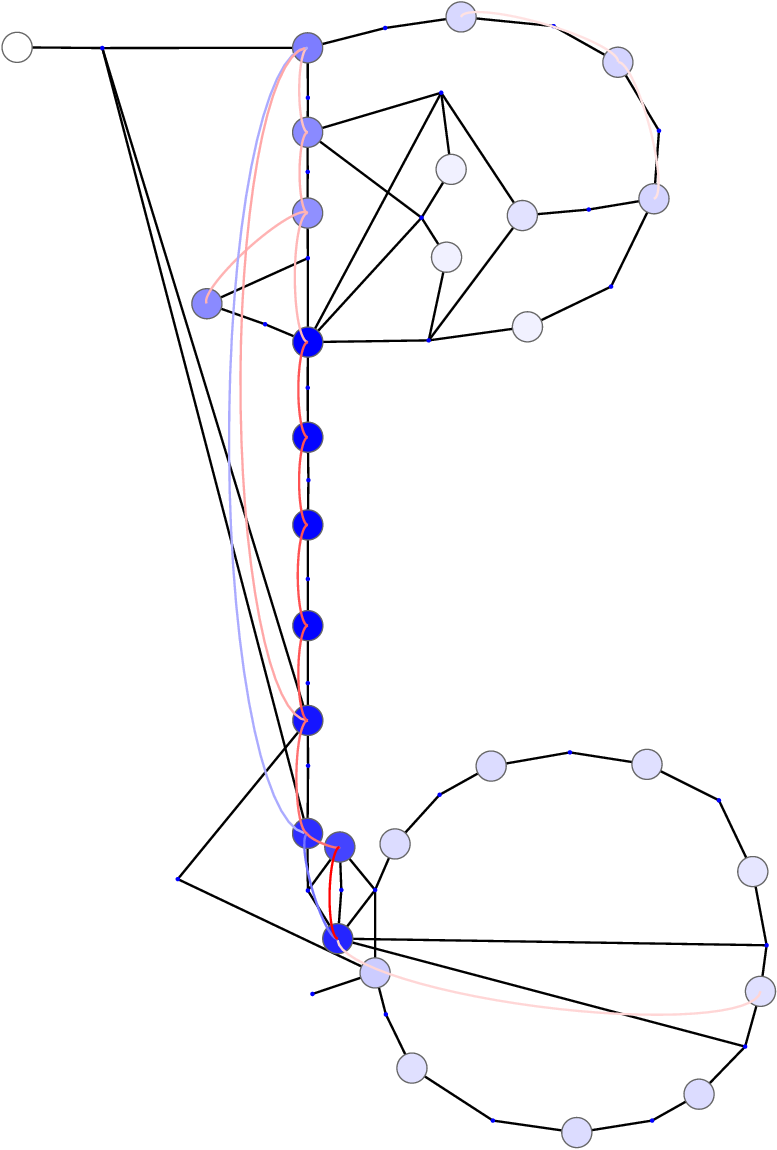}
 \end{tabular} 
 \end{center} 
 \caption{\textbf{Static changes and
     dynamic fluctuations of metabolite concentrations.} \co{was sind
     die stoerungen? bei scenario only\_enzymes (in b und c)
     unkorrelierte varianz aller enzyme mit geom std dev 2; in e and
     f: chemical noise, assuming an E coli cell volume} \co{oder extra-abbildung (aehnlich der hier und abb 7):
     unsicherheit / variation in einem einzelnen enzym: varianz und
     korrelation von metaboliten; das gleiche mit langsamen und
     schnellen (chem noise?) fluktuationen fuer EIN enzym! (das
     gleiche auch mit linearer kette mit ``dioden''-reaktion (wie
     schon in SI; und gestoertem enzym entweder vor oder hinter der
     diode), um einfache effekte zu zeigen! amliebsten das gleiche
     erst mit linearer kette, dann E coli-netzwerk)} (a)
   Variability of metabolic states caused by perturbed reaction rates
   (schematic). An enzyme inhibition (left) changes a single
   reaction rate (solid arrow), which 
   changes the   metabolite concentrations
   and fluxes in steady state (dotted arrows). Similarly, variability in enzyme activities (quasi-static
   random changes) causes slow correlated metabolite variability
   (arcs on the right, blue: positive; red: negative). (b) Variation
   of metabolite concentrations in \emph{E.~coli} central metabolism
   (Figure \ref{fig:ATPrephosphorylation}) caused by an uncorrelated
   variation of enzyme concentrations (geometric standard deviation of
   enzyme levels: 2).  Standard deviations of log concentrations shown
   by colours.  (c) Correlated metabolite variations (covariances of
   log concentrations) shown as coloured arcs (values below 10\%
   of the maximal value were removed for clarity). Metabolite variances (in
   (b)) and covariances  (in (c)) computed from first-order
   response coefficients.  Similarly, local enzyme fluctuations lead
   to network-wide metabolite fluctuations: the
   frequency spectra are related by spectral response coefficients.
   (d) Thermodynamic forces (same data as in
   Fig.~\ref{fig:ATPrephosphorylation} (d)). Effects of chemical noise
   in a a cell volume of 10000 $\mu$m$^{3}$ and with a glycolytic flux of 1
   mM/min. Reactions close to equilibrium (small thermodynamic forces)
   produce strong chemical noise because of their large forward and
   backward fluxes. Spectral power density of the original noise.
   (e) Resulting metabolic fluctuations at fast fluctuations a
   frequency of 1 s$\inv$. Formulae for all quantities are given in
   Supplementary Materials.}
 \label{fig:staticVariation} 
  \end{adjustwidth}
\end{figure*} 

\begin{figure*}[t!]
  \begin{adjustwidth}{-\extralength}{0cm}
 \begin{center}
 \begin{tabular}{llll}
 Fast noise ($f=$1 s$\inv$) & Frequency dependency & Slow noise ($f=$0.001 s$\inv$)\\[1mm]
 \includegraphics[height=5.8cm]{\ecoliccmpsfiles/energetics_control_and_noise_ecoli_ccm_spectral_corr_conc_fast_network.eps}&
 \parbox[b]{5cm}{
 \includegraphics[height=3cm]{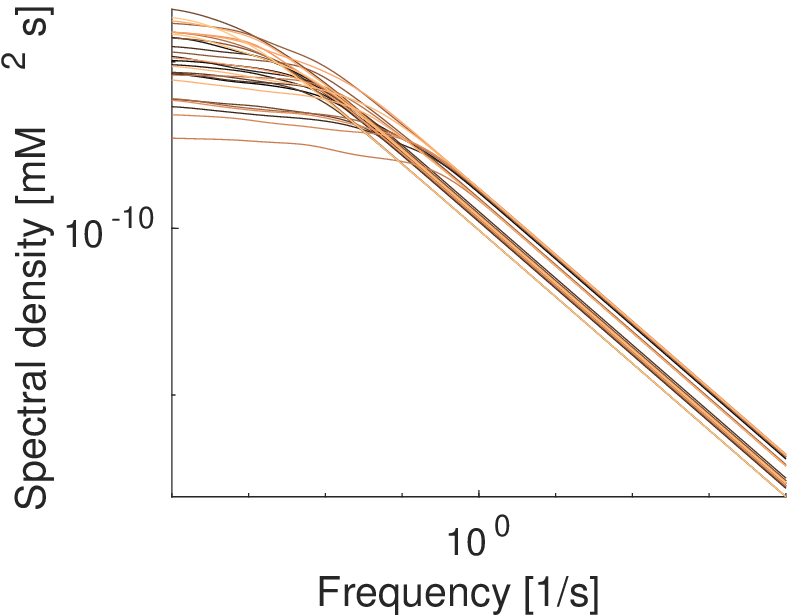}\\
 \includegraphics[height=3cm]{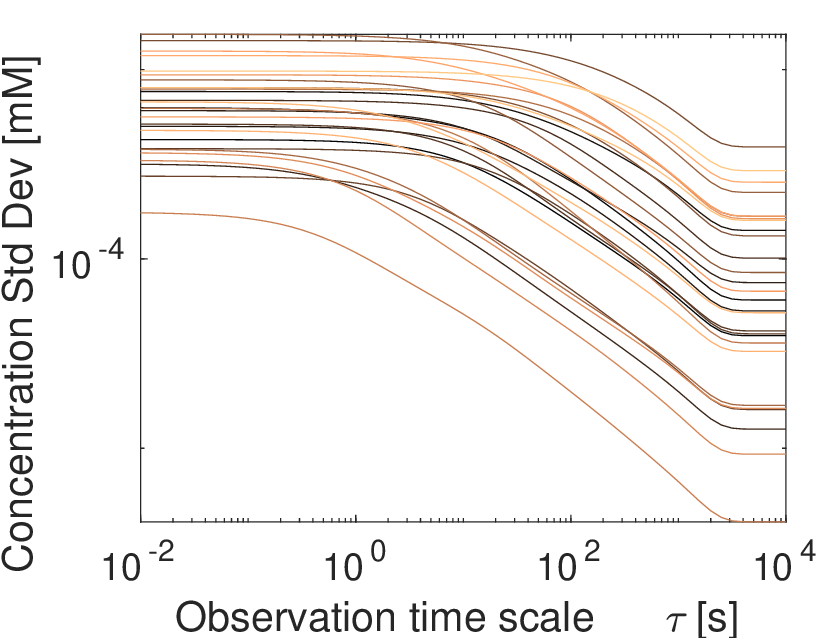}}&
 \includegraphics[height=5.8cm]{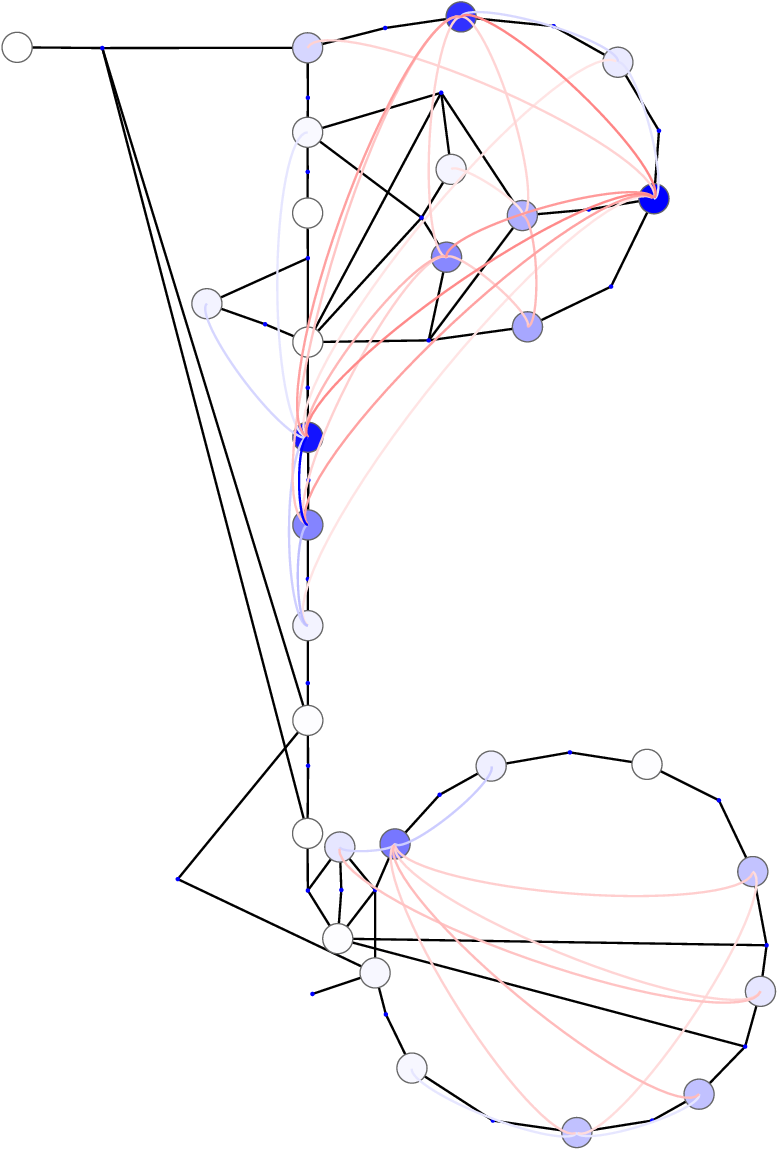}\\[3mm]
 \end{tabular}
 \end{center} 
 \caption{\textbf{Metabolite fluctuations due to chemical noise.}
   Chemical noise
   causes random fluctuations of metabolite levels and fluxes. Using the chemical Langevin equation,
   noise can be  modelled by adding white-noise terms to the reaction rates.
 \co{JA!  wie sind die
     groessenordnungen? ist das rauschen relevant?  was ist mit
     relativem rauschen?} \co{CCM model as in fig 2. UEA!}  Fast and
   slow fluctuations are damped differently on their way through the network. Left: High-frequency
   metabolite fluctuations (spectral densities at oscillation freuency
   f = 1 s$\inv$) occur only close to the noise source. \co{anderes
     bild, mit rauschen nur von einer reaktion?}  Circle colors show
   the spectral densities at 1 s$\inv$, arc colors show covariances
   (blue: positive; red: negative). Right: Low-frequency metabolite
   fluctuations (oscillation period of 17 minutes) are correlated
   along the entire pathway.  Centre: metabolite fluctuations decrease
   with the frequency (see Fig.~\ref{fig:staticVariation}). Top:
   spectral densities (variances), each curve corresponds to one
   metabolite. Bottom: standard deviations of metabolite
   concentrations (square root of spectral density, y-axis), computed
   for different time resolutions of observation (x-axis). For
   details, see Supplementary Materials. 
  \co{explain? explain in text?} \co{kurze erklaerung, was das
     heisst. NICHT GLEICH spektrale dichte!  klarstellen, was in abb
     gezeigt wird!}  Results for flux fluctuations are shown in
   Supplementary Materials.
   Smoothing at different time resolutions
   changes the variance of metabolite fluctuations. At low time
   resolutions, high-frequency fluctuations are filtered out (see Supplementary Materials).
 }
\label{fig:chemicalnoise}
  \end{adjustwidth}
\end{figure*}

\co{\textbf{Correlated variation} In the context of the
 discussion on synergies between cooperating enzymes on page 6 it
 would be of interest to also include Flux Coupling Analysis (FCA),
 \cite{bnsm:04} \coout{cf. Burgard et al.~Genome
 Res. 14(2004)301-312. In this paper, we introduce the Flux
 Coupling Finder (FCF) framework for elucidating the topological
 and flux connectivity features of genome-scale metabolic
 networks. The framework is demonstrated on genome-scale metabolic
 reconstructions of Helicobacter pylori, Escherichia coli, and
 Saccharomyces cerevisiae. The analysis allows one to determine
 whether any two metabolic fluxes, v(1) and v(2), are (1)
 directionally coupled, if a non-zero flux for v(1) implies a
 non-zero flux for v(2) but not necessarily the reverse; (2)
 partially coupled, if a non-zero flux for v(1) implies a non-zero,
 though variable, flux for v(2) and vice versa; or (3) fully
 coupled, if a non-zero flux for v(1) implies not only a non-zero
 but also a fixed flux for v(2) and vice versa. Flux coupling
 analysis also enables the global identification of blocked
 reactions, which are all reactions incapable of carrying flux
 under a certain condition; equivalent knockouts, defined as the
 set of all possible reactions whose deletion forces the flux
 through a particular reaction to zero; and sets of affected
 reactions denoting all reactions whose fluxes are forced to zero
 if a particular reaction is deleted. The FCF approach thus
 provides a novel and versatile tool for aiding metabolic
 reconstructions and guiding genetic manipulations.} }

\myparagraph{Metabolic fluctuations resulting from chemical noise}
\co{sort! mention noise in reaction rates! also mention enzyme
  fluctuations and other noise sources.} For an example, remember that
individual reaction events happen randomly, which is one of the causes
of metabolic fluctuations. The resulting random fluctuations in
single reaction rates are called chemical noise.  A typical enzyme
(with 1000 copies in a bacterial cell and $\kcat$ = 10 s$\inv$)
catalyses around $10^{4}$ net reaction events per second. If we
assume strongly driven reactions (i.e.~neglect backward rates), count
the reaction events within one-second intervals, and aproximate their
numbers by a Poisson distribution, we can expect $10^{4}$ events per
second on average with a standard deviation of
$\sqrt{10^{4}} = 10^{2}$, i.e.~a relative standard deviation of one
percent. If we average over  larger time intervals, the relative standard deviation is
smaller (because fluctuations average out).  How does this chemical noise
translate into metabolite and flux fluctuations?  In 
metabolism, noise from different reactions propagates through
the network, adds up, becomes damped (or sometimes amplified), and
leads to observable fluctuations, correlated between metabolite concentrations and fluxes.
\co{irgendwo sollte dann explizit gesgt werden, dass man die
  fluktuationen schwer direkt verfolgen kann, aber ueber fouriertrafo
  ihre eigenschaften rausbekommt} To model all this, we can add a
white noise term to each rate law, which leads to a chemical
Langevin equation \cite{gill:00}. The  white noise spectrum contains all frequencies
with uniform amplitudes. Tracing the stochastic fluctuations would be
complicated, but their noise spectrum is easy to obtain: the linearised
model acts as a linear filter that translates the original white
spectrum into a coloured noise spectrum of the resulting metabolite
fluctuations. \co{formula?}

\myparagraph{Chemical noise and thermodynamics} Thermodynamics plays a
role in the generation and transmission of noise. The noise amplitude
at the source reaction depends on the individual rates $v_{+}$ and
$v_{-}$. In reactions near chemical equilibrium, but at a given net
flux, these rates become large and contribute strongly to chemical
noise. In contrast, strongly driven reactions produce less chemical
noise and can serve as rectifiers (``diodes'') for noise propagation.
To model this, we need to know the reaction elasticities, and we can
obtain them by STM. \co{Berechnung der amplituden auf zeitskala
  erwaehnen (see SI \ref{sec:SIvaryingTheAffinities}} Figure
\ref{fig:chemicalnoise} shows correlated metabolite fluctuations in
our \emph{E.~coli} example model (see Figure
\ref{fig:ATPrephosphorylation}). In the simulations, fast fluctuations
are strongly damped outside their source reactions, while slow
fluctuations can propagate further, causing network-wide flux and
metabolite variations.

\co{  But what we are actually intersted in is the uncertainty in a
  variable, averaged within a certain time window of length $\Delta t$
  (or time-smoothing kernel of width $\Delta t$) of interest. For ex:
  take the true time course, take averages within time windows of
  lenght $\Delta t$, and consider the distribution of all the results
  (from the different time windows) GOOD GRAPHICS!. This is NOT just
  the spectral density at frequency $\omega = 2\pi/\Delta t$, but an
  integral over all frequencies! For example, white noise has a unit
  spectral density for all frequencies, but the resulting variance
  decreases with the time scale!  generally, consider a random time
  series $F(x)$; to compute the variance at time scale $\Delta t$,
  compute the convoluation with a Gaussian kernel (width $\Delta t$),
  then take the (time-dependent) one-time variance of the resulting
  stochastic process; for a stationary process (ergodicity), this is
  time-independent and equal to the resulting variance OF ONE instance
  of the process, where the variance is taken over time.  With a
  gaussian kernel, the convolution is a multiplication in Fourier
  space; that is, we take the spectral density, multiply it by a
  gaussian squared, and integrate over all frequencies?}

\co{that resembles a
  statistical ensenble of quasi-steady states.}  \co{JA!  Discuss
  whether chemical noise actually plays a role; discuss propagation of
  expression noise; noise in comprehensive (metabolic and expression)
  models (kurz und nur verbal) } \co{amplitudes very small!! say:
  effects of (correlated) enzyme fluctuations / variability can be
  studied in the very same way}

\co{klar, dass behandlung anderer quellen von fluktuationen aehnlich laeuft?} 
\co{klar, dass
 behandlung von unsicherheiten aehnlich laeuft? cite uncertain
 params}

\subsection{Network structure and thermodynamic forces shape metabolic
 dynamics}

\co{Zeit für eine runde stunde bzgl thermodynamic forces}
\co{hier die ergebnisse aus results zu sammenfassen und mathematisch klarmachen, wie man von kraeften direkt zu den genannten eigenschaften kommt; formeln im  anhang zitieren}

\myparagraph{\mysmallbreak Metabolic dynamics and control reflect network
  structure} \co{klarer schreiben // mehr durch den gedankengang
  fuehren // also: was lernen wir? wie sind dynamische eigenschaften
  durch netzwerk, thermodyn und elast bestimmt, und welche auswrkungen
  haben thermodyn kraefte} The structure, dynamics, regulation, and
function of metabolic systems are closely related.  If proteins
cooperate (e.g.~if they belong to the same complex or pathway),  this results in 
synergisms and can be  reflected in  epistasis (i.e.~synergisms towards a fitness-relevant
variable), co-expression (or a temporal order of
activation along the  pathway
\cite{tlbd:04}), and shared regulation mechanisms (e.g.~enzymes
encoded in an operon).  Similarly, network structure is reflected in metabolite fluctuations and
their correlations \cite{skfw:03}. Of course,
all these patterns portray network structure indirectly: while
network edges are ``sparse'', connecting only adjacent metabolites and
reactions, the resulting dynamic effects link elements across the
entire network (but reflecting, for example, closeness in
pathways). \co{explain?} To understand this in detail, we need to relate (local)
network structure to (global) metabolic dynamics and how it is shaped
by individual thermodynamic forces or enzyme parameters. Once we
understand dynamic effects (e.g.~enzyme levels affecting metabolite
concentrations and fluxes), we may invert the relationship and  ask
what enzyme profiles are required (or best suited) to achieve a desired metabolic
behaviour \cite{klhe:99,lksh:04,lieb:14a,nfbd:16}, and what regulation
mechanisms can realise these enzyme profiles.

\myparagraph{Local and global properties of metabolic systems} How are
network structure, dynamics, and function related precisely? And how
can we study the effects of varying network structure or enzyme
kinetics in practice? In metabolic systems, a perturbation (of an
enzyme, a metabolite concentration, or a reaction rate) has immediate
local effects, which then lead to network-wide, long-term effects.
The immediate local dynamics is determined by the kinetics of single
reactions (at given metabolite levels), while the network-wide
dynamics emerges from them as metabolite levels change dynamically
after the perturbation. Metabolic Control Theory describes these two
stages separataly by elasticitiy and control (or response)
coefficients. Local network structure and kinetics are encoded in the
stoichiometric matrix $\Nint$ and elasticity matrix $\Eunc$ (shaped by
thermodynamics, enzyme saturation, and regulation as considered in
STM). The matrix product $\Nint\,\Eunc$ \co{FN on moety conservation?}
yields the Jacobian matrix, which is local and sparse.  The step to a
network-wide dynamic description requires a matrix inversion,
\co{REF?} which leads from a sparse (Jacobian) matrix to (full)
control and response matrices that describe network-wide changes of
the steady state.  \co{WD?  To explain the link between local and
  network-wide effects, we need to remember that enzyme control and
  synergies, metabolite fluctuations, and possibly optimal
  differential expression patterns \cite{lksh:04} reflect response
  coefficients, and that these coefficients depend on local network
  structure and on local fluxes, concentrations, and reaction
  elasticities. Stoichiometric matrix and elasticity matrix determine
  the control properties only indirectly: their product
  $\Amat = \Nint \,\Eunc$ is the Jacobian matrix which, by its
  sparsity structure, links neighbouring compounds and therefore
  represents the local network topology.} As an example of global
effects, we may consider the propagation of metabolic fluctuations: as
we saw above, chemical noise originates in each single reaction and
propagates in the entire network. On their way through the network,
high-frequency components are strongly damped (so they are not visible
far from their source reaction), while low-frequency components travel
through the entire network, leading to slow or quasi-static variations
of the flux distribution.

\co{EDIT Another way to relate (network-wide) metabolic behaviour to
(local) network structure and kinetics is given by the summation and
connectivity theorems of MCT \cite{Heinrich1996}: the control
coefficients along a stationary flux distribution (which only depend
on network structure, but not on kinetics) must have a fixed sumsum ,
while the control coefficients around metabolites are constrained by
local elasticities (connectivity theorem). Thus, the control
coefficients are constrained by network structure via the summation
theorem, but to know them precisely we need to know the elasticities
(in the connectivity theorem) -- which can be explored by STM.}

\myparagraph{Expected effects of thermodynamic forces} We saw that
thermodynamic forces can have various effects on dynamics.  Within a
reaction, large forces entail a large energy dissipation (per flux),
and due to the relatively small one-way fluxes (given by
$v_{+}\approx v, v_{-} \approx 0$) , there is little chemical noise.
Since the backward flux is small, the enzyme works efficiently and the
enzyme demand per net flux is low. At the same time, the net rate is
insensitive to the product concentration, so the product elasticity is
low. On the contrary, if a thermodynamic force is large, flux
reversals require large concentration changes, which makes them
physiologically difficult or impossible.  Thermodynamic forces,
saturation values, and fluxes define typical relaxation times for
metabolites.  Local metabolite perturbations $\delta c_{i}$ tend to be
dampened, as described by the linearised dynamics
$\delta \dot{c_{i}}=\sum_{l} n_{il}\,\Eun_{li}\,\delta c_{i}$. If the
initial state is stable, the dynamics will decrease the initial
perturbation, and $c_{i}$ will return to its reference value
\cite{here:91}. \co{bezug zu kontrollkoeffizienten fuer dynamische
  zeitreihen?}  The relaxation time
$\tau_{i} =1/(\sum_{l} n_{il}\,\Eun_{li})$ can be estimated from
thermodynamic forces and enzyme saturation values. Finally, due to
their influence on elasticities, the thermodynamic forces have an
impact on flux control in the entire network.

\myparagraph{Example: Linear pathway} In a linear pathway, these
effects are easy to see: a strongly driven reaction has a low product
elasticity and is not affected by downstream processes: therefore, it
has full flux control and leaves no flux control to the downstream
enzymes. Therefore, a change in downstream enzyme levels has no effect
on the flux: it just changes the metabolite levels. Dynamically, the
forward-driven reaction would act as a rectifier for metabolic
fluctuations (like a diode in electrical circuite): fluctuations can
pass only in one direction. In reactions with large but finite
thermodynamic forces, similar tendencies can be expected (see
Supplementary Materials).
Due to their high flux
control, such reactions \co{FN?  sagen, dass all das wiederum einfluss
auf den GEBRAUCH (opt adaptation and regulation) der enzyme hat!} are
likely targets for regulation in linear pathways
\cite{nbfr:14}. \co{POSITIONING OF FORWARD-DRIVEN STEPS. alves und
savageau \cite{alsa:01}

\coout{It has been observed experimentally that most unbranched
biosynthetic pathways have strongly driven reactions near their
beginning, many times at the first step. If there were no functional
reasons for this fact, then one would expect strongly driven reactions
to be equally distributed among all positions in such pathways. Since
this is not the case, we have attempted to identify functional
consequences of having a strongly driven reaction early in the
pathway. We systematically varied the position of the strongly driven
reaction in model pathways and compared the resulting systemic
behavior according to several criteria for functional effectiveness,
using the method of mathematically controlled comparisons. This
technique minimizes extraneous differences in systemic behavior and
identifies those that are fundamental. Our results show that a pathway
with a strongly driven reaction located at the first step, and with
all other reactions reversible, is on average better than an otherwise
equivalent pathway with all reactions reversible, which in turn is on
average better than an otherwise equivalent pathway with a strongly
driven reaction located at any step other than the first. Pathways
with a strongly driven first reaction and low concentrations of
intermediates (one of the primary criteria for functional
effectiveness) exhibit the following profile when compared to fully
reversible pathways: changes in the concentration of intermediates in
response to changes in the concentration of initial substrate are
equally low, the robustness of the intermediate concentrations and of
the flux is similar, the margins of stability are similar, flux is
more responsive to changes in demand for end product, intermediate
concentrations are less responsive to changes in demand for end
product, and transient times are shorter. These results provide a
functional rationale for the positioning of strongly driven reactions
at the beginning of unbranched biosynthetic pathways.}}  \co{ref aus
SI FIG S4} \co{CITE MDF SI}

\todo{All this can also been shown mathematically. While a pathway
  flux is described by a closed formula -- which exists for linear
  chains -- this is not possible in larger networks. So even though we
  expect similar behaviours in larger networks, we cannot prove them
  analytically. This is where STM is particularly useful.}  \co{alle
  allgemeinen beweise hier} \co{FN: Whether or not a reaction can
  serve as a rectifier depends only on the thermodynamic force. Its
  scaled flux control, in contrast, depends also on thermodynamic
  forces of other reactions!} \co{say that the remaining effect comes
  from saturation, s etc etc} \co{weniger bockloser satz! add ref to
  section \ref{sec:SIEffectOfDriving} (effect of forces in linear
  chain) in article}

\section{Discussion}

\myparagraph{\mysmallbreak STM: usage for ensemble modelling, model
  fitting, and optimisation} \co{am ende nochmal kurz sagen, dass STM
  MCT praktisch verwendbar macht.} The STM framework can be used for
ensemble modelling, model fitting, and optimisation.  STM can help us
understand the interplay of enzyme concentrations, metabolite
concentrations, and fluxes, simulate periodic random perturbations and
chemical noise \cite{lieb:2005}, and predict enzyme adaptation to
changing metabolic supply and demand or to enzyme-inhibiting drugs
\cite{lksh:04}.  Ensemble models may either reflect biological
variability in a population or our subjective uncertainty due to
missing data, and can help us assess the probabilities of different
types of metabolic behaviour.  \co{FN note that fitting (max
  likelihood or bayes) is based on parameter distriutions, and that
  optimisation can be seen as a limiting case of ensemble modelling,
  namely boltzmann distributions (simulated annealing) or of
  population modelling (genetic algorithms)} \co{Say a few words about
  optimisation: loop around STM; optimisation or metropolis (similar
  to sampling) // with global optimisation, gradient descent,
  simulatated annealing, genetic algorithms. // also Bayesian,
  posterior sampling usw; alsobriefly mention new (roberto!)  bayesian
  method // for a target (approx: sampling and choosing the sample
  that maximises the target); multi-objective optimisation
  (approximate by wide sampling and just plotting the targets)}
\co{fragen, ob auch OPTIMALE zustaende konstruiert werden koennen. ja!
  ECM und CBA I erwaehnen} STM is useful whenever precise models are
not available or when general possibilities or variances in a cell
population should be explored.  \co{By linearisation and model
  reduction, we may obtain simple black-box models
  \cite{libk:05}. Such models could be coupled to pathway models of
  interest to provide them with realistic dynamic boundaries.}
\co{Using kinetic constants of the original model, one could convert
  given kinetic models into a form with generic reversible rate laws.}
In global variability analysis, model variables are sampled from broad
distributions, reflecting plausible parameter ranges
\cite{lieb:05,likl:06a,bnsl:11}. For model fitting, available kinetic,
thermodynamic, or metabolic data \cite{gold:99,sceg:04} are directly
inserted into a model, or be used to define bounds or probability
distributions for sampling. Even if automatically constructed models
are less reliable as hand-curated ones, they may still be useful, for
example, as scaffold networks into which hand-curated pathway models
can be inserted and which provide a dynamic environment for them.  On
the contrary, once a model with standard rate laws has been
constructed by STM, some rate laws may be replaced by laws obtained
from enzyme assays to make the model more realistic \cite{bghs:09}. To
ensure a consistent model, the fluxes, concentrations, and equilibrium
constants between reconstructed model and new rate laws must be
matched. To guarantee this, rates and equilibrium constant from the
``real'' rate law can imposed upfront during STM.

\co{STM has a three major  use cases: \co{for metabolic engineering}
  
1. Modelling a metabolic system based on available data. If data are
available, a broad sampling is not necessary. Nevertheless, we can
insert known values directly, we may sample around them, or get model
results with uncertainties. In bioengineering, this may be useful if
data are missing or uncertain,.  Predicting the an expected
performance may help choose the most promising pathways.

2. Understand the effects and affordances of model features, e.g. structural variants, regulation, thermodynamic forces, or saturation. vary certain features and observe the effects: what is changing, what is not changing? and infer global dependencies (averaged over other model features)

3. Optimising  different objectives (scoring different types of
variables) can be optimised subsequently and independently (not useful
for bioengineering, since / if kinetic constants are determined last,
from others, and the kinetic constants cannot be engineered).  }

\co{klarmachen: welche methoden
 stehen jetzt zur verfuegung? wie verhalten sie sich zu frueheren
 methoden? was kann man jetzt damit machen? was haben wir biologisch
 oder allgemein gelernt? welche modelle stelle ich zur verfuegung?}

\myparagraph{Prediction of flux changes by flux analysis and MCT} How
can we predict the effects of enzyme perturbations such as
single-enzyme inhibitions or differential expression of many enzymes?
Metabolic responses may be quasi-static or dynamic and may concern
metabolite concentrations and fluxes. Constraint-based models would
ignore kinetics and replace them by heuristic rules: FBA predicts the
most favourable flux changes, while MoMA predicts the smallest
possible changes. In both cases, enzyme repression can be modelled by
restricting a reaction to smaller (knock-down) or zero fluxes
(knock-out), and the perturbed flux distribution is obtained by
maximising a metabolic objective (in FBA) or minimising the necessary
changes in flux (in MoMA). If multiple enzymes are repressed, the flux
in a linear pathway is set by the lowest flux bound, while the other
repressions have no effect. Therefore, in linear pathways, the two
methods predict buffering interactions. MCT, in contrast, predicts
metabolic responses by kinetic models and control coefficients, which
arise from thermodynamics, enzyme saturation, and regulation and can
be determined by STM. Reactions are not simply ``reversible'' or
``irreversible'' like in FBA, but vary gradually between
``near-equilibrium'' and ``forward driven'' reactions, depending on
thermodynamic forces.  Hence, the quantitative effects of
thermodynamic forces, e.g.~on the transmission of fluctuations, can be
studied systematically. Moreover, an enzyme inhibition may
change the control coefficients, which implies synergies between
enzymes. MCT captures these synergies by synergy coefficients.

 \coout{ (an example is given in \todo{SI
     \ref{sec:SIhepatocyteSI}}).}
 
\myparagraph{Limitations} Of course, MCT has its limitations.  First,
it holds only for small perturbations: for large perturbations,
predicted concentrations may become negative. In usual MCT, negative
concentrations can be avoided by considering log-concentrations, and
the corresponding scaled control coefficients. In dynamic simulations
of the linearised model, this is not possible: here we can compute
with absolute concentrations and avoid negative values by transforming
the time courses by heuristic correction functions. \co{explanation?
  ref roberto?} Second, STM assumes that enzyme changes are known and,
by default, does not predict enzyme adaptations. In fact, the
metabolic extra effects of such adaptations are second-order effects
and can be neglected if perturbations are small, \co{ausser an
  bifurktionspunkt!  das auch in opt enz rhythms sagen!}  or if we
care about dynamics much faster than the time scale of enzyme
changes. However, there are ways to consider enzyme adaptationsin our
predictions: if they are known from experiments, they can be inserted
as enzyme perturbations; otherwise, they may be predicted based on a
principle of optimal adaptation and using control coefficients from
STM \cite{lksh:04}.

 \co{mention anwendungen (effect of differential expression and
 anwendung model fitting) again in discussion} \co{What else to do
 with it? hint at dynamic simulations with matrix exponential:
 future paper!}

\myparagraph{Dependency schemas and model construction} STM provides
not only a framework for model construction, but also clarifies more
generally physical and statistical dependencies. A dependency schema,
as in Figure \ref{fig:samplingscheme}, summarises constraints between
model variables and shows how these variables can be sampled, fitted,
optimised, or manually chosen.  To construct a dependency schema, we
choose a suitable set of physically independent ``basic'' model
variables and treat all other variables as dependent. Aside from the
strict physical dependencies, statistical dependencies can be
implemented by defining distributions for basic variables, which
define a joint distribution of all model variables (where strict
dependencies between variables arise from the dependency schema, and
other soft dependencies arise from the probability distributions).
While a schema, endowed with a prior, determines statistical
dependencies between all variables, the opposite is not the
case. Moreover, the same physical and statistical dependencies can be
expressed by different choices of basic variables. Thus, in practice
it can be important to choose a good dependency schema allowing us to
use simple priors. Our dependency schema can be used for model
construction, fitting, optimisation, and statistical analysis, and
show how model variables determines metabolic behaviour. Known model
constraints allow us to discard infeasible models (e.g.~describing a
\emph{perpetuum mobile}), to determine possible ranges of unknown
model variables, and thus to restrict the model results.  By combining
a schema with probability distributions, plausible or measured data
values can be inserted into the model and the remaining uncertainties
can be assessed.  Altogether, STM serves for translating metabolic
networks into realistic kinetic models even when data are scarce.
\co{use ``pseudo prior'' instead of pseudo values, and explain!}
\co{clarify dependence, priors, pseudo values etc (same ethoughts also
  in MB} \co{NEW FN: A simple choice would be an uncorrleated prior
  for the basic variables \co{should pseudo values be part of the
    prior?}  \co{conversely, the dependency schema itself helps us
    define meaningful correlated priors for basic variables and
    pseudopriors for dependent ones. This has been used in PB, a
    schema-based approach to model fitting.}  \co{explain more clearly
    physical vs statistical dependencies \co{for example, by assuming
      uncorrelated priors for independent variables \emph{and}
      uncorrelated pseudo values for dependent variables, which we
      then combine into a correlated prior for the independent
      variables}.} \co{note that there are two questions: how are
    parameters interdependent, and how can we describe these
    dependencies conveniently by our choice of ``independent''
    parameters (knowing that later, we will typically choose
    independent distributions for them -- so our ``innocent'' choice
    has an effect, in practice, on the multivariate distributions we
    use!)}

  \co{say that dependency schema contains matrices and
    functions. zusammensetzen von schema part und network stucture
    part. see ``RBA systematic''!} }

\myparagraph{Covariation of kinetic constants in evolution}
Understanding the dependencies and covariances of variables is not
just a modelling question, but one that concerns cells in reality. It
arises in different contexts: kinetic constants or state variables may
co-vary between the states of a cell, between cells in a population,
during evolution, or between species. Some of this covariation is
caused by physical constraints. For example, $\kcat$ and $\km$ values
are constrained by Haldane relationships that limit the ways in which
these parameters can vary (via changes in the enzyme's amino acid
sequence) \cite{klhe:94}. If constants co-vary, can we see some of
them as independent, and the others are dependent on them?  For
example, can we assume that Michaelis constants and forward $\kcat$
values vary independently, and see backward $\kcat$ as dependend
variables? \todo{Likewise,} should we assume that enzyme
concentrations correlate positively with catalytic constants (because
both quantities are under a selection pressure for high enzyme
capacity) or negatively (because if an enzyme is highly efficient,
less of this enzyme is needed)? In fact, there are no ``true'' sets of
basic parameters, only pragmatic choices: if variables have low
statistical dependencies and large effects in the model, it can be
convenient to treat them as (physically) independent.  Moreover,
remember that dependencies can be conditional on a third variables:
two variables may be dependent, but independent \emph{given another
  variable}. Here we used this fact, for example, to sample dependent
elasticities ``independently'' by fixing the thermodynamic forces. But
this means: whether two variables are ``dependent'' will also depend
on assumptions about other variables (e.g.~whether they are are fixed
or co-vary), as well as on the scenario consided (dynamic variation in
a cell, variability in a cell population, changes in evolution). In
any case, to study dependencies between metabolic variables
\cite{bnsl:11}, we need meaningful dependency schemas, and
comprehensive kinetic and metabolic data.

\co{besser erklaeren;
  klarmachen, warum das keine absolute unterscheidung ist, sondern vom
  modellierer abhangt} \co{As an example, consider again the Haldane
  relationship for uni-uni MM kinetics; assume keq is fixed. if KM
  were also fixed, kcat+- would be proportional; in reality, nobody
  knows how single mutations will affect the four kinetic constants,
  let alone how ongoing mutation and selection will affect them}
\co{logik besser erklaeren}

\co{More general thoughts about dependency schemas\\
  algebra of models\\
  layered models\\
 (1) explain that this (distributions, dependencies) can be seen as
 an empirical statistical question or as a question about imagined
 distributions, priors, and model assumptions.\\ (2) a ``science of
 dependencies'' beyond simple values and variations. refer to
 Bayesian networks and graphical models\\ (3) the ``algebra of
 models'' (gespraech mit dimitrij) laesst sich schon mal ganz gut mit
 der trennung network / dependency schema behandeln. das weiter
 durchdenken und evtl im ``RBA network graphics''-artikel andeuten,
 spaeter (in CBA field??) mathematisch ausfuehren. // ebenfalls
 relevant fuer die algebra von modellen: die hierarchie von model
 ensembles fuer ein gegebenes modell! ``zoomen'' im genauigkeitsraum
 (nicht wie genau dinge in der wirklichkeit aufgeloest sind, sondern
 wie genau zwischen modellvarianten aufgeloest wird)}

\co{Was man mit verstaendnis von elastizitaeten machen kann} \co{ELASTICITIES and ENZYME
 Efficiencies as connection between flux analysis models and kinetic
 models} \co{sagen: jetzt aktuell zb wegen rba-modellen //
 easlticities and enzyme efficiencies in anderen
 modellierungsansaetzen} \co{important insight: how elasticities can
 be split into terms of different origin in the scheme! discuss
 relationship between elasticity formula splitting in modular rate
 laws and in factorised rate laws (some paper by elad?)}
\co{discussion? fluctuation formula also apply to models that capture
 enzyme production as part of the network! (what about whole-cell
 models with growth? ref laurens krah?)}

\co{was lernen wir biologisch?}

\myparagraph{Retromodelling: defining the state first, then control
  properties} What makes STM and SKM useful in practice is their
retromodelling approach: we first choose a metabolic reference state
and then the elasticities, rate laws, and control properties in this
state. Retromodelling allows us to define state variables and kinetic
constants separately and to study the effects on dynamics. For
instance, by varying the thermodynamic forces, we may change how
reactions act as rectifiers for metabolic fluctuations and adjust the
stability and control of metabolic states. In unbranched pathways --
the pathway flux can be computed analytically \co{ref} -- these
effects can be easily understood. In larger networks, this is
impossible: here sampling is helpful, and the formulae of STM relate
thermodynamics to elasticities and elasticities to control
coefficients can help us understand these tendencies. \co{emphasise
  that also reference concentrations may play an important role (also
  they are to be sampled or optimised)!  // Not only the saturation
  values play a role for the dynamics: also the metabolite
  concentrations are important!  so instead of only sampling
  elasticities also met conc should be sampled!  STM alte notiz in
  matlab: Another important parameter (big consequences) is the
  minimal allowed A value .. currently 3 kJ/mol, like the worst value
  in Avi's EMP model}

\co{NACH ERSTEM Review?:

\co{komplett neu schreiben:

  in evolution, we can assume that metabolism must optimise, first, a typical standard state (most of all fluxes; but at good metabolite levels, and enzyme levels)), and second, how cells can deal with deviations and fluctuations; can be caused by environment or by internal fluctuations; dealing with can mean homeostasis or using opportunities (all this is encoded in sensitivites ).

  beispiel:
\co{mention convergent evolution} \co{say: the
  metabolic state is the outcome, which usually deternmines fitness
  (or measurable quantities). In STM, we deliberately START from this
  outcome as a given (or desired) fact.}  

this is exactly reflexted in STM:

also her, we first choose a state etc etc

In normal models, this would be more difficiult:

...

thus, the ``retromodelling'' approach, while being seemingly ``unrealistic'', follows exctly the order in which OPTIMALITY may be considered (assuming that the the OUTCOME - the metabolic state - is most important, then the sensitivities, and that the kinetic constants can be evolutionarily adapated.)
}
}

\myparagraph{Optimisation of several cell objectives}
\co{``teleological modelling'': for ex, the way small-molecule
  regulation would be included here (AT the same given flux, and
  therefore reducing the enzyme efficiency), reflects exactly this
  thinking, by which the fluxes are ``chosen first'', and dynamic
  sensitivities are formed around them, if necessary, by increasing
  the enzyme demand.}  Retromodelling can be helpful for optimisation,
with several objective, e.g.~a desired metabolic state and desired
control properties in this states (e.g.~homeostasis, robustness, and
adaptability). We begin with a desired flux distribution and
construct, step by step, other state variables and kinetic constants
that support this flux distribution. In each step we can account for
different constraints and objectives, e.g.~optimal production rates
(when choosing the flux distribution).  This reverse approach may help
us think differently about optimisation in evolution or biotechnology,
how metabolic states can be tuned, and what are the protein costs. For
example, we may imagine that cell need to optimise, first and
foremost, their fluxes and concentrations under standard conditions,
and secondly stabilise this state (or make it homeostatic or better
controllable) by tuning its elasticities.  \co{If fluxes and reactant
  concentrations (defining the thermodynamic forces) are
  evolutionarily optimised, evolution may still modify the
  elasticities by tweaking the saturation levels or adding regulation
  between virtually any reaction and any compound.}  To achieve this,
evolution may vary thermodynamic and kinetic properties of reactions,
leading to different control patterns. For example, if a reaction is
driven by an ATP-ADP cofactor pair, which increases its driving force,
this ATP investment can allow the cell to use the reaction as a
rectifier for fluxes and metabolic signals (i.e.~perturbations and
noise): at a fixed net flux $v$, this reduces noise production in the
reaction. To keep the fluxes unchanged, the lower enzyme efficiency
must be compensated by higher enzyme concentrations. This makes
small-molecule regulation of enzymes effectively costly. STM supports
this way of thinking.  During the phases of model construction, a
series of objectives (separate objectives for metabolic states and for
their control properties) can be applied. This shows again that STM is
not just a tool for model construction, but also a framework for
thinking about variability and optimality in cell populations, in
evolution, or in engineered cells.

\co{Understanding this room for variation, and trade-offs between
  variables, can help us better understand other modelling approaches
  and cells in reality.}  \co{auf runde stunde achten! dh auf kraefte
  zurueckkommen}

\co{\myparagraph{MATERIAL: Retromodelling for engineering metabolic models around
  metabolic objectives} Retromodelling, as practiced
in STM, turns model fitting or optimisation around.  In traditional
modelling, kinetic constants and enzyme levels are regarded as tunable
quantities that determine the metabolic state.  To fit or optimise
these parameters numerically, the metabolic state has to be computed
repeatedly, which can be very time-consuming. In this two-level
approach, the step from kinetic constants (and possible enzyme
concentrations) to metabolic states reflects causality on a short time
scale (the time scale of metabolic dynamics in a single individuum),
while an optimisation of the kinetic constants themselves may resemble
evolution by mutation and selection. \co{(while the optimisation of
  enzyme concentrations can be related either to evolution or an
  optimal transcriptional regulation).} While this makes physically
and biologically sense, it does not show how (or to what extent)
different objectives -- fluxes, concentrations, controllability and
homeostasis -- can be optimised simultaneously and how much freedom
there is for varying these objectives -- given that each parameter
change will not only change a metabolic state, but also it control
properties. To understand how variables can be varied, we need to
understand their dependencies. In retromodelling, instead of seeing
enzyme levels as ``tunable'' and metabolite levels and fluxes as the
resulting outcome, we can think teleologically, i.e.~starting from a
desired end: we predefine fluxes and metabolite levels, and ask what
enzyme levels are needed to realise this state: therefore, we treat
the enzyme levels as functions of metabolite levels and fluxes!
\co{ \co{FN note that
    for constructing something, it is not practical to leave the
    kinetic constants open; they should not be sampled but
    predetermined, while metabolite concentrations should be sampled;
    but this leads rather to ECM ..}}  \co{If we think in terms of
  evolutionary optimisation, we may imagine that enzyme concentrations
  are not ``free'', but also dependent on metabolite concentrations?
  so they are only ``free parameters'' if we disregard transcriptional
  regulation, ie on a timescale of a few minutes; for all longer
  timescales, it makes sense to see everything as entangled - so we
  can well start from requiring fluxes and metabolite concentrations
  and then walk our way back!}
}

\co{WO? ANOTHER INSIGHT from retromodelling: directly show that
  (state-neutral) enzyme regulation is costly and how much it will
  cost}

\section*{Code and data availability}

The STM algorithm is described in Supplementary
Materials. 
A MATLAB implementation with code for reaction elasticities is
available at \url{github.com/liebermeister/stm}. \co{move to gitlab?
  need to zenodo link?}  SBML and SBtab formats are used for models
and data.  The \emph{Escherichia coli} model, a modified version of
the model from \cite{nfbd:16}, is included. 

\section*{Acknowledgements}
I thank Edda Klipp, Elad Noor, Ronan Fleming, Andreas Hoppe, Hermann-Georg Holzh\"utter and
Mattia Zampieri for inspiring discussions.  This work was funded by
the German Research Foundation (Ll 1676/2-1 and Ll 1676/2-2).

\bibliographystyle{unsrt}
\bibliography{biology}


\newcommand{\psfilesSupp}{ps-files_supplement}
\newcommand{\linchainpsfiles}{\psfilesSupp/linear_chain}
\newcommand{\threonine}{ps-files/threonine/}
\newcommand{\hepatocyte}{ps-files/hepatocyte_glycolysis}

\newcommand{\thermcond}{C}

\clearpage
\centerline{\LARGE Supplementary Materials}

\begin{appendix}
\section{Mathematical symbols}

\coout{check if symbols are good + consistent with CBA}

\begin{table}[h!]
\begin{center}
 \begin{tabular}{ll}
  \hline
  \textbf{Network}\\ \hline
  Stoichiometric matrix (all metabolites) & $\Ntot$ \\
   Stoichiometric matrix (internal metabolites) & $\Nint = \Lmat\,\NR$ \\
   Link matrix and reduced stoichiometric matrix &  $\Lmat, \NR$\\
  Stoichiometric matrix (external metabolites) & $\Next$ \\
  Cooperativity coefficient & $h_{l}$ \\
  Stoichiometric coefficient & $n_{il}$ \\
  Reactant molecularity & $m^{\pm}_{li} = h_{l}\,|n_{il}|$ \\
   Activation coefficient & $m^{A}_{li}$\\
   Inhibition coefficient & $m^{I}_{li}$\\[2mm]
   \hline
  \textbf{Metabolic variables} \\
   \hline
   Flux & $v_{l}$ \\
   Internal metabolite level & $c_{i}$ \\
   External metabolite level & $x_{j}$ \\
   Enzyme level & $e_{l}$ \\[2mm]
  \hline
  \textbf{Thermodynamic variables} \\ 
  \hline
  Mass-action ratio &
  $q^{\rm ma}_{l} = \prod_{i} c_{i}^{n_{il}}$ \\
  Equilibrium constant & 
      $\keq_{l} = \prod_{i} \left( c_{i}^{\rm eq}\right)^{n_{il}}$ \\
  Standard chemical potential & $\mu^\circ_{i}$ \\
  Chemical potential & $\mu_{i} = \mu^\circ_{i} + RT\,\ln c_{i}$ \\
  Thermodynamic  force & $\theta_{l} = -\DeltaR \mu_{l}/RT$ \\
  One-way flux ratio $\zeta_{l}$ & $\zeta_{l} = v_{+l}/v_{-l} = \e^{h_{l}\,\theta_{l}}$\\[2mm]
   \hline
  \textbf{Rate laws} \\
  \hline
  Rate law & $\rate_{l}(\cv,\ev,\xv) = e_{l} \,\ratelaw_{l}(\cv,\xv)$ \\
  Michaelis-Menten constant & $\kM_{li}$ \\
  Activation constant & $\kA_{li}$ \\
  Inhibition constant & $\kI_{li}$ \\
  Catalytic constant & $\kcatpml$ \\
  Maximal velocity & $v^{\rm max}_{\pm l} = e_{l}\,\kcatpml = \max_{\cv,\xv} \rate_{l}(\cv,\ev,\xv)$ \\
  Velocity constant & $\kV_{l} = \sqrt{\kcatpl\,\kcatml}$ \\[2mm]
  \hline
  \textbf{Elasticity sampling} \\
  \hline
  Saturation value & $\beta^{\rm M}_{li}, \beta^{\rm A}_{li}, \beta^{\rm I}_{li}$ \\
  Unscaled elasticity & $\Euns^{v_{l}}_{c_{i}} = \frac{\partial \rate_{l}}{\partial c_{i}}$\\
  Scaled elasticity &  $\Escs^{v_{l}}_{c_{i}} = \frac{\partial \ln |\rate_{l}|}{\partial \ln c_{i}}$\\[2mm]
  \hline 
  \textbf{Metabolic control theory} \\
  \hline
  Steady-state flux & $v_{l} = \vsteady_{l}(\ev,\xv)$ \\
  Steady-state concentration & $c_{i} = \csteady_{i}(\ev,\xv)$ \\
  Jacobian matrix (independent metabolites) & $\Amat = \NR\,\Eunc\,\Lmat$ \\
  Unscaled response coefficient & 
     $\Run^{c_{i}}_{e_{l}} = \frac{\partial c_{i}}{\partial e_{l}}, 
     \Run^{v_{j}}_{e_{l}} = \frac{\partial v_{j}}{\partial e_{l}}$\\
  Unscaled control coefficients & 
   $\Cun^{c_{i}}_{v_{l}} = \Run^{c_{i}}_{e_{l}}/\Euns^{v_{l}}_{e_{l}},\,
   \Cun^{v_{j}}_{v_{l}} = \Run^{v_{j}}_{e_{l}}/\Euns^{v_{l}}_{e_{l}}$\\
  Scaled response/control coefficient & 
     $\Rsc^{c_{i}}_{e_{l}} = \Csc^{c_{i}}_{v_{l}} = \frac{\partial \ln c_{i}}{\partial \ln e_{l}}, 
     \Rsc^{v_{j}}_{e_{l}} = \Csc^{v_{j}}_{v_{l}} = \frac{\partial \ln |v_{j}|}{\partial \ln e_{l}}$\\
  \hline
 \end{tabular}
\end{center}
\caption{Symbols used in STM. Network elements are denoted by $i$
  (metabolites) and $l$ (reactions). Second-order elasticities,
  response coefficients (called synergy coefficients), and control
  coefficients are defined similarly to first-order
  coefficients.}
\label{tab:quantitylist}
\end{table}

\clearpage 

\section{Kinetic models and reaction elasticities}

\subsection{Kinetic models}
\label{sec:SIAlgorithm}

\begin{figure}[t!]
  \parbox{7.5cm}{\includegraphics[width=7cm]{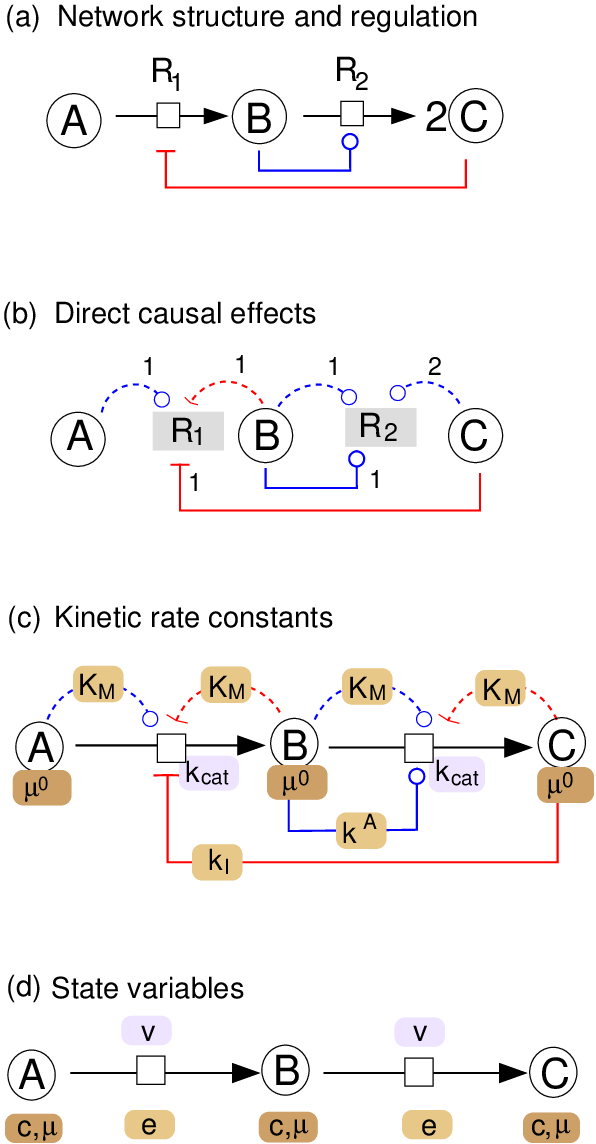}}\hspace{5mm}
  \parbox{8.5cm}{ \parbox{3cm}{\includegraphics[width=2cm]{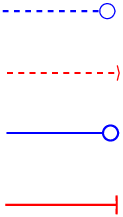}} \parbox{4cm}{
    $\Mmat^{\rm S} =   \left(\begin{array}{lll} 1&0&0\\0 & 1&0  \end{array}\right)$\\[2mm]
    $\Mmat^{\rm P} = \left(\begin{array}{lll} 0&1&0\\0 & 0&2 \end{array}\right)$\\[2mm]
    $\Mmat^{\rm A} =   \left(\begin{array}{lll} 0&0&0\\0 & 1&0 \end{array}\right)$\\[2mm]
    $\Mmat^{\rm I} = \left(\begin{array}{lll} 0&0&1\\0 &
      0&0 \end{array}\right)$
   }\ \\
   \caption{\textbf{Structure and kinetic description of metabolic networks.}
     (a) Metabolic network and structure matrices. Our metabolic
     pathway consists of two reactions $A \leftrightharpoons B$ and
     $B \leftrightharpoons 2\, C$.  Reaction R$_{1}$ is inhibited by
     the end product C, while reaction R$_{2}$ is activated by its own
     substrate B. (b) Pathway shown as a bipartite network. Each arrow
     carries a number: dotted arrows show stoichiometric coefficients
     for substrates and products, regulation arrows (solid) carry
     coefficients of 1. Right: the arrows and coefficients can be
     represented by network matrices $\Mmat^{\rm S}$, $\Mmat^{\rm P}$,
     $\Mmat^{\rm A}$, and $\Mmat^{\rm I}$ whose rows and columns
     correspond to reactions and metabolites,
     respectively. $\Mmat^{\rm S}$ and $\Mmat^{\rm P}$ follow directly
     from the stoichiometric matrix. (c) Kinetic constants in modular
     rate laws. The standard chemical potentials $\mu^\circ_{i}$ of
     metabolites determine the equilibrium constants. All
     stoichiometry or regulation arrows are associated with
     dissociation constants $\kM_{li}$, $\kA_{li}$, or $\kI_{li}$. The
     velocity constants $\kV_{l} = \sqrt{\kcatpl\,\kcatml}$ are
     assigned to reactions. Together, the constants determine the
     catalytic constants $\kcatpml$. (d) A metabolic state is
     characterised by fluxes $v_{l}$, metabolite levels $c_{i}$,
     chemical potentials $\mu_{i}=\mu_{i}^{\circ}+RT\,\ln c_{i}$, and
     enzyme levels $e_{l}$.}
 }\\[4mm]
  \label{fig:structurematrices}
\end{figure}

\myparagraph{\ \\Metabolic networks and kinetic models} A metabolic
network is defined by a set of chemical reactions and regulatory
arrows pointing from metabolites to reactions (Figure
\ref{fig:structurematrices} (a)).  The molecularities\footnote{The
molecularities resemble stoichiometric coefficients, but with a slight
difference: while stoichiometric cofficients (in a reaction sum
formula) can be rescaled, the molecularities are actual molecule
numbers in the enzyme mechanism, and therefore uniquely
determined. They can be written as the (absolute) stoichiometric
cofficients multiplied by a cooperativity exponent $h_{l}$ for each
reaction (typically $h_{l}=1$).) \cite{liuk:10}. In the formula for
thermodynamic forces, $\thetav={\Ntot}\trans \muv/RT$, we tacitly
assume that stoichiometric coefficients are given by
molecularities. Otherwise this formula must contain $h_{l}$ as a
prefactor.} $m^{\rm S}_{li}$
(for substrates) and $m^{\rm P}_{li}$ (for products) are given by the
stoichiometric coefficients $n_{li}$ between metabolite $i$ and enzyme
$l$, multiplied by the reaction's cooperativity coefficient $h_{l}$
(i.e.~$h_{l}\,n_{li} = m^{\rm S}_{li}-m^{\rm P}_{li}$)
\cite{liuk:10}. The stoichiometric coefficients and regulation
coefficients $m^{\rm A}_{li}$ and $m^{\rm I}_{li}$ (activation:
$m^{\rm A}_{li}=1$; inhibition: $m^{\rm I}_{li}=1$; zero values
otherwise) are collected in matrices defining the network.  Kinetic
models \cite{hesc:96} describe reaction rates by rate laws
$\rate_{l}(\cv,\ev,\xv) = e_{l}\,\ratelaw_{l}(\cv,\xv)$ (Figure
\ref{fig:structurematrices} (b)). Modular rate laws \cite{liuk:10}
(see below) contain two types of kinetic constants: catalytic
constants $\kcatpm$ (in s$\inv$) describe the speed of the forward and
backward rates, while dissociation constants $\kM$ for reactants,
activation constants $\kA$ for activators, and inhibition constants
$\kI$ for inhibitors (in mM).  For each reaction, a certain ratio of
these parameters must be equal to the equilibrium constant (Haldane
relationship). Given stoichiometric matrix and rate laws, we obtain
the dynamic rate equations
$\md c_{i}/\md t = \frac{1}{V_{i}} \sum_{l}
n_{il}\,\rate_{l}(\cv,\ev,\xv)$ for internal metabolite concentrations
$c_{i}$, with external metabolite levels $x_{j}$ and enzyme levels
$e_{l}$ as parameters. We assume that all metabolites $i$ are
homogeneously distributed within cell compartments of constant size
$V_{i}$. Metabolite concentrations are given in mM = mol/m$^{3}$,
reaction rates as amounts per time (mol/s), enzyme levels as amounts
(mol), and volumes in m$^{3}$. In single-compartment models, we may
choose a compartment size of 1 (dimensionless) and measure reaction
rates in mM/s and enzyme levels in mM. If we analyse steady
states, the choice of flux units does not play a role.

\myparagraph{Modular rate laws} Modular rate laws \cite{liuk:10} are
generic reversible rate laws that capture various reaction
stoichiometries, enzyme mechanisms, and types of regulation by
effector molecules. Formulae for different rate laws and their
elasticities can be found in \cite{liuk:10} (supplementary material).
As an example, let us consider a reaction $A+B \leftrightharpoons 2C$
without effectors. The common modular (CM) rate law is a reversible
Michaelis-Menten kinetics, generalised for arbitrary
stoichiometries. With two substrates A and B and one product C, it
reads
\begin{eqnarray}
 \label{eq:CSkineticsexample}
 \rate(a,b,c,e) = e\, \frac{ \kcatp \,(a/\kM_{\rm A}), (b/\kM_{\rm B}) 
  -\kcatm \,(c/\kM_{\rm C})^{2}}
 {(1+a/\kM_{\rm A})(1+b/\kM_{\rm B}) + (1+c/\kM_{\rm C})^{2} -1}
\end{eqnarray}
with reactant constants $\kM_{\rm A}$, $\kM_{\rm B}$, and
$\kM_{\rm C}$ (in mM) and catalytic constants $\kcatp$ and $\kcatm$
(in s$^{-1}$) for forward and backward direction.  Modular rate laws
can be adapted to different types of reactions and enzymes: if an
enzyme is regulated by an effector molecule, this can be described,
for example, by prefactors
$\beta^{\rm A}_{\rm X}=\frac{x/\kA_{\rm X}}{1+x/\kA_{\rm X}}$ for
activators X or $\alpha^{\rm I}_{\rm Y} = \frac{1}{1+y/\kI_{\rm Y}}$
for inhibitors Y. Moreover, with a Hill-like exponent $h$ the rate
laws can capture sigmoidal kinetics. Other types of rate laws use the
same parameters, but another denominator, for example, the saturable
modular (SM) rate law
\begin{eqnarray}
 \label{eq:MSkineticsexample}
 \rate(a,b,c,e) = e\, \frac{ \kcatp \,(a/\kM_{\rm A})\, (b/\kM_{\rm B}) 
  -\kcatm \,(c/\kM_{\rm C})^{2}}
 {(1+a/\kM_{\rm A})(1+b/\kM_{\rm B})(1+c/\kM_{\rm C})^{2}}.
\end{eqnarray}
In the denominator, the substrate and product terms are simply
multiplied. Modular rate laws assume an enzyme mechanism in which
substrates and products bind rapidly, independently, and in random
order.  The reactant constants $\kM_{li}$ are dissociation constants
of the elementary binding steps. Like the $\kM$ values in
Michaelis-Menten kinetics, they denote reactant concentrations that
would yield a half-maximal saturation (or $1/|n_{il}|$-maximal
saturation if $|n_{il}| > 1$). The catalytic constants $\kcatpm$ stem
from the slow conversion step between substrate and product molecules.
In all modular rate laws, the kinetic constants are required to
satisfy Haldane relationships. One way to ensure this is to predefine
the equilibrium constants, to treat the velocity constants $\kV$ as
free parameters, and to compute the catalytic constants by using
Eq.~(\ref{eq:turnoverrates}).

\myparagraph{Thermodynamic laws} The kinetics and steady states
of metabolic systems are constrained by  thermodynamic, which defines relationships between
metabolite concentrations $c_{i}$, reaction rates $v_{l}$,
equilibrium constants $\keq_{l}$, and chemical potentials
$\mu_{i}$. The chemical potentials $\mu_{i}$ are defined as the
derivatives $\mu_{i}= \partial G/\partial n_{i}$ of the system's total
Gibbs free energy by the metabolite mole numbers $n_{i}$. For ideal
mixtures (metabolites at low concentrations in an aqueous solution, no
mixture effects, activity coefficients of 1), the chemical potential of metabolite $i$
is given by the
formula
\begin{eqnarray}
 \label{eq:muideal}
 \mu_{i} &=& \mu_{i}^\circ + R T \ln c_{i}/c_{\rm std}
\end{eqnarray}
with Boltzmann gas constant $R$, absolute temperature $T$, and
chemical potential $\mu_{i}^\circ$ of metabolite $i$ at
standard concentration\footnote{For kinetic models, it
  is convenient to use a standard concentration of 1 mM, equal to the
  measurement unit for concentrations. In thermodynamic flux analysis,
  the common standard concentration is 1M. The conversion between the
  two conventions requires an adjustment of reaction Gibbs free
  energies, standard Gibbs free energies of formation,
  and equilibrium constants.} $c_{\rm std}$. Here I omit
the division by $c_{\rm std}$, assuming that all concentrations are
given in units of the standard concentration.  The ratio
$\kma_{l} = \prod_{i} (c_{i})^{n_{il}}$ of product and substrate
concentrations for a reaction is called mass-action ratio.  In
chemical equilibrium states, this ratio has always the same value,
called equilibrium constant $\keq_{l}$, which can be written
as $\keq = \e^{-\Delta \mu^\circ/RT}$. The thermodynamic force $\theta_r$,
in a (possibly non-steady) metabolic state is defined as
\begin{eqnarray}
 \label{eq:definitionA}
\theta_{l} &=& -\frac{1}{RT}\DeltaR \mu_{l} = - \frac{1}{RT} \sum_{i} \mu_{i}\,n_{il}.
\end{eqnarray}
It describes the Gibbs free energy dissipation (in kJ/mol) associated
with a reaction event and can be computed from equilibrium constant
$\keq_{l}$ and mass-action ratio $q^{\rm ma}_{l}$ (for reaction $l$):
\begin{eqnarray}
 \label{eq:reactionaffinity}
 \theta_{l} = -\frac{1}{RT}\sum_{i} n_{il}\, \mu_{i} 
 =  \ln \keq_{l} / q^{\rm ma}_{l}.
\end{eqnarray}
In generalised Michaelis-Menten rate laws, the ratio of  forward and
backward rates $v_{l}^{\pm}$ of reaction $l$ is given by 
\begin{eqnarray}
 \label{eq:velocityratio}
 \zeta_{l} = v_{+l}/v_{-l} = \e^{h_{l}\,\theta_{l}} 
 = \keq / \prod c_{i}^{n_{i}}.
\end{eqnarray}
The thermodynamic-kinetic formalism \cite{edgi:07}  defines the thermokinetic potential
$\xi_{i} = \e^{\mu_{i}/RT}$ and splits it into
$\xi_{i} = C_{i}\, c_{i}$, where $C_{i}$ is called capacity. For ideal
mixtures (satisfying Eq.~(\ref{eq:muideal})), the thermokinetic
potential is given by $\xi_{i} = \e^{\mu^\circ_{i}/RT}\, c_{i}$ with a
capacity $C_{i}=\e^{\mu^\circ_{i}/RT}$, and related to  $\zeta_{l}$ via
  \begin{eqnarray}
    \zeta_{l} = \e^{h_{h}\,\theta_{l}} = \e^{-h_{h}\,\DeltaR \mu_{l}} =
    \prod_{i} (\e^{\mu_{i}/RT})^{-h_{l}\,n_{il}} = \prod_{i}
    \xi_{i}^{-h_{l}\,n_{il}}.
  \end{eqnarray}
In kinetic models, thermodynamic laws impose three sorts of constraints:
a relation between flux directions and thermodynamic forces, Wegscheider
conditions for equilibrium constants, and Haldane relationships
between equilibrium constants and kinetic parameters. They arise as
follows.  (i) To carry a non-zero flux, chemical reactions must show a
positive production of entropy per volume and time,
$\sigma_{l} = v_{l}\, A_{l}/T = R v_{l}\,\theta_{l}$ (with the
reaction affinity $A=-\DeltaR G$). As a consequence, a (non-zero)
reaction rate $v_{l}$ and the corresponding thermodynamic force $\theta_{l}$
must point in the same direction. (ii) The vector $\ln \kv^{\rm eq}$
of logarithmic equilibrium constants ca be written as
$\ln \kv^{\rm eq} = {\Ntot}\trans \ln \cv^{\rm eq}$, with a vector
$\cv^{\rm eq}$ of metabolite concentrations in an equilibrium
state. For any such vector $\ln \kv^{\rm eq}$, the Wegscheider
conditions $\Kmat\trans \ln \kv^{\rm eq} = 0$ have to be satisfied
\cite{wegs:02,scsc:89} , where $\Kmat$ is a null space matrix
satisfying $\Nint\,\Kmat=0$.  Similar Wegscheider conditions hold for
all vectors of the form $\xv = \Ntot\,\yv$, including logarithmic
mass-action ratios $\ln q^{\rm ma}_{l}$ and thermodynamic forces
$\theta_{l}$. (iii) The fact that reaction rates
$v_{l}(\cv^{\rm eq},\ev)$ vanish at chemical equilibrium implies a
Haldane relationship between equilibrium constant and kinetic  constants
\cite{likl:05,edgi:07}. For all modular rate laws, the Haldane relationship
reads
\begin{eqnarray}
 \label{eq:haldanerelationapp}
 \keq_{l} &=& \kcatpl/\kcatml \prod\limits_{i} (\kM_{li})^{n_{il}}.
\end{eqnarray}
To construct parameter sets that satisfy this relation, we define the
turnover constants $\kV_{l} = \sqrt{\kcatpl\, \kcatml}$ as the
geometric means of forward and backward catalytic constants. By
rewriting Eq.~(\ref{eq:haldanerelationapp}), we can now express the
forward and backward catalytic constants as
\begin{eqnarray}
 \label{eq:turnoverrates}
 \kcatpml &=&  \kV_{l} ( \keq_{l} \prod\limits_{i} 
 (\kM_{li})^{-n_{il}} )^{\pm 1/2}.
\end{eqnarray}
These values satisfy the Haldane relationship by construction.

\subsection{Reaction elasticities and thermodynamics}
\label{sec:SIelastAndThermodynamics}

\myparagraph{\ \\Scaled and unscaled reaction elasticities}
\label{sec:SIelasticityconversion}
The derivatives between kinetic laws $\rate_{l}(\cdot)$ and enzyme
concentrations $e_{p}$, metabolite concentrations $c_{j}$, or other
function arguments are called reaction elasticities (see Figure
\ref{fig:elasticitiesthermoandcontrol} (a)). Given a rate law
$\rate_{l}(e_{l},\cv)$, the unscaled reaction elasticities are defined
by derivatives
\begin{eqnarray}
 \label{eq:unscaledelasticitiessupp}
 \Euns^{v_{l}}_{c_{i}} = \frac{\partial \rate_{l}}{\partial c_{i}}, \quad
 \Euns^{v_{l}}_{c_{i} c_{j}} = \frac{\partial^{2} \rate_{l}}{\partial c_{i}\, \partial c_{j}},
\end{eqnarray}
while the corresponding scaled elasticities are defined by logarithmic
derivatives
\begin{eqnarray}
 \label{eq:scaledelasticitiessupp}
 \Escs^{v_{l}}_{c_{i}} = 
 \frac{\partial \ln |\rate_{l}| }{ \partial \ln c_{i}}, \quad
 \Escs^{v_{l}}_{c_{i} c_{j}} 
 = \frac{\partial^{2} \ln |\rate_{l}|}{\partial \ln c_{i}\, \partial \ln c_{j}}.
\end{eqnarray}
Elasticities for other arguments of the rate law function (e.g.~the
enzyme level $\esymbol_{l}$) are defined accordingly.  Scaled
elasticities are dimensionless and can be seen as  effective reaction
orders: for mass-action kinetics, they are given by the substrate
molecularities; for an enzyme that is fully saturated  with the
metabolite in question, they vanish. Scaled and unscaled elasticities
can be interconverted by
\begin{eqnarray}
 \label{eq:elasticityscaling}
 \Escs^{v_{l}}_{c_{i}} =
 \frac{c_{i}}{v_{l}} \Euns^{v_{l}}_{c_{i}}, \qquad
 \Escs^{v_{l}}_{c_{i} c_{j}} 
 = \frac{c_{i}\,c_{j}}{v_{l}} \Euns^{v_{l}}_{c_{i} c_{j}} 
 - \frac{c_{i}\,c_{j}}{v_{l}^{2}}
 \Euns^{v_{l}}_{c_{i}} \Euns^{v_{l}}_{c_{j}} 
 + \delta_{ij} \frac{c_{i}}{v_{l}} \Euns^{v_{l}}_{c_{i}}
\end{eqnarray}
and
\begin{eqnarray}
 \label{eq:elasticityunscaling}
 \Euns^{v_{l}}_{c_{i}} = \frac{v_{l}}{c_{i}}\, \Escs^{v_{l}}_{c_{i}}, \qquad
 \Euns^{v_{l}}_{c_{i} c_{j}} 
 = \frac{v_{l}}{c_{i}\,c_{j}} 
 \left[ 
  \Escs^{v_{l}}_{c_{i} c_{j}} 
  + \Escs^{v_{l}}_{c_{i}}\, \Escs^{v_{l}}_{c_{j}} 
  - \delta_{ij} \Escs^{v_{l}}_{c_{i}}
 \right].
\end{eqnarray}
Analogous conversion formulae hold for all types of sensitivities,
including elasticities with respect to other parameters, control
coefficients, and response coefficients \cite{rede:88,hofm:01}.

\myparagraph{Elasticities and thermodynamic force}  Elasticities depend on the rate laws, but also on
thermodynamics. In reversible rate laws, the net reaction rate
$v_{l} = v_{+l}-v_{-l}$ is the difference of forward and backward
 rates, whose ratio $v_{+l}/v_{-l} = \e^{\theta_{l}}$ is
determined by the thermodynamic force $\theta_{l}$.  The thermodynamic force, in
turn, depends on reactant concentrations and equilibrium constant as
$\theta_{l} = - \DeltaR G_{l}/RT = \ln \frac{\keq}{\prod_{i}
  c_{i}^{n_{il}}}$ (see Figure \ref{fig:elasticitiesthermoandcontrol}
(b)). If the thermodynamic force is large, the forward flux dominates and
the net rate becomes sensitive to substrate fluctuations, but less
sensitive to product fluctuations; therefore, the substrate elasticity
increases and the product elasticity decreases. Near chemical
equilibrium, where thermodynamic forces come close to zero, the scaled
elasticities go to infinity.

\myparagraph{Elasticities of modular rate laws}
\label{sec:SImodularratelawselasticities}
The scaled elasticities of the SM rate law
Eq.~(\ref{eq:MSkineticsexample}) contain the thermodynamic term as
well as four terms that correspond to substrates, products,
activators, and inhibitors:
\begin{eqnarray}
 \label{eq:totalelasticities}
 \Escs^{v_l}_{c_i} 
 = 
 \frac{ \zeta_l\, m^{\rm S}_{li} - m^{\rm P}_{li}}{\zeta_l-1}
 -  m^{\rm S}_{li} \alpha^{\rm M}_{li}
 - m^{\rm P}_{li} \beta^{\rm M}_{li}
 + m^{\rm A}_{li} \, \alpha^{\rm A}_{li} 
 - m^{\rm I}_{li}\,  \beta^{\rm I}_{li}.
\end{eqnarray}
For near-equilibrium reactions (small $\theta_{l}$) and for strongly
driven reactions ($|\theta_{l}| \rightarrow \infty$), the first
three terms can be approximated by
\begin{eqnarray}
 \label{eq:netelasticities2}
 |\theta_{l}|\,\approx 0 &:&
 \frac{1}{\theta_{l}} (m^{\rm S}_{li} - m^{\rm P}_{li})
 + m^{\rm S}_{li} \alpha^{\rm M}_{li}
 - m^{\rm P}_{li} \beta ^{\rm M}_{li} 
 \nonumber \\
 \theta_{l} \rightarrow \infty &:&
 (m^{\rm S}_{li}-m^{\rm P}_{li}) \, \e^{-\theta_{l}}
 + m^{\rm S}_{li} \alpha^{\rm M}_{li}
 - m^{\rm P}_{li} \beta ^{\rm M}_{li} 
 \nonumber \\
 \theta_{l} \rightarrow -\infty &:&
 (m^{\rm P}_{li}-m^{\rm S}_{li}) \e^{-|\theta_{l}|}
 -  m^{\rm S}_{li} \, \beta^{\rm M}_{li}
 + m^{\rm P}_{li} \alpha^{\rm M}_{li}.
\end{eqnarray}
The last two terms in these formulae represent exactly the formula
used in SKM \cite{sgsb:06}, while the first term employs a
thermodynamic correction. The scaled elasticities of the common
modular rate law are a bit more complicated: 
{\small
 \begin{eqnarray}
  \Escs^{v_{l}}_{c_{j}} = 
  \beta_{lj} 
  \frac { \zeta_{l}\, m^{\rm S}_{lj} - m^{\rm P}_{lj} } { \zeta_{l} - 1 }
  -   \beta_{lj} \frac
  { m^{\rm S}_{lj} \psi^{+}_{l} + m^{\rm P}_{lj}  \psi^{-}_{l} }
  { \psi^{+}_{l} + \psi^{-}_{l} -1}
  +  m^{\rm A}_{li} \, \alpha^{\rm A}_{li} - m^{\rm I}_{li}\,  \beta^{\rm I}_{li},
 \end{eqnarray}} 
where $\psi^{\pm}_{l}=\prod_{l}(1+c_{i}/\kM_{li})^{m^{\pm}_{li}}$ (see \cite{liuk:10}).
Formulae for second-order elasticities, unscaled elasticities,
parameter elasticities, and other types of modular rate laws can be
found in \cite{liuk:10}.

\myparagraph{Elasticities of factorised rate laws (derivation
    of Eq.~(\ref{eq:FacRateLawElast}))} For the factorized rate laws,
  and assuming a positive flux $v>0$ for simplicity, we obtain
\begin{eqnarray}
\Escs^{v_{l}}_{c_{j}} &=& \frac{\partial \ln v_{l}}{\partial \ln c_{i}} = \frac{\partial \ln (1-\e^{-\theta_{l}})}{\partial \ln c_{i}} + \frac{\partial \ln \eta^{\rm kin}}{\partial \ln c_{i}}
\end{eqnarray}
where
\begin{eqnarray}
  \frac{\partial \ln (1-\e^{-\theta_{l}})}{\partial \ln c_{i}} = \frac{1}{1-\e^{-\theta_{l}}}(-\e^{-\theta_{l}}) (-) \partial \theta_{l}/\partial \ln c_{i} = \frac{\e^{-\theta_{l}}}{1-\e^{-\theta_{l}}}(-n_{il}) = \frac{-1}{\e^{-\theta_{l}}-1}n_{il}.
\end{eqnarray}

\myparagraph{Any set of saturation values yields a consistent model}
\label{sec:SIElSampConsistent}
STM relies on the fact that, given consistent fluxes and thermodynamic
forces, any choice of the saturation values yields a consistent
kinetic model and that any consistent model can be constructed in this
way.  This can be proven as follows. Consider a kinetic model with
modular rate laws and a thermodynamically consistent flux distribution
$\vv$. For simplicity, enzyme levels are subsumed in the catalytic
constants $\kcatpml$.  As shown in \cite{liuk:10}, a consistent set of
parameters, realising $\vv$, can be obtained by the following
procedure:
\begin{enumerate}
\item Choose standard chemical potentials $\mu_{i}^\circ$  (free choice) and
 determine the equilibrium constants.
\item Determine concentrations $c_{i}$ such that the signs of thermodynamic
  forces agree with the flux directions. If the metabolite
  concentrations are bounded, this may not always be possible, even if
  the flux distribution is loopless.
\item Choose Michaelis constants $\kM_{li}$ and activation and
  inhibition constants $\kA_{li}$ and $\kI_{li}$ (free
  choice). Of course, given the previously chosen metabolite concentrations, this is equivalent to
  choosing the saturation
    constants (free choice in the range between 0 and 1) and computing
    the $\kX_{li}$ values from them.
\item Set all velocity constants to preliminary values
  ${\kV_{l}}'$. Compute the catalytic constants ${\kcatpml}'$ from the
  Haldane relationships (\ref{eq:haldanerelationapp}). Use the rate
  laws to compute the reaction rates $v_{l}'$. By construction (due to
  thermodynamically feasible metabolite levels and thermodynamically
  consistent rate laws), these rates have the same signs as the
  predefined fluxes. To match reaction rates and fluxes exactly, we
  just need to adjust the velocity constants, setting
  $\kV_{l} = (v_{l}/v'_{l})\,{\kV}'_{l}$.
\item If our flux distribution contains inactive reactions, we can
 decide, for each of them, whether we assume a vanishing
 thermodynamic force, a vanishing velocity constant, or a vanishing
 enzyme level. In the first case, we need to apply the strict
 energetic feasibility criterion for this reaction (i.e.~require
 that the thermodynamic force vanishes); in the other cases, there is
 no feasibility criterion for the reaction, and the $\kV$ value
 or enzyme level is set to zero.
\end{enumerate}

If this procedure yields correct models, this also holds for all
models constructed by STM. In STM, we first determine consistent
fluxes $v_{l}$ and thermodynamic forces $\theta_{l}$ that can be realised by
a choice of standard chemical potentials $\mu_{i}^\circ$ and
concentrations $c_{i}$. Thus, when  choosing the saturation constants, any
choice is equivalent to a choice of $\kM_{li}$, $\kA_{li}$, and
$kI_{li}$ in the algorithm above. Thus, the quantities chosen until
this point correspond exactly to the results of step 3. Thus, steps 4
and 5 will yield a unique, consistent set of parameters. Therefore,
models obtained by STM satisfy all relevant constraints.

\myparagraph{Independently sampled elasticities would yield inconsistent models}
\label{sec:SISKMinconsistent}
A main problem with SKM is that elasticities are sampled
independently, which means that the resulting kinetic models may
violate important constraints. If the forward and backward one-way
rates of reactions were independent, and nont constrained by
thermodynamics, an independent sampling would be justified: the
elasticities could be directly translated into kinetic constants, and
each sampled elasticity matrix would define a specific kinetic
model. However, in models with reversible rate laws, independently
sampled elasticities lead to inconsistent results. For example,
consider a reaction A $\leftrightharpoons$ B with reversible
mass-action kinetics $v=k^{+}\,a-k^{-}\,b$: the scaled reaction
elasticities read $\Escs_{\rm A}=k^{+}\,a/v$ (for substrate A) and
$\Escs_{\rm B}=k^{+}\,b/v$ (for product B), so their difference
$\Escs_{\rm A}-\Escs_{\rm B}=1$ is fixed. If we sample these
elasticities independently, this relationship is violated and our
sampled values cannot be realised by reversible mass-action rate
laws. Similar constraints hold for all thermodynamically consistent
reversible rate laws.

There can also be inconsistencies between the elasticities of
different reactions.  Here is a simple example: the reaction A
$\rightarrow$ B is catalysed by two isoenzymes with reversible
mass-action kinetics:
\begin{eqnarray}
 \label{eq:dependentelasticities1}
 v_{1} &=& \rate_{1}(a,b) = k^{+}_{1}\, a - k^{-}_{1}\, b \nonumber \\
 v_{2} &=& \rate_{2}(a,b) = k^{+}_{2}\, a - k^{-}_{2}\, b.
\end{eqnarray}
The symbols $a$ and $b$ denote the concentrations of A and B, and
$k^{\pm}_{1}$ and $k^{\pm}_{2}$ denote the rate constants. In each
reaction, forward and backward kinetic constant must have the same
ratio given by the equilibrium constant:
\begin{eqnarray}
 \label{eq:dependentelasticities2}
 \keq &=& 
 \frac{k^{+}_{1}}{k^{-}_{1}} = \frac{k^{+}_{2}}{k^{-}_{2}}.
\end{eqnarray}
Therefore, the scaled elasticity matrix can be written as 
\begin{eqnarray}
 \label{eq:dependentelasticities3}
 \Ematsc &=& 
 \begin{pmatrix} \frac{\partial \ln \rate_{1}}{\partial \ln a} & 
  \frac{\partial \ln \rate_{1}}{\partial \ln b}\\
  \frac{\partial \ln \rate_{2}}{\partial \ln a} & 
  \frac{\partial \ln \rate_{2}}{\partial \ln b}
 \end{pmatrix} 
 =
 \begin{pmatrix} k^{+}_{1}\,\frac{a}{v_{1}} & - k^{-}_{1}\,\frac{b}{v_{1}}
  \\ k^{+}_{2}\,\frac{a}{v_{2}} & - k^{-}_{2}\,\frac{b}{v_{2}} \end{pmatrix} 
 = \begin{pmatrix} \frac{\keq\,a}{\keq\,a -b} &
  \frac{- b}{\keq\,a -b} \\
  \frac{\keq\,a}{\keq\,a -b} &
  \frac{-b}{\keq\,a -b} 
 \end{pmatrix}
 = \begin{pmatrix} \frac{\zeta}{\zeta-1 } & \frac{-1}{\zeta-1 } \\ \frac{\zeta}{\zeta-1 } & \frac{-1}{\zeta-1 } \end{pmatrix} 
\end{eqnarray}
with $\zeta = \keq/(b/a)$. All four elasticities are determined by
the same parameter $\zeta$, so sampling them independently  leads
to a contradiction.

\myparagraph{Dependencies between first- and second-order elasticities}
\label{sec:SIcorrfirstsecelast} 
For a given modular rate law, the scaled elasticities can be computed
from stoichiometric coefficients, thermodynamic forces, and saturation
values. The fact that the same model details (e.g.~the thermodynamic force
of a reaction) have an influence on different reaction elasticities
leads to dependencies between these elasticities: if these factors are
varied, the resulting elasticities will be statistically
dependent. Here is an example.  When a thermodynamic force becomes larger,
the substrate elasticities tend to increase and the product
elasticities tend to decrease. When comparing all elasticities in a
network, or when comparing the different instances of an ensemble
model, this relationship between thermodynamic force and elasticities leads
to positive correlations among substrate elasticities, positive
correlations among product elasticities, and negative correlations
between substrate and product elasticities in each single reaction. On
the contrary, the unscaled elasticities in a pathway tend to increase
with the pathway flux, leading to positive correlations between them
(when comparing model instances with different flux
distributions). Just like first-order elasticities are dependent, also
the second-order elasticities are dependent on them.  This entails
statistical dependencies, both between the different sampled instances
of a model (e.g.,~assuming different values of the thermodynamic forces) and
between the reactions in a single model.  Second-order elasticities
$\Escs^{v_{l}}_{c_{i}c_{j}}$ tend to be negatively correlated with the
product $\Escs^{v_{l}}_{c_{i}}\,\Escs^{v_{l}}_{c_{j}}$. To see this,
consider a simple mass-action or power-law rate law without
regulation: the elasticities are directly given by the thermodynamic
terms
\begin{eqnarray}
 \Escs^{v_{l}}_{c_{i}} &=&  \frac{m^{\rm S}_{li}\,\zeta_{l} - m^{\rm P}_{li}}{\zeta_{l}-1} 
 = \left\{ 
  \begin{array}{lll}
   i\;\mbox{is a substrate}&:&
   \frac{\zeta_{l}}{\zeta_{l}-1}\, m^{\rm S}_{li}\\
   i\;\mbox{is a product} &:&
   \frac{-1}{\zeta_{l}-1}\, m^{\rm P}_{li}
  \end{array}
 \right.\\
 \Escs^{v_{l}}_{c_{i}c_{j}} &=& - \frac{\zeta_{l}\,h_{l}^{2}\, n_{il}\, n_{jl}}{(\zeta_{l}-1)^{2}}
 = \left\{ 
  \begin{array}{rlrlr}
   \mbox{$i$, $j$ are substrates}&:&
   - \frac{\zeta_{l}\, m^{\rm S}_{li}\, m^{\rm S}_{lj}}{(\zeta_{l}-1)^{2}} 
   &\approx& -\frac{1}{\zeta}\,\Escs^{v_{l}}_{c_{i}}\,\Escs^{v_{l}}_{c_{j}}\\
   \mbox{one substrate, one product} &:&
   \frac{\zeta_{l}\, m^{\rm S}_{li}\, m^{\rm P}_{lj}}{(\zeta_{l}-1)^{2}}
   &\approx& - \,\Escs^{v_{l}}_{c_{i}}\,\Escs^{v_{l}}_{c_{j}}\\
   \mbox{$i$, $j$ are products} &:&
   - \frac{\zeta_{l}\, m^{\rm P}_{li}\, m^{\rm P}_{lj}}{(\zeta_{l}-1)^{2}}
   &\approx& - \zeta\,\,\Escs^{v_{l}}_{c_{i}}\,\Escs^{v_{l}}_{c_{j}}
  \end{array}
 \right..
\end{eqnarray}
In this formula, the distinction between substrates and product is not
made based on the actual flux direction, but on their roles in the
reaction formula; the flux direction enters the formulae via
$\zeta_{l}$, which may be larger or smaller than 1, depending on the
sign of the thermodynamic force and thus on the flux direction.  The
second-order elasticities read
\begin{eqnarray}
 \label{eq:secondorderelasticitysplitting}
 \Escs^{v_{l}}_{c_{i}c_{j}} = \vartheta^{l}_{ij}\,\Escs^{v_{l}}_{c_{i}}\,\Escs^{v_{l}}_{c_{j}}
\end{eqnarray} 
where, in the present case (non-regulated mass-action rate law), the
prefactor $\vartheta^{l}_{ij}$ reads
\begin{eqnarray}
 \label{eq:splittingprefactor}
 \vartheta^{l}_{ij} = \left\{ 
  \begin{array}{rlrlr}
   \mbox{$i$, $j$ are substrates}&:& -1/\zeta_{l}\\
   \mbox{one substrate, one product} &:& -1\\
   \mbox{$i$, $j$ are products} &:&-\zeta_{l}
  \end{array}
 \right.
\end{eqnarray}
Due to this negative prefactor, we can expect a negative statistical
correlation between the second-order elasticity
$\Escs^{v_{l}}_{c_{i}c_{j}}$ and the product
$\Escs^{v_{l}}_{c_{i}}\,\Escs^{v_{l}}_{c_{j}}$ of first-order
elasticities.  In particular, close to equilibrium (where
$\theta_{l}\approx 0$ and therefore $\zeta_{l}\approx 1$), we obtain
$\vartheta^{l}_{ij} \approx -1$ and thus the general formula
$\Escs^{v_{l}}_{c_{i}c_{j}} \approx -
\Escs^{v_{l}}_{c_{i}}\,\Escs^{v_{l}}_{c_{j}}$, which is symmetric
between substrates and products. For completely forward-driven reactions
(with $\theta_{l} \rightarrow \infty$ and
$\zeta_{l} \rightarrow \infty$), in contrast, we obtain
$\Escs^{v_{l}}_{c_{i}c_{j}} \approx 0 $ because the factor
$\zeta_{l}/(\zeta_{l}-1)^{2}$ is close to 0. Can we expect the saxme
relationship also for other rate laws? For generic saturable rate
laws, a splitting as in Eq.~(\ref{eq:secondorderelasticitysplitting})
is formally possible, but there is no simple formula for
$\vartheta^{l}_{ij}$. Therefore, a tendency for negative correlations
between $\Escs^{v_{l}}_{c_{i}c_{j}}$ and
$\Escs^{v_{l}}_{c_{i}}\,\Escs^{v_{l}}_{c_{j}}$ may remain, but the
negative correlation will be weaker.

\subsection{Metabolic control theory} 

\myparagraph{\ \\Steady states and metabolic control} A steady state
is a metabolic state in which metabolite levels and fluxes are
constant in time.  Steady-state fluxes $v_{l}(\ev,\xv)$ and
concentrations $c_{i}(\ev,\xv)$ depend on enzyme levels $e_{l}$ and
external metabolite levels $x_{j}$. These dependencies may be
complicated and not explicitly known (see Figure
\ref{fig:structurematrices} (c)).  However, if steady-state
concentrations and fluxes are given, their sensitivities to parameter
changes can be computed from the elasticities.  The sensitivity
$\Run^{y}_{p_{m}} = \partial y/\partial p_{m}$ between an target
variable $y$ -- e.g.~a stationary concentration $c_{i}$ or a flux
$v_{l}$ -- and model parameters $p_{m}$ is called response
coefficient. If each reaction has one reaction-specific parameter
$p_{l}$, for example the enzyme level $e_{l}$, then we can divide the
response coefficients $\Run^{\rm y}_{il}$ by the elasticities
$\Euns^{v_{l}}_{p_{l}}$ and obtain the control coefficients
$\Cun^{y}_{v_{l}} = \Run^{y}_{p_{l}} / \Euns^{v_{l}}_{p_{l}}$ (Figure
\ref{fig:elasticitiesthermoandcontrol} (c)). Control coefficients
describe how local perturbations of a reaction rate affect the
network-wide steady state. They are defined in such a way that they
depend on the perturbed reaction, but not on which parameter caused
the perturbation. Thus, response coefficients refer to perturbed
parameters and control coefficients to perturbed reactions. The
effects of global parameters such as temperature, which affect many
reactions, are described by response coefficients
$\Run^{y}_{p_{m}} = \sum_{l} \Cun^{y}_{v_{l}} \,
\Euns^{v_{l}}_{p_{m}}$ (for more details, see SI
\ref{sec:SIMCAcalculation} and
\cite{rede:88,hohe:93,hofm:01}). Elasticities, response coefficients,
and control coefficients can be defined in their unscaled form
$\partial y/\partial x$ (denoted by a bar $\bar{X}$) or in their
scaled form $\partial \ln y/\partial \ln x$ (denoted by a hat
$\hat{X}$) (see SI \ref{sec:SIelasticityconversion}). If an enzyme
catalyses a single reaction, the enzyme level appears as a prefactor
in the rate law and its scaled response and control coefficients are
identical. The summation and connectivity theorems \cite{hesc:96}, a
central finding in Metabolic Control Theory (MCT), entail linear
dependencies among the control coefficients along a stationary flux
distribution or in the reactions surrounding a common
metabolite. Quantities and formulae related to fluctuations in time
(such as spectral response coefficients, spectral power density, and
variability on different time scales) are described in the SI.

\myparagraph{Enzyme synergies} The synergy
effect of an enzyme pair on a flux $v$ (or on some other steady-state
variable) can be approximated by second-order response coefficients,
called synergy coefficients. Assume that two enzymes are inhibited,
thus decreasing their levels $e_{a}$ and $e_{b}$ to small values
$u^{*}_{a}$ and $u^{*}_{b}$, and that this leads to relative flux
changes $w_{\rm a} = v^{\rm a}/v$, $w_{\rm b} = v^{\rm b}/v$ for the
single inhibitions and $w_{\rm ab} = v^{\rm ab}/v$ for the double
inhibition. Based on these numbers, we define the synergy effect
\begin{eqnarray}
 \label{eq:defepistasis}
 \etasc^{\rm v}_{\rm ab} &=& \ln \frac{w_{\rm ab}}{w_{\rm a} w_{\rm b}}.
\end{eqnarray}
A positive value of $\etasc^{\rm v}_{\rm ab}$ indicates a buffering synergy ($w_{\rm
 ab}> w_{\rm a}\,w_{\rm b}$), while a negative value indicates an aggravating synergy
($w_{\rm ab} < w_{\rm a}\, w_{\rm b}$). If $w_{\rm ab}= w_{\rm a}\,
w_{\rm b}$, there is no synergy. In   a second-order expansion  around
the unperturbed state, the synergistic effect can be written as 
(see SI \ref{sec:SIproofinterference})
\begin{eqnarray}
 \label{eq:apprepistasis2}
 \etasc^{\rm v}_{ab} &\approx& \Rsc^{\rm y}_{e_{a} e_{b}} \DeltaR \ln e_{a} \cdot\DeltaR \ln e_{b},
\end{eqnarray}
so the scaled synergy coefficients $\Rsc^{\rm y}_{e_{a} e_{b}}$
quantify synergisms between two enzymes. If perturbations and target
variables are measured on non-logarithmic scale, the synergistic
effects $\etaun^{\rm v}_{\rm ab} = w_{\rm ab} - w_{\rm a} - w_{\rm b}$
can be approximated by
$\etaun^{\rm v} \approx \Run^{\rm y}_{e_{a} e_{b}} \Delta
e_{a}\,\Delta e_{b}$ with the unscaled response coefficient
$\Run^{\rm y}_{e_{a} e_{b}}$.

\subsection{Model construction by STM} 
\label{sec:SIconstructionalgo}

\myparagraph{\ \\Constraints between parameters and variables in  kinetic steady-state
  models} A kinetic steady-state model is  determined by its
network structure, rate laws, state variables, and saturation
values: these values define the elasticities and kinetic
constants. For a consistent model, these variables have to satisfy a
number of constraints (\thermcond1-\thermcond5), explained by  the dependency schema.
\begin{itemize}
\item \textbf{Wegscheider conditions (\thermcond1).} Some
 biochemical quantities, for example Gibbs free
 energies of reactions, can be written as differences $\DeltaR x_{l}$ along
 reactions, or in vector form
 $\DeltaR \xv = {\Ntot}\trans \xv$, where ${\Ntot}$ is the
 stoichiometric matrix including both internal and external
 metabolites. Such quantities must obey the Wegscheider condition
 $\Kmat\trans \DeltaR \xv= 0$, where $\Kmat$  is  a null-space matrix
 satisfying ${\Ntot}\,\Kmat=0$. Wegscheider conditions must hold, for
 example, for equilibrium constants
 ($\ln {\bf k}^{\rm eq} = {\Ntot}\trans \ln \cv^{\rm eq}$),
 mass-action ratios ($\ln {\rm q}^{\rm ma} = {\Ntot}\trans \ln \cv$),
 and thermodynamic forces
 (${\bf A} = - \DeltaR {\boldsymbol \mu} = - {\Ntot}\trans
 {\boldsymbol \mu}$).
\item \textbf{Haldane relationships (\thermcond2).} In a chemical
 equilibrium state, all metabolic fluxes $v_{l}$ must vanish. If we 
 consider a single reaction in equilibrium, set its rate to
 zero ($\rate_{l}(\cv^{\rm eq},...)=0$), and solve for the equilibrium
 constant, we obtain a equation between equilibrium constant and
 kinetic constants, the so-called Haldane relationship
 \cite{hald:30}. For example, for a
 reversible mass-action law $v_{l}=\kcatpl a - \kcatml b$ the Haldane relationship  reads
 $\keq_{l}=\kcatpl/\kcatml$. For the modular rate laws, it reads
 $\keq_{l}=\frac{\kcatpl}{\kcatml}\, \prod_{i}
 (\kM_{li})^{h_{l}\,n_{il}}$ (see \cite{liuk:10}).
\item \textbf{Equilibrium constant and chemical standard potentials
    (\thermcond3).} In chemical equilibrium, the Gibbs free energies of reaction
  $\DeltaR G_{l} = \DeltaR \mu_{l}$ must vanish. With the formula for chemical potentials $\mu_{i} =
 \mu^\circ_{i} + R T \ln c_{i}$ (i.e.~assuming an activity
 coefficient and a standard concentration equal to 1), this leads to the formulae $\ln \keq_{l}
 = - \frac{1}{R T} \sum_{i} \mu_{i}^\circ n_{il}$ and $\theta_{l} = -
 \frac{1}{RT}\DeltaR \mu_{l} =  \ln \frac{\keq_{l}}{\kma_{l}}$.
\item \textbf{Signs of fluxes and thermodynamic forces (\thermcond4).}
 According to the second law of thermodynamics, all chemical
 reactions must dissipate Gibbs free energy. This implies that rates
 and thermodynamic forces have the same signs ($v>0 \Rightarrow A>0$
 and $v<0 \Rightarrow A<0$), in agreement with the relationship $\ln
 A/RT = \frac{v_{+}}{v_{-}}$. A stricter
 version of this constraint, excluding near-equilibrium reactions,
 is given below.
\item \textbf{Steady-state fluxes (\thermcond5).} For applying MCT,
  the metabolic reference state must be a steady state, i.e.~a state
  in which the metabolic fluxes satisfy the stationarity condition
  $\Nint\, \vv = 0$. In addition, we may impose bounds
  $\vv^{\rm min} \le \vv \le \vv^{\rm max}$ on the reaction rates and
  bounds
  $\vv_{\rm ext}^{\rm min} \le \Next \vv \le \vv_{\rm ext}^{\rm max}$
  on the production or consumption of external metabolites. Such
  bounds can be used to predefine reaction directions or to keep
  fluxes close to measured values.
\item \textbf{Stability (\thermcond6).} For applying MCT to a metabolic reference
  state, this state must be asymptotically stable, i.e.~its Jacobian
  matrix must not have eigenvalues with positive real parts. This
  constraint depends on all model details, and we cannot guarantee it
  be the dependency schema. Following \cite{sgsb:06}, stable states can
  be obtained by generating a model ensemble and omitting all model
  instances with unstable states.
\item \textbf{Amounts or concentrations.}  In compartmentalised
  models, we need to distinguish between metabolites amounts and
  metabolites concentrations, which are related by compartment
  volumes. Generally, mass balances concern amounts while rate laws
  depend on concentrations.  In STM, fortunately, amounts as such do
  not play a role. While concentrations appear as model variables,
  amounts are modelled only implicitly (e.g.~if stationarity is
  imposed on fluxes).
\end{itemize}
If an algorithm for kinetic steady-state modelling violates these
constraints, it leads to inconsistent models or metabolic states.

\begin{figure}[t!]
 \begin{center}
  \includegraphics[width=16cm]{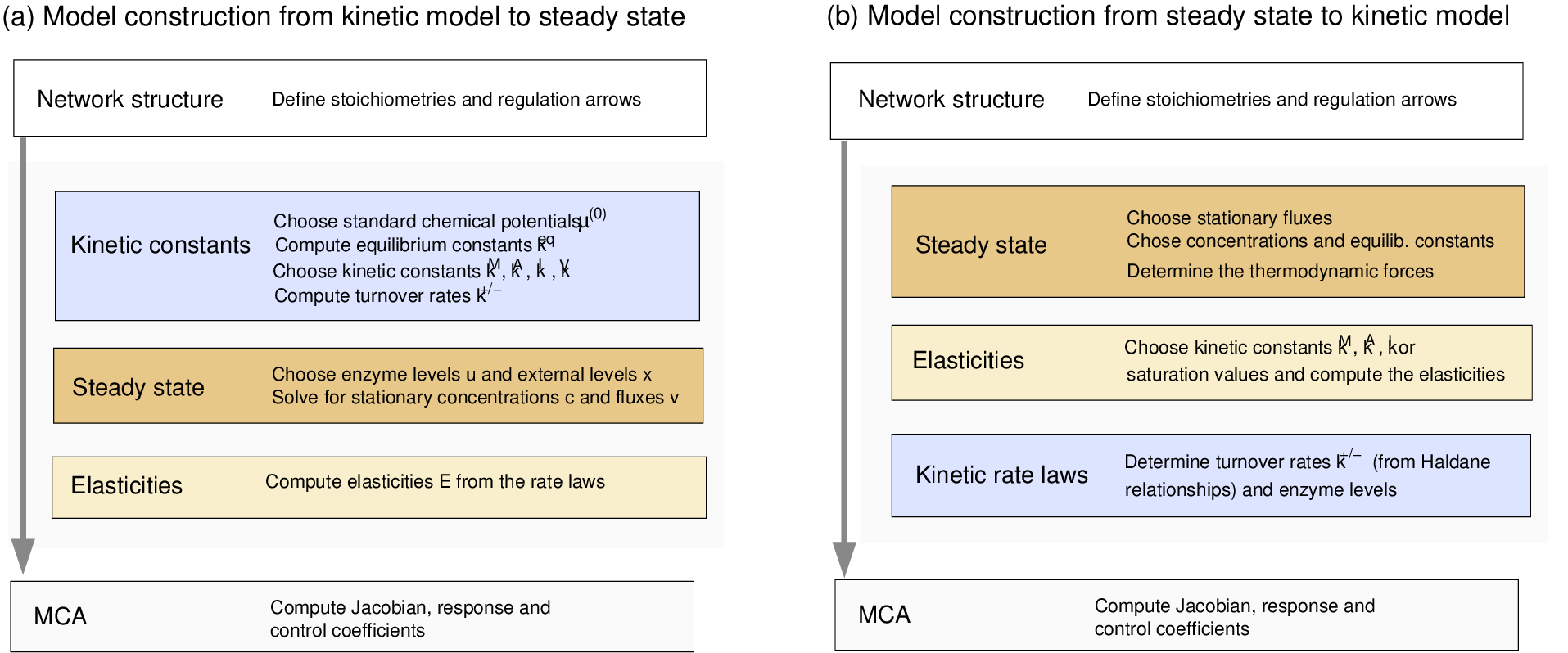}
 \end{center}
 \caption{\textbf{Constructing kinetic metabolic models in steady
     state.} (a) In
   ``causal'' kinetic modelling, \coout{gegenseitig referenzieren mit CBA
     opt} we start with rate equations and determine a steady state. A
   kinetic model is defined by network structure and rate laws. Given
   the kinetic constants, we solve for a steady state and compute the
   elasticities in this state by taking the derivatives of the rate
   laws. The elasticities determine the linearised dynamics around our
   steady state and are central to Metabolic Control Theory. (b)
   Retromodelling, starting from metabolic fluxes. First, the
   steady-state variables (metabolite levels, fluxes, and equilibrium
   constants) are chosen under thermodynamic constraints. Then,
   saturation values or dissociation constants are chosen, and kinetic
   constants and elasticities are computed. The free model variables
   (state variables and saturation values or dissociation constants)
   can be independently chosen, sampled, or optimised based on
   predefined values, bounds, or probability distributions. Prior
   knowledge about kinetic constants can be employed in the choices of
   saturation values.}
 \label{fig:samplingalgorithm}
\end{figure}

\myparagraph{Model construction algorithm} In a model with predefined
kinetic constants, enzyme levels, and external metabolite
concentrations, steady-state fluxes and concentrations can be
determined numerically (Figure \ref{fig:samplingalgorithm}
(a)). However, to construct models with physiolocally plausible
states, it may be safer to start from reasonable metabolic fluxes and
to construct metabolite levels and kinetic rate laws around them in
such a way that they yield the predefined fluxes (Figure
\ref{fig:samplingalgorithm} (b)). Model construction based on STM
combines ideas from thermodynamic flux analysis
\cite{bbcq:04,kuph:06,hohh:07} (in the metabolic state phase), SKM
\cite{sgsb:06} and thermodynamically consistent model parametrisation
\cite{likl:06a,edgi:07} (in the kinetics phase). Like in SKM,
steady-state concentrations and fluxes can be predefined. To satisfy
Wegscheider conditions, Haldane relationships, and the sign constraint
between reaction rates and thermodynamic forces, steady state and kinetic
constants are chosen sequentially. The network structure, our starting
point, is defined by stoichiometric matrix, regulation matrix, and the
list of external metabolites. In the algorithm, free variables
(steady-state variables and saturation values) are determined step by
step based on known values, constraints, or probability distributions:
they can be chosen manually, by optimisation, by fitting them to data,
or by random sampling.  Dependent variables (e.g.~kinetic constants)
are computed from variables chosen previously, as prescribed by the
dependency schema in Figure \ref{fig:samplingscheme}. Finally, models
can be checked for a stable steady state (if such a state is required
for the further analysis, e.g.~Metabolic Control Analysis). 

\myparagraph{Reconstruction of kinetic rate laws} The scaled
elasticities in STM resemble the saturation values as
in SKM, but depend on saturation values and thermodynamic forces. Each
sampled elasticity matrix corresponds to one instance of the model
ensemble, that is, a particular kinetic model with consistent kinetic
constants. In a variant of the algorithm, we do not sample the
saturation values directly ($\beta^{\rm M}_{li}$,
$\beta^{\rm A}_{li}$, $\beta^{\rm I}_{li}$), but compute them from
sampled dissociation constants ($\kM_{li}, \kA_{li}$, and
$\kI_{li}$). In both cases, the catalytic constants $\kcatpm$ are
computed from two constraints: the ratio $\kcatpl/\kcatml$ is
determined by the Haldane relationships while their geometric mean
$\sqrt{\kcatpl \cdot \kcatml}$, treated as a model parameter $\kV$,
depends on their absolute scaling. To choose this scaling, we may
either predefine the enzyme levels $e_{l}$ and scale the catalytic
constants $\kcatpm$ such that the reaction rate matches the predefined
flux; or we predefine the $\kcatpm$ values and solve for the enzyme
levels.

\myparagraph{Combining modular and hand-curated rate laws} Once the rate laws have been reconstructed, some of
them may be replaced by laws obtained from enzyme assays to make the
model more realistic \cite{bghs:09}. To ensure a consistent model, the
fluxes, concentrations, and equilibrium constants in these reactions
must be equal between the reconstructed model and teh new rate laws to
be inserted. There are different ways to guarantee this: either
reaction rates and equilibrium constant, taken from the rate law,
imposed as constraints when sampling the elasticities, or kinetic
constants and enzyme level in the rate law are adjusted to the network
model\footnote{If an irreversible rate law is given, the equilibrium
  constant can be ignored in the network model}.

\section{Model ensembles}

\subsection{Sampling the model variables}

\myparagraph{\ \\Model ensembles} Our models represent samples from an
ideal, infinitely large model ensemble defined by three types of
information: (i) restrictions on model structure and steady state, set
by the modeller (e.g.~the choice of a fixed flux distribution); (ii)
the independency between free variables, and dependencies of other
variables on them, as encoded in the dependency schema (Figure
\ref{fig:samplingscheme}); (iii) the random distributions from which
the free variables are sampled. Together, these choices define the
distributions and statistical dependencies of all model variables.
Any kinetic model satisfying the constraints can be obtained by the
algorithm.

\myparagraph{Sampling free variables randomly or based on data} In the
metabolic state phase, a flux distribution may be chosen by flux
minimisation. Metabolite concentrations, Gibbs free energies of
formation, and thermodynamic forces may be determined by parameter
balancing \cite{lskl:10}, using upper and lower bounds for
concentrations and thermodynamic forces, known values (for
concentrations) and predicted values (for Gibbs free energies of
formation) as data, and flux directions as inequality constraints.
Upper and lower bounds, signs, predefined values, and distributions
used for sampling reflect model assumptions and available data; by
choosing them, we can adjust the model to specific metabolic states
and to kinetic or metabolic data. Fluxes and thermodynamic forces may
be sampled uniformly, under linear constraints and with predefined
sign patterns. Metabolite levels, enzyme levels, and dissociation
constants can be sampled from log-normal or gamma distributions, and
saturation values can be sampled, e.g., from uniform or beta
distributions. In all phases of model
construction, instead of sampling the variables freely, experimental
data can be inserted, or distributions centred around data values can
be used. To use experimental data in a more solid way, the entire
model construction procedure can also be integrated into a Bayesian
framework, in which a posterior for the model parameters is determined
from data and priors.

\coout{A beta distribution for saturation values is an approprioate
 assumption. In general, if two random variables $X$ and $Y$ follow
 gamma distributions with the same scale parameter, the random
 variable $X/(X+Y)$ will follow a
 $\mbox{Beta}(\frac{a}{2},\frac{b}{2})$ distribution. With the
 definitions $\alpha = \kX /(x+\kX)$ and $\beta = x/(x+kX)$, whenever
 $x$ and $\kX$ are independently $\chi^{2}$ distributed, the
 saturation values will follow a beta distribution.}

\myparagraph{Varying the thermodynamic forces at given fluxes}
\label{sec:SIvaryingTheAffinities}
To systematically study the impact of thermodynamic forces on model
dynamics, one may vary the thermodynamic forces at a fixed metabolic
flux distribution. Varying the thermodynamic forces (e.g.~doubling all
their values) requires a variation of metabolite levels, but these
metabolite variations are not uniquely defined. For a simple
procedure, we start from a metabolic state with concentration vector
$\cv^{\rm orig}$ and force vector $\thetav^{\rm orig}$ (which must
agree with the flux directions). To realise a different force vector
$\thetav$ (with the same signs as $\thetav^{\rm orig}$), we choose the
new concentration vector
\begin{eqnarray}
 \cv &=& \mbox{argmin}_{\cv}\; || \ln \cv - \ln \cv^{\rm orig} ||^{2} \nonumber\\
\mbox{s.t.} \quad 
\thetav-{\thetav^{\rm orig}} &=& - {\Nint}\trans [\ln \cv - \ln \cv^{\rm orig}]
\end{eqnarray}
The idea is to apply a minimal overall change in metabolite
concentrations (on logarithmic scale, and in the sense of an Euclidan
distance).  In this procedure, we may impose upper and lower bounds on
$\cv$. However, depending on the bounds there may be no solution.

\subsection{Do model variants differ in their behaviour? Some useful statistical tests}
\label{significance}

Among the dynamic features of a metabolic model, which of them result
from network structure and which depend mostly on quantitative factors
like kinetic constants? To pose this question more generally, we may
consider two model variants that differ in some aspect (e.g.~network
structure or flux distribution), while other aspects (e.g.~rate
constants) can be varied for each of the variants; then we ask whether
a (qualitative or quantitative) model output differs significantly
between the variants. To prove or disprove such differences, we
describe each model variant by a model ensemble, sample instances from
both ensembles, compute their target variables, and compare them for
significant differences. This allows us, for example, to compare two
variants of a kinetic model (with different network structures,
synergies, steady-state fluxes, or expression patterns) and to check
whether these differences lead to typical differences in their synergy
patterns, irrespective of a variation of kinetic parameters. The test
described below have been implemented in matlab (see
\url{github.com/liebermeister/stm}).

\myparagraph{Significant differences in binary target variables} If we ask about  a
qualitative model property (e.g.~is the steady state stable or
unstable?), each model ensemble can be characterised by the fraction $p$
of ``positive'' model instances. Given a set of sampled model
instances (with count numbers $n_{+}$ and $n_{-}$ for ``positive'' and
``negative'' model instances), the true fraction $p$ can be estimated
by Bayesian estimation. If $p$ were the true fraction, the number
$n_+$ of positive model instances (out of $N=n_{+} + n_{-}$ model
instances sampled) would be binomially distributed with mean $p\,N$
and maximal value $N$.  Knowing this, we can estimate the value of $p$ 
 from the given number $n_+$ by using Bayesian estimation.
Assuming a flat prior, the posterior of $p$ is a beta distribution
$\mbox{Prob}(p) \sim p^{\alpha-1}(1-p)^{\beta-1}$ where $\alpha=n_++1$
and $\beta=N-n_++1$. The mean value of this distribution, $\langle p
\rangle = \alpha /(\alpha+\beta)= (n_++1)/(N+2)$ can be used as an
estimator for $p$. The corresponding variance reads $\sigma_{p} =
\frac{\alpha \beta}{(\alpha+\beta)^{2}\,(\alpha + \beta + 1)} =
\frac{(n_++1)(N-n_++1)}{(n_{\rm perm}+2)^{2}\,(N+3)}$.

\myparagraph{Significant differences in the occurrence of  positive
 and negative synergies.}
\label{sec:SIsignScore}
Given a matrix of enzyme synergies (where small-magnitude values have
been removed by thresholding), we count the positive and negative
synergies (numbers $n_{+}$ and $n_{-}$) between two groups of enzymes,
e.g.~between enzymes involved in two metabolic pathways. As a null
hypothesis, we assume that synergies can be positive or negative with
equal probabilities. Under this null hypothesis, and assuming that
only very few synergies remain after thresholding, $n_{+}$ and $n_{-}$
would be independently binomially distributed with the same unknown
mean value $n$. Given $n$, the difference $n_{+}-n_{-}$ would have a
mean value of 0 and a standard deviation of $\sqrt{2\,n}$. Thus, the
ratio between an observed difference $n_{+}-n_{-}$ and this standard
deviation can be used as a score for sign bias. Since the value of
$n$ is unknown, we approximate it by $\frac{n_{+}+n_{-}}{2}$ and
obtain the empirical sign bias score
$z_{\rm sign} = \frac{n_{+}-n_{-}}{\sqrt{n_{+}+n_{-}}}$.

\myparagraph{Quantitative target variables} Finally, we consider
quantitative target variables and their differences between model
variants as seen in model ensembles. As an example, we consider
predicted enzyme synergies. The general idea is as follows.  To see
whether a synergy (between two enzymes $i$ and $j$) differs
significantly between two model variants, we compute the synergies for
many instances of the two model variants, take the mean value for each
variant, and compare the two values using a p-value (obtained from a
permutation test) as a criterion for significant differences. Since we
run many such tests in parallel (namely, for many different enzyme
pairs), we expect a certain amount of false positives. To account for
multiple testing, we choose a false discovery rate and select
significant enzyme pairs based on their p-values \cite{beho:95}.  This
is how the procedure works in detail:
\begin{enumerate}
\item \textbf{Sample synergy values} We sample $n_{\rm model}$ model
  instances for each of the two model variants. For each model
  instance, we compute the synergies of all enzyme pairs. Altogether,
  we obtain a collection of synergy values $\etaun_{ijk}$, indexed by
  $i \in 1,..,n_{\rm pairs}$ for enzyme pairs, $j \in \{1,2\}$ for the
  two model variants, and $k \in 1, .., n_{\rm inst}$ for the sampled
  model instances of each variant. If the network contains $n_{\rm r}$
  enzymes, there are $n_{\rm pairs} = n_{\rm r} (n_{\rm r}-1)/2$
  enzyme pairs, i.e.~possible synergies to be computed. The synergy
  data $\etaun_{ijk}$ are now tested for significant differences.
\item \textbf{Quantify large (positive or negative) synergies by
    p-values} For each enzyme pair $i$, we first test whether this
  pair shows a significantly large (or small) synergy value caused by
  the model structure, i.e.~if it does, the synergy values will stand
  out from the general distribution of synergy values, even if we
  average over many random choices of the kinetic parameters. We apply
  the following statistical test: for each enzyme pair $a$, we test
  whether the mean value $\etaun_{i \cdot \cdot}$ (from the two model
  variants and all Monte Carlo samples) is significantly larger (or
  smaller) than other mean synergy values. We use a permutation test:
  the actual mean value $\etaun_{i \cdot \cdot}$ for our enzyme pair
  is compared to mean values obtained from batches of resampled
  $\etaun_{ijk}$ values. In each permutation run $l \in 1,..,N$, we
  resample $2 \cdot n_{\rm inst}$ of the $\etaun_{ijk}$ values with
  replacement and compute their mean value
  $\langle\etaun_{il}\rangle$. Let $n_{i}$ denote the number of
  resampled mean values $\langle\etaun_{il}\rangle$ larger than
  $\etaun_{i \cdot \cdot}$. Whether $\etaun_{i \cdot \cdot}$ is
  significantly large is indicated by a p-value $p_{i}$, estimated by
 \begin{eqnarray}
  p_{i} = \frac{n_{i}+1}{N+2}\label{eq:pvalue}
 \end{eqnarray}
 (for a justification, see the above treatment of binary variables).
 Small values $p_{i}\approx 0$ indicate that $\etaun_{i \cdot \cdot}$
 is  larger than expected by chance (i.e.~significantly large),
 large values $p_{i}\approx 1$ indicate that it is  smaller
 than expected by chance (i.e.~significantly small).
\item \textbf{Quantify differences in synergies (between model
  variants) by p-values} Next, for each enzyme pair $i$, we  ask
 whether the synergy values differ significantly between the two
 model variants. In the test, we consider the mean values
 $\etaun_{i 1 \cdot}$ and $\etaun_{i 2 \cdot}$ of the two model variants,
 averaged over all kinetic parameter samples, and check whether they
 differ significantly. Again, we use a permutation test.
 This time, we compute the mean difference
 $\Delta \etaun_{ij} = \etaun_{i 1\cdot}- \etaun_{i 2\cdot}$. In each run
 $d$ of the the permutation test, we randomly permute the values
 $\etaun_{ijk}$ for the pair $a$ under study, divide them into two
 batches of size $n_{\rm inst}$, and compute the mean difference
 $\Delta \langle\etaun_{il}\rangle$ between the two batches. A p-value, stating
 whether $\Delta \etaun_{ij}$ is large, for is computed as above, by
 counting how many of the permutation samples lead to larger values.
\item \textbf{Select significant synergies or differences based on
    p-values} Given the previously computed p-values, we determine
  which of their mean values $\etaun_{\cdot j \cdot}$ and of their
  mean differences $\Delta \etaun_{\cdot j \cdot}$ are significantly
  high (or low). Since we test this for many $n_{\rm pairs}$ gene
  pairs, we need to account for multiple testing: we fix a false
  discovery rate of 5\% and choose the confidence level $\alpha=0.05$
  for the individual tests. With this choice,
  $\alpha \cdot n_{\rm pairs}$ of the apparently significant values
  (for each of the four tests, high or low mean synergy and difference
  in synergy) are expected to be false positives. \coout{ist das schon alles?}
\end{enumerate}

\subsection{Extensions of STM}
\label{Sec:SIextensions}

The algorithm for model construction can be extended in many ways:
\coout{reihenfolge; untergruppierung nach modellaenderungen, numerik,
  statistik usw}
\begin{itemize}
\item \textbf{Cell compartments} In kinetic models with compartments,
  the compartment sizes appear in the balance equations and may follow
  differential equations themselves. In our model construction,
  compartments are not modelled explicitly, but a compartment
  structure can be added to the reconstructed kinetic model. This,
  however, changes control properties such as Jacobian, response, and
  control matrices.
\item \textbf{Dilution by cell growth.} In models with dilution (due
  to growth rate $\lambda$), intracellular metabolites will be
  effectively ``consumed'' by a dilution flux $\lambda\,c_{i}$. This
  changes the stationarity condition and directly couples fluxes to
  metabolite levels. Given a steady state, the elasticities can be
  computed as normally, but the Jacobian, response, and control
  coefficients will be affected. The elasticities for the dilution
  reactions are directly given by dilution rate and metabolite levels.
\item \textbf{Thermodynamically infeasible fluxes.}
  Eq.~(\ref{eq:modelasticities}), the main formula in elasticity
  sampling, requires that flux directions follow the thermodynamic
  forces, that is, fluxes must lead from higher to lower chemical
  potentials. In practice, even valid flux distributions may violate
  this assumption, for example, if cofactors or protons are omitted in
  the model. To apply STM regardless, we may choose to ignore
  thermodynamics and treat some reactions as irreversible -- that is,
  we ignore the thermodynamic term in the substrate elasticity and set
  the product elasticity to zero. Alternatively, we may adjust the
  thermodynamic forces to the given flux directions: whenever
  $\sign(-\DeltaR \mu_{l})$ differs from the flux direction, we add a
  virtual substrate and choose its chemical potential $\mu_{x}$ such
  that the thermodynamic force
  $-\DeltaR \mu^{*} = -(\DeltaR \mu_{l} + \mu_{x})$ has the correct
  sign. The virtual substrate changes the equilibrium constants and
  thermodynamic forces, but can otherwise be ignored in the kinetic rate
  law.
\item \textbf{Avoiding divergencies close to chemical equilibrium.} In
  reactions close to chemical equilibrium, with a thermodynamic force
  $\theta_{l} \approx 0$, the one-way rates $v^{\pm}_{l}$ and scaled
  elasticities become very large.  This does not only cause numerical
  problems, but also implies a very fast or very abundant enzyme, to
  sustain a finite flux at an enzyme efficiency close to 0. To avoid
  this in our models, we set a constraint $v^{\pm}_{l}<\rho\,|v_{l}|$
  on the forward and backward rates in each reaction. With a threshold
  $\rho=100$, for example, forward fluxes can be at most 100 times as
  large as the net flux. This translates into a constraint for
  thermodynamic forces in the metabolic state phase: the flux sign
  constraint $\thermcond4$
  ($v_{l} \ne 0 \Rightarrow \mbox{sign}(v_{l})\,\theta_{l} \ge 0$) is
  replaced by the stricter constraint\footnote{The constraint can be
    derived as follows (assuming $v_{l}>0$ without loss of
    generality): the ratio between forward and net reaction rate is
    given by $v_{+l}/v_{l} = \zeta_{l}/(\zeta_{l}-1)$. Close to
    equilibrium, we can approximate
    $1/\zeta_{l} \approx 1- \theta_{l}$ and obtain
    $\rho \ge \frac{v_{+l}}{v_{l}} = \frac{1}{1-1/\zeta_{l}} \approx
    1/\theta_{l}$.}
  $v_{l} \ne 0 \Rightarrow \mbox{sign}(v_{l})\,\theta_{l} \ge 1/
  \rho$. It will prevent extreme values in the thermodynamic
  elasticity term $\Escs^{\rm rev}_{li}$. The unscaled elasticities,
  in contrast, do not diverge in chemical equilibrium and can be
  computed from non-divergent formulae \cite{liuk:10}.
\item \textbf{Enzyme reactions composed of elementary steps.}  If we
  think of enzyme mechanisms as composed of elementary mass-action
  steps, we also represent them in this way in a model, replacing each
  reaction by a more fine-grained description. In the resulting model,
  there is a much larger number of (elementary) reactions. Since all
  rate laws are mass-action kinetics, the elasticities are directly
  given by reversibility terms and completely determined by thermodynamic
  forces. It sounds surprising: by knowing the fluxes and thermodynamic
  forces, we completely know the enzme kinetics!  But we should not
  forget that we are now talking about the thermodynamic forces of
  \emph{elementary steps} and that the equilibrium constants of these
  steps corresponds to $\kM$ values (of saturable modular rate laws)
  in the more coarse-grained model.
\item \textbf{Prior distributions for saturation values.} Saturation
  values $\beta=k/(k+x)$ can be set to fixed values (e.g.~$\beta=0$
  for enzymes believed to be in the linear range, $\beta=1/2$ for
  enzymes in half-saturation, or $\beta = 1$ for enzymes in full
  saturation), or they can be sampled independently from the range
  $]0,1[$. If $\beta$ is drawn from a uniform distribution (as
  suggested by the principle of minimal information \cite{jayn:57})
  and the metabolite levels are fixed, the resulting dissociation
  constant $k$ is randomly distributed with
  probability density $p(k)=\frac{k}{(c+k)^{2}}$ (see SI
  \ref{sec:SIprobabilitydistributions}).  If a saturation value is
  approximately known, we can use a beta distribution instead, with
  density $p(\beta) \sim \beta^{a-1} (1-\beta)^{b-1}$ with a mean
  value $a/(a+b)$ given by the known value. This yields a distribution
  $p(k)=\frac{k}{(c+k)^{2}} (\frac{c}{k+c})^{a-1}
  (\frac{k}{k+c})^{b-1} = \frac{k^{b-1} c^{a-1}}{(k+c)^{a+b}}$ for the
  dissociation constant. Saturation values can also be sampled from
  dependent distributions: this may be necessary if enzymes bind to
  different reactants, for example NAD$^{+}$ und NADH, with unknown
  but similar binding affinities, leading to correlations between
  their saturation values in physiological states.
\item \textbf{Multiple steady states.}  Under different choices of
  external metabolite levels and enzyme levels, a kinetic model will
  show different steady states. Each of these states is characterised
  by different metabolite levels, fluxes, saturation values, and
  elasticities. For constructing a model with multiple states
  directly, STM to be modified: we need to ensure that the saturation
  values and elasticities in different states correspond to the same
  set of kinetic constants. However, this is simple. In the
  metabolic state phase, we sample a set of equilibrium constants
  and different sets of state variables for the multiple steady
  states; in the kinetics  phase, we sample a single set of
  kinetic constants, which then determine the saturation values,
  elasticities, and enzyme levels for each steady state. Again, enzyme
  levels are determined last from the other variables; the only way to
  adjust them -- e.g.~to proteome data -- is by surrounding the
  algorithm into another layer of parameter fitting or posterior
  sampling.
\item \textbf{Adaptation of enzyme levels.} In our metabolic models,
  enzyme levels appear as parameters. In reality, they are controlled
  by transcriptional regulation, an important mechanism for shaping
  metabolic behaviour. To include transcriptional regulation into our
  metabolic models, the model must be extended to describe the
  production of enzymes. Alternatively, we may treat enzyme levels as
  choice variables and attempt to derive plausible enzyme adaptation
  profiles from optimality considerations \cite{lksh:04}. \co{show
    formula for diff exp or refer to opt enzyme rhythms for these
    formulae} To improve metabolic efficiency, enzymes should be
  expressed in the right proportions, to be adapted continuously to
  the current metabolic tasks. For instance, a rising demand for a
  certain metabolite may lead to an induction of a biosynthesis
  pathway for this metabolite. If such feedback systems are in place,
  changes in enzyme levels will affect the metabolic state, and induce
  secondary adaptations of other enzymes, and so on. For predicting
  the effects of, e.g.~gene knock-downs, we need to consider the
  global interplay of such adaptations. Given a kinetic model with a
  metabolic objective function and an enzyme cost function, optimal
  enzyme adaptations to external changes or single-enzyme knock-downs
  can be predicted with the help of synergy coefficients
  \cite{lksh:04}. Again, the predicted enzyme adaptations reflect
  network structure and elasticities.
\item \textbf{Analysis of sampled target variables.} A model ensemble
  can be seen as a statistical model with independent (``free'') and
  dependent (``determined'') variables. The dependencies are described
  by a schema like the one from Figure \ref{fig:samplingscheme}. Each
  binary property (e.g.~the sign of a control coefficient) has a
  certain probability in the ensemble. In practice, we can only
  estimate this probability from a limited number of model instances.
  If $n$ out of $N$ sampled models show the property $P$, the
  probability of $P$, called $q$, can be estimated as follows:
  assuming that $q$ has a uniform prior, its posterior mean and
  variance read $\mu_{q} = \frac{n+1}{N+2}$ and
  $\sigma_{q}^{2} = \frac{(n+1)(N-n+1)}{(N+2)^{2}(N+3)}$.  For each
  target variable (e.g.~a control coefficient), we obtain a number of
  sampled values. Their distribution can be characterised by 
   mean value, variance, probabilities of signs, and correlations
  with other model variables.
\item \textbf{Significant differences between model variants.} By
  choosing network structure, fluxes, thermodynamic forces, and enzyme
  saturation step by step, we can realise a nested sampling that leads
  to a hierarchy of model variants. In this hierarchy, each model
  variant is represented by a model ensemble with a specific
  distribution of target variables. To make them comparable, model
  variants should at least have the same metabolite and reaction
  lists, but they can differ in their network structures and in the
  values assigned to any of the free variables. Using statistical
  tests, we can determine significant differences between the
  distributions of model targets, if necessary with corrections for
  multiple testing (see section \ref{significance} for details). By
  comparing the distributions of model variables between subensembles,
  we can study how structural model features affect the target
  variables: for example, whether certain regulation arrows can
  enhance the stability of steady states. More generally, we can
  systematically study the effects of network structure, regulation,
  thermodynamic forces, enzyme saturation, and different rate laws on
  our model outputs.
\item \textbf{Choosing the distributions of target variables.}  To see
  how specific values or ranges of target variables can be obtained,
  we can build a model ensemble and then filter it for models that
  show these values or ranges.  Again, we obtain a subensemble of
  models with different distributions and correlations of the model
  variables. Even free variables that were chosen independently can
  become dependent by the subselection. Alternatively, we can set
  the distribution our target variables during model construction by
  applying a Bayesian posterior sampling. In this approach the
  ``free'' variables are not sampled freely but by a Metropolis Monte
  Carlo procedure: as prior distributions we can choose the same
  probability distributions as usually; for the likelihood function,
  we compare the resulting target variables to the prescribed
  distribution, for example, a distribution defined by experimental
  data.
\end{itemize}

\subsection{Example model: glycolysis in human hepatocytes}
\label{sec:SIhepatocyteSI}

In the original publication \cite{gbhb:10}, thermodynamically feasible
fluxes were determined by flux minimisation with various different
flux objectives. I first focused on aerobic rephosphorylation of ATP
on glucose. Optimising this objective leads to a sparse flux
distribution that uses only a small part of the network, containing
glycolysis and TCA cycle. With this flux distribution, I first
determined a standard model in which all saturation values were set to
values of 1/2, assuming half-saturated enzymes. In the resulting
state, the steady ATP rate (our model output) is strongly controlled
by glucose import; all control coefficients are positive, i.e.~a small
increase of any enzyme would always increase the metabolic target (ATP
level). Although the flux distribution was chosen to support ATP
rephosphorylation, this is not a trivial finding: first, flux analysis
can capture the ATP rephosphorylation rate, but not the ATP level as a
target function; second, it describes which fluxes -- and in which
proportions -- are optimal to realise a certain metabolic objective,
but it does not capture the marginal effects of enzyme levels,
i.e.~how the objective would change upon small enzyme changes. By
sampling the saturation values, we obtain a model ensemble, and the
statistical distribution and correlations of these control
coefficients can be studied.  \coout{briefly list models in repo, say
  that data are available?}  \coout{Control analysis of ATP
  rephosphorylation flux} \coout{the glycolysis model is based on ..}

\begin{figure}[t!]
\parbox{10.5cm}{\includegraphics[width=10cm]{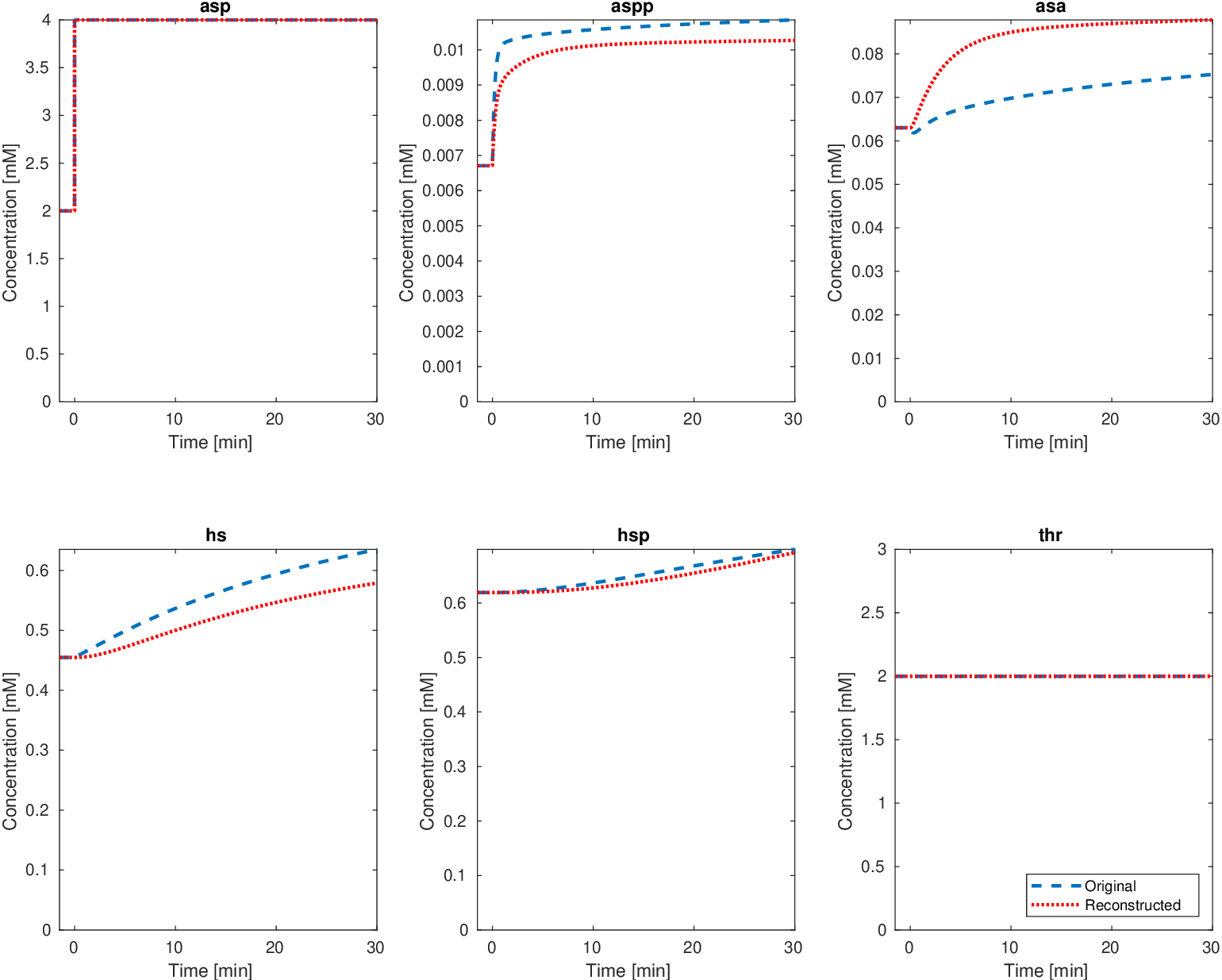}}
\parbox{6cm}{
  \caption{\textbf{A kinetic model reconstructed by STM.}
    \coout{netzwerkmodell zeigen?  nicht unbedingt noetig} The
    threonine synthesis pathway in \emph{E.~coli} converts aspartate
    into threonine.  A kinetic model from \cite{crqf:01} was used to
    simulate the metabolic dynamics after a sudden increase of the
    external aspartate level (dashed blue lines): the internal
    metabolite levels increase with different time delays. Aspartate
    and threonine are treated as external metabolites with predefined
    concentrations.  Based on the network structure, steady state
    fluxes, and concentrations as the original model, a model was
    reconstructed by STM. The reconstructed model shows the
    qualitative behaviour (solid red lines). Abbreviations asp
    (aspartate); aspp (aspartyl phosphate); asa (aspartate
    beta-semialdehyde)); hs (homoserine); hsp (O-phospho-homoserine);
    thr (threonine).}}
  \label{fig:dynamicsimulations}
\end{figure}


\section{Metabolic synergies and fluctuations}

\begin{figure}[t!]
 \hspace{1cm} (a) Flux analysis \hspace{2cm} (b) Metabolic control theory\\[-2mm]
  \begin{center}
    \begin{tabular}{lllll}
      FBA 0.1  & 
      MoMA 0.1  & 
      CM  Half-saturation& 
      SM  Half-saturation\\
      \includegraphics[width=3.5cm]{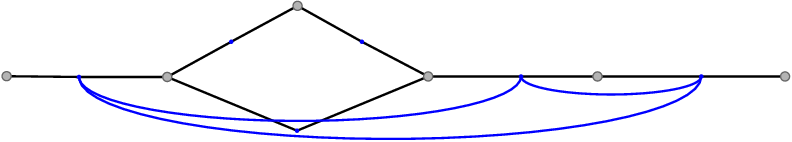}&
      \includegraphics[width=3.5cm]{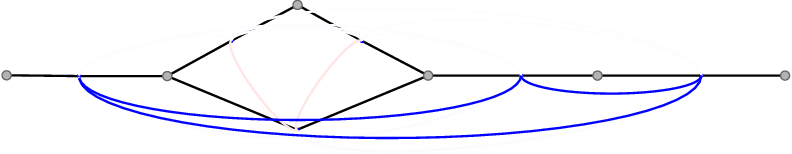}& 
      \includegraphics[width=3.5cm]{\looppsfiles/es_model_loop_chain_cs_inter_RFuu_network.eps}&
      \includegraphics[width=3.5cm]{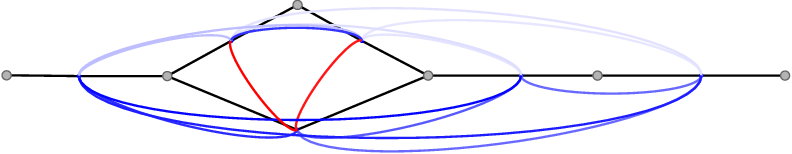}\\
      \includegraphics[width=3.5cm]{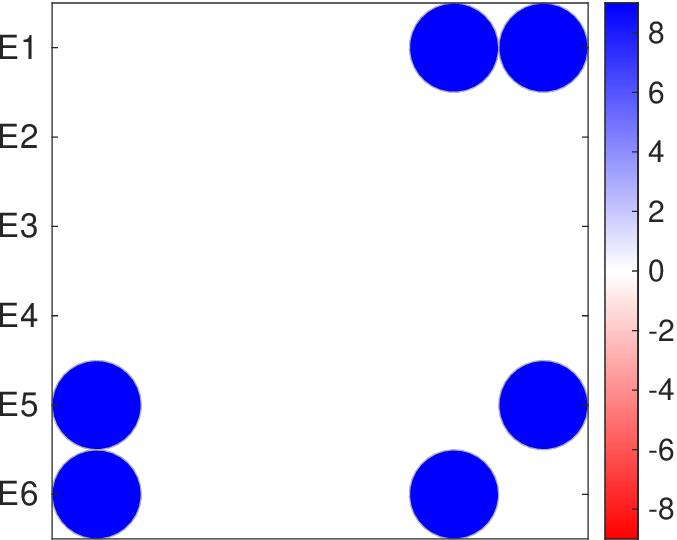}&
      \includegraphics[width=3.5cm]{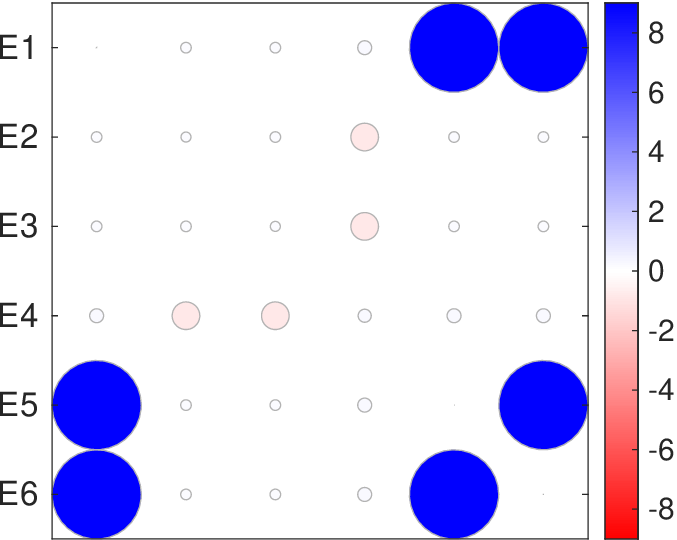}&
      \includegraphics[width=3.5cm]{\looppsfiles/es_model_loop_chain_cs_inter_RFuu_matrix.eps}&
      \includegraphics[width=3.5cm]{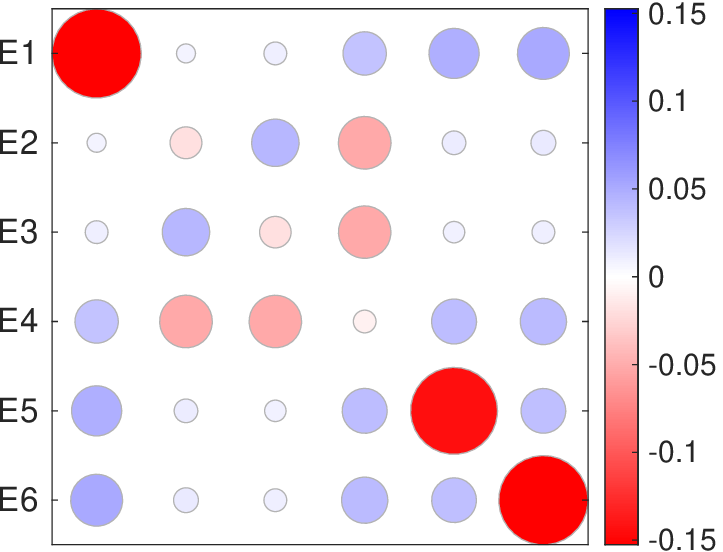}\\
      FBA 0.9  & 
      MoMA 0.9  &
      CM Ensemble mean& 
      SM Ensemble mean\\
      \includegraphics[width=3.5cm]{\looppsfiles/es_model_loop_chain_cs_inter_FBA_epistasis_0_9_network.eps}&
      \includegraphics[width=3.5cm]{\looppsfiles/es_model_loop_chain_cs_inter_moma_epistasis_0_9_network.eps}&
      \includegraphics[width=3.5cm]{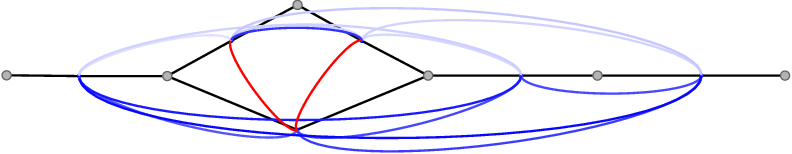}&
      \includegraphics[width=3.5cm]{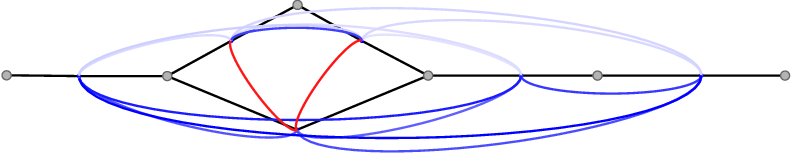}\\
      \includegraphics[width=3.5cm]{\looppsfiles/es_model_loop_chain_cs_inter_FBA_epistasis_0_9_matrix.eps}&
      \includegraphics[width=3.5cm]{\looppsfiles/es_model_loop_chain_cs_inter_moma_epistasis_0_9_matrix.eps}& 
      \includegraphics[width=3.5cm]{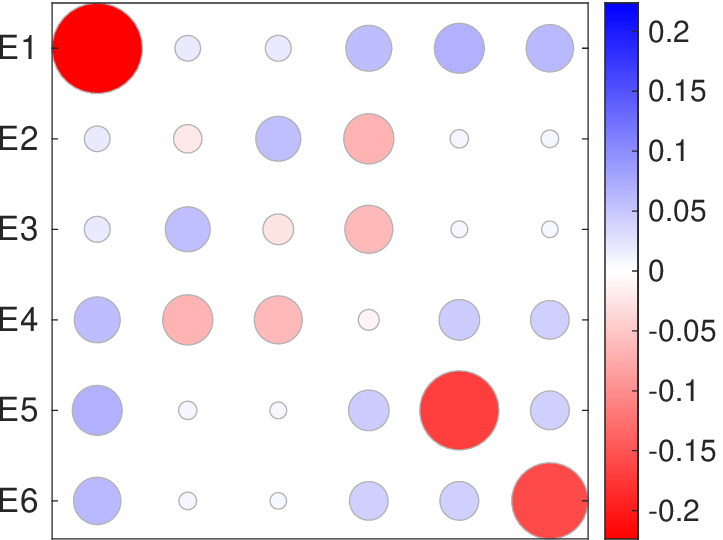}&
      \includegraphics[width=3.5cm]{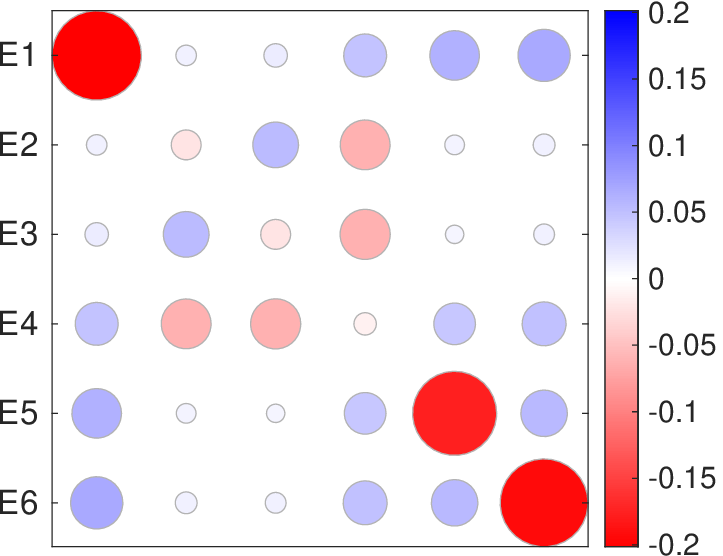}
    \end{tabular}  
  \end{center}
  \caption{\co{put ref to this figure in the SI text?} \textbf{Enzyme
      synergies in a schematic metabolic pathway (extended version of
      Figure \ref{fig:FBAsynergiesExample}).}  Enzyme synergies in a
    linear pathway with alternative routes between the intermediates
    S1 and S3.  (a) Synergies for double inhibitions, predicted by
    constraint-based methods.  Synergies are shown by arc colours
    (red: aggravating, blue: buffering).  Different panels show
    synergies computed by different methods (FBA and MoMA) and
    inhibition strengths (flux decrease by applying scaling factors of
    0.1 or 0.9).  In the calculation the ``double inhibition'' of a
    single enzyme is realised by applying the relative inhibition
    twice, i.e.~leading to inhibition strengths of 0.01 and 0.81,
    respectively.  Colour scales differ between panels, spanning the
    range of synergy values in each case. Small values (below one
    percent of the maximal absolute value) are not shown. (b)
    Synergies computed by Metabolic Control Theory. The panels show
    results for different rate laws (CM: common saturable rate law;
    SM: simultaneous-binding modular rate law). Results based on
    half-saturated enzymes (top) are compared to mean results from a
    model ensemble with random saturation values (bottom).  The two
    results are almost identical.}
  \label{fig:SIFBAsynergiesExample} 
\end{figure}

\subsection{Synergies between static perturbations}

Synergy effects between enzymes can be important in medical
applications, e.g.~to model drug interactions or patient-specific side
effects, to plan combination therapies, and to avoid the emergence of
bacterial resistance \cite{chck:07}. Epistasis, an important concept
in genetics, denotes the synergistic effects of gene knockouts on cell
viability. Epistasis can shape genetic variability in populations and
the evolvability of genetic features. \coout{REF evolvability?}
Moreover, as shown by FBA simulations and experiments, epistatic
interactions can indicate functional associations between proteins,
for example the cooperation or alternative usage of enzymes in
metabolic pathways \cite{sdck:05}.  Importantly, while synergies are
described here for enzyme perturbations, they can also be computed and
used for any other parameter perturbations, including synergistic
effects between concentrations in the growth medium.

\myparagraph{Epistasis}
\label{sec:SIepistatic}

Epistatis describes synergy effects of gene deletions on Darwinian
fitness. In buffering epistasis between two genes (double-deletion
phenotype is less severe than expected), the loss of one gene lowers
the selection pressure on the other one: in evolution, such genes will
tend to co-occur in genomes, a phenomenon called phylogenetic
correlation \cite{pmte:99}. In th eopposite case, called aggravating
epistatis, the double-deletion phenotype is more severe than expected:
the loss of one gene increases the selection pressure on the other
one, leading to phylogenetic anti-correlation. In \cite{sdck:05},
epistatic synergies in the yeast \emph{S.~cerevisiae} were computed by
FBA. The maximal biomass production rate was used as a quantitative
output function and enzyme deletions were simulated by setting the
corresponding reaction rates to zero. The predicted epistasis pattern
showed a modular structure \cite{sdck:05}. \coout{WEG? genes could be
  grouped reflecting the involvement of genes in metabolic pathways:
  epistatic synergies across such enzyme groups showed consistent
  signs and were called ``monochromatic'' .}

\myparagraph{Epistasis measure by Segr\`e et al.}
\label{sec:SIsegremeasure}
Segr\`e et al.~introduced a synergy score for double enzyme deletions
\cite{sdck:05}, in which a special weighting makes buffering synergies
better detectable. Let $v_{\rm wt}$ denote an target variable
(e.g.~maximal biomass production rate computed by FBA) observed in the
wildtype network, and let $v_{\rm a}$, $v_{\rm b}$, and $v_{\rm ab}$
denote the values in mutatnt networks in which enzyme a, enzyme b, or
both have been deleted. After a scaling by the wildtype value, the
target values read $w_{\rm wt}=1$,
$w_{\rm a}= v_{\rm a} / v_{\rm wt}$,
$w_{\rm b}= v_{\rm b} / v_{\rm wt}$ and
$w_{\rm ab}= v_{\rm ab} / v_{\rm wt}$. Since deletions can decrease,
but cannot increase the target in FBA, the values must satisfy
$w_{\rm a}\le 1$, $w_{\rm b}\le 1$, and
$w_{\rm ab}\le \mbox{min}(w_{\rm a},w_{\rm b})$.  The effect
$w_{\rm ab}$ of a double deletion is compared to the effects
$w_{\rm a}$ and $w_{\rm b}$ of single deletions, yielding an epistasis
score. To obtain a clear distinction between neutral
($w_{\rm ab} = w_{\rm a}\,w_{\rm b}$), aggravating
($w_{\rm ab} < w_{\rm a}\,w_{\rm b}$), and buffering
($w_{\rm ab}>w_{\rm a}\,w_{\rm b}$), gene pairs, Segr\`e \emph{et al}
introduced a heuristic epistasis measure with the following
definition:
\begin{eqnarray}
 \label{eq:segrecorrectionformula}
 \varepsilon^{\rm Segre}_{\rm ab} = \left\{
  \begin{array}{lll}
   \mbox{neutral} &:& 0 \\
   \mbox{aggravating} &:& \frac{w_{\rm ab}}{w_{\rm a}\,w_{\rm b}}-1\\ 
   \mbox{buffering} &:& 
   \frac{\frac{w_{\rm ab}}{w_{\rm a}\,w_{\rm b}}-1}{\frac{1}{w_{\rm b}}-1}
  \end{array}
 \right.
\end{eqnarray}
In the formula for buffering epistasis, we assume
$w_{\rm a}\le w_{\rm b}$ without loss of generality. In summary,
neutral and aggravating synergies are defined ``normally'', but
buffering synergies are given special weights: if one of the
single-deletion effects is mild, then $w_{\rm b}$ is close to 1 and
the buffering synergy is increased by this definition. If both single
deletions are already severe, then the buffering synergy gets a lower
weight.

\myparagraph{Synergies predicted by FBA and
  MCT} \label{sec:SIsynergiesFBAMCA} FBA, MoMA, and MCT all predict
enzyme synergies, but based on different assumptions and input
data. For double inhibitions, all three methods predict, not
surprisingly, that cooperating enzymes (i.e.~enzymes in one linear
pathway) tend to show buffering synergies, while alternative enzymes
(e.g.~enzymes in alternative pathways) show aggravating
synergies. However, the reasons for these predictions, their details,
and other predictions differ between the methods. An example is shown
in Figure \ref{fig:FBAsynergiesExample}. As expected, a double
inhibition that block both alternative routes has an aggravating
effect, while enzymes within the same linear pathway show buffering
synergies.  MCT explains this by control coefficients: in the first
case (across alternative pathways), inhibition of one branch increases
the flux control of the other branch, while in the second case (within
one linear pathway), the first enzyme inhibition decreases the flux
control of all other enzymes (see Fig.~(\ref{fig:SIFBAsynergiesExample})).
\coout{NOT SHOWN In the graphics,
  significance values were thresholded (values below one percent of
  the maximal synergy value are not shown).}

\subsection{Metabolic fluctuations} 

\myparagraph{\ \\Computing the fluctuations caused by chemical noise}
To model a metabolic pathway under dynamic external perturbations, we
may model these perturbations as a random process. The perturbations
themselves, and the resulting metabolite fluctuations are described by
spectral power density matrices. These matrices resemble the static
covariance matrices, but are frequency-dependent. If the noise
amplitudes are small, we can use a linearised model and compute the
spectral densities of concentration fluctuations from the spectral
response coefficients \cite{lieb:2005}
\begin{eqnarray}
 \label{eq:spectraldensity}
 {\mathcal S}_{\rm c}(\omega) = \RS(\omega) \,{\mathcal S}_{\rm p}(\omega) \, \RSdag(\omega).
\end{eqnarray}
${\mathcal S}_{\rm c}(\omega)$ and ${\mathcal S}_{\rm p}(\omega)$
denote the spectral power densities of metabolites and of perturbation
parameters at circular frequency $\omega$, the symbol $\dag$ indicates
the adjoint matrix, and the unscaled first-order spectral response
matrices $\RS(\omega)$ and $\RJ(\omega)$ can be computed from the elasticities
\cite{inga:04,lieb:2005}. For fluctuations of reaction
rates, there is a similar formula. 

\myparagraph{Chemical noise} An important example of random
fluctuations is chemical noise. On a microscopic scale, chemical
reactions do not run continuously as assumed in kinetic models, but as
discrete random events, converting individual molecules. The resulting
random dynamics can be described by a Langevin equation, i.e.~a
kinetic model with additive noise and separate forward and backward
rates \cite{gill:00}. The fluctuations spread in the network, leading
to fluctuations of molecule numbers in the macroscopic steady
state. In the Langevin equation, the noise term scales with the square
root of the mean reaction rate (in units of reaction events per
second). Therefore, the smaller the particle numbers, the bigger the
relative noise. If the average rates become very small, the
approximation breaks down and a more detailed model with discrete
reaction events must be used. The fluctuations can be described by
Eq.~(\ref{eq:spectraldensity}), setting
$E_{p_{l^{*}}} = \sqrt{\frac{v_{l^{*}}}{N_{\rm A} \Omega}}$ (where $*$
marks once-way fluxes) and ${\mathcal S}_{\rm p}(\omega) = \Imat$,
because the chemical fluctuations originate from white noise (see
\cite{lieb:2005}). In practice, the spectral power density of the
original noise in reaction $l$ is given by
\begin{eqnarray}
 \label{eq:originalnoisepower}
 {\mathcal S}_{\rm p}(\omega) = \frac{v_{+l} + v_{-l}}{N_{\rm A}\,
  \Omega} = \frac{\coth(h_{l}\,\theta_{l})\,v_{l}}{N_{\rm A}\, \Omega}
 \approx \frac{1}{N_{\rm A}\, \Omega} \,\frac{v_{l}}{h_{l}\,\theta_{l}}
\end{eqnarray}
where the approximation holds close to equilibrium (small thermodynamic
force $\theta_{l}$). Mediated by the metabolic dynamics, these input
fluctuations lead to fluctuations of metabolite levels and fluxes (for
an example, see Fig \ref{fig:linchain1a}). Fast fluctuations are
strongly damped: the noise spectrum of the metabolite levels decreases
at high frequencies, and the system acts as a low-pass filter. Since
each reaction rate is also directly affected by its own noise, fluxes
also fluctate at high frequencies. If a stable metabolic state is
close to a Hopf bifurcation, it will show a tendency towards
oscillations. This becomes visible apparent in the way it transmits
random fluctuations, and in its noise spectrum: noise will be
amplified around a resonance frequency close to the oscillation
frequency after the Hopf bifurcation. All this can be seen from the
eigenvalue spectrum of the Jacobian matrix \cite{lieb:2005}.

\myparagraph{Limiting behaviour of spectral power densities at high or
  low frequencies} For low or high frequencies, the noise amplitudes
can be understood through simple approximations. The spectral power density matrix for
metabolite fluctuations has the form
$\Rmatun^{\rm s}(\omega)\, \Mmat \,\RSdag(\omega)$ with a diagonal
matrix $\Mmat$. The noise variances for individual metabolites, at specific frequencies, are given by diagonal
values
\begin{eqnarray}
 \label{spectralpowerapprox}
 \sum_{p} ||\Run^{\rm c}_{\rm p}(\omega)||^{2} \, m_{p}.
\end{eqnarray}
The spectral response matrix itself is given by
$\Rmatun^{\rm c}_{\rm p}(\omega) = \Cmat\,(\Amat- i \omega
\Imat)\inv\,\Bmat$, with matrices $\Cmat$ and $\Bmat$ and the Jacobian
matrix $\Amat$.  For large frequencies ($\omega$ much larger than any
eigenvalue of the Jacobian), the term $i\omega\,\Imat$ dominates and
the entire expression (\ref{spectralpowerapprox}) becomes proportional
to $\frac{1}{\omega^{2}}$.  High-frequency noise is dominated by
direct effects of chemical noise on the adjacent reactant levels,
i.e.~fast, non-stationary fluctuations around the stationary fluxes.
For low frequencies, in contrast, the spectral power density
approaches the variability expected for static variability, and the
correlations reflect slow, stationary variations of the stationary
fluxes.

\myparagraph{Fluctuations on different timescales} The amplitude of
random fluctuations at specific frequencies are described by the
spectral power density. However, in reality we are usually not
interested at the noise level at a precise frequency, but at noise
affecting processes on a certain time scale, where much faster noise
averages out and much slower noise can be seen as quasi-static. To
measure the relevant noise on a time scale of interest, we consider a
noisy curve from our model, e.g.~of a metabolite level, compute a
sliding average with a Gaussian kernel (of width $\tau$, e.g.~one
second), and study how much this average varies (across the
statistical ensemble at one point in time, or along time in one
realisation of the process). Alternatively, we can also consider a
sliding average with. By changing the width of the kernel, we obtain
the variance of our metabolite curve on different time scales. This
measure of concentration fluctuations at different time scales can be
computed from the spectral power densities using a Fourier
transformation (see section \ref{sec:timescalevariation}).

\begin{figure}[t!]
(a)\\
  \begin{center}
  \includegraphics[width=10.5cm]{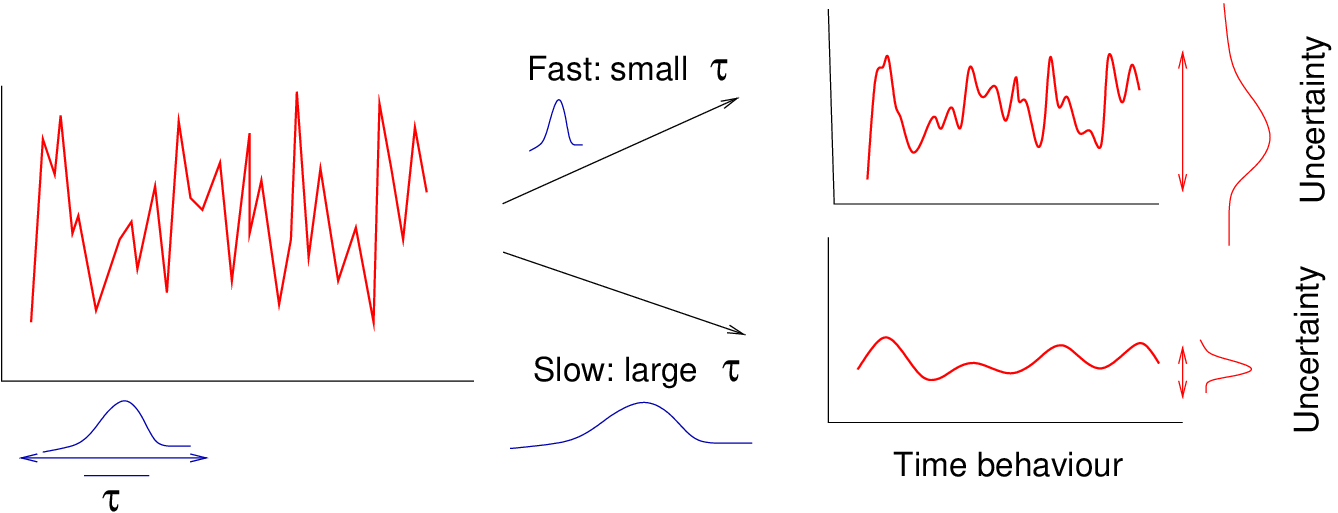}\\
  \end{center}
(b)\\
  \begin{center}
   \begin{tabular}{llll}
 Fast noise ($f=$1 s$\inv$) & Frequency dependency & Slow noise ($f=$0.001 s$\inv$)\\[1mm]
 \includegraphics[height=5.5cm]{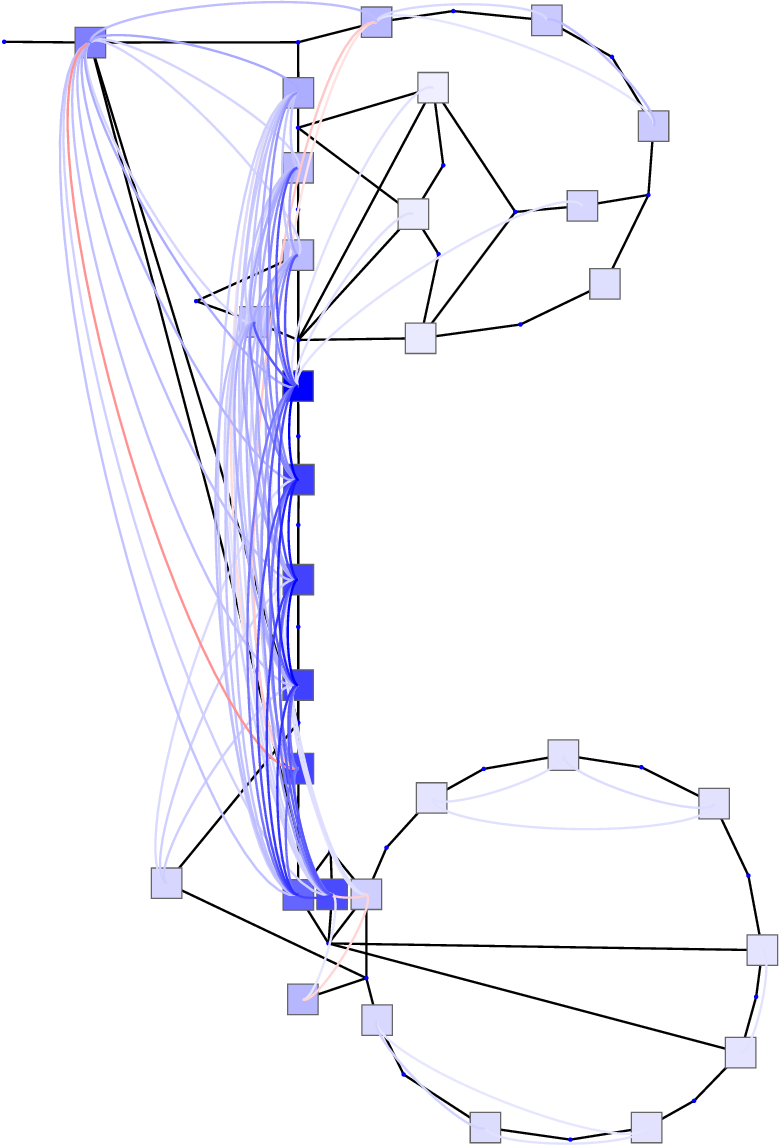}&
 \parbox[b]{5cm}{
 \includegraphics[height=3cm]{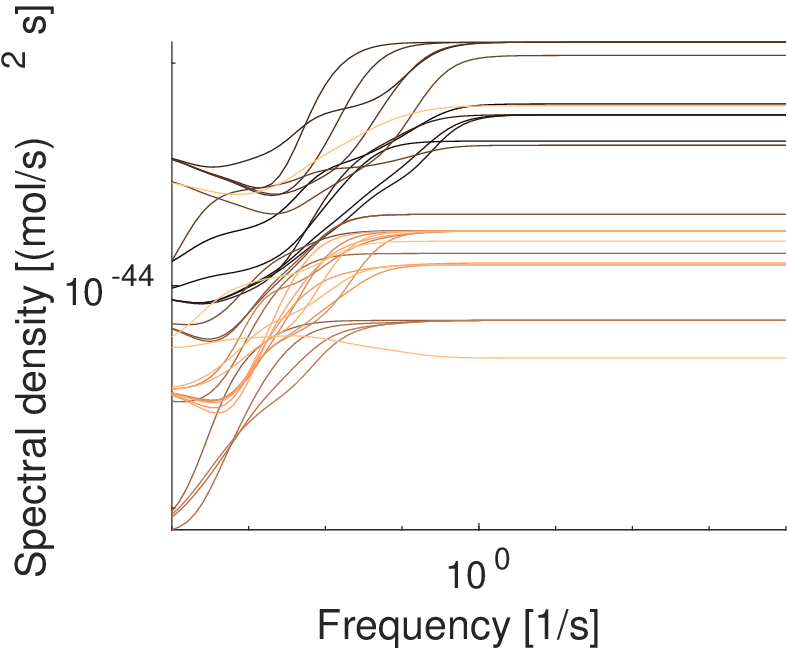}\\
 \includegraphics[height=3cm]{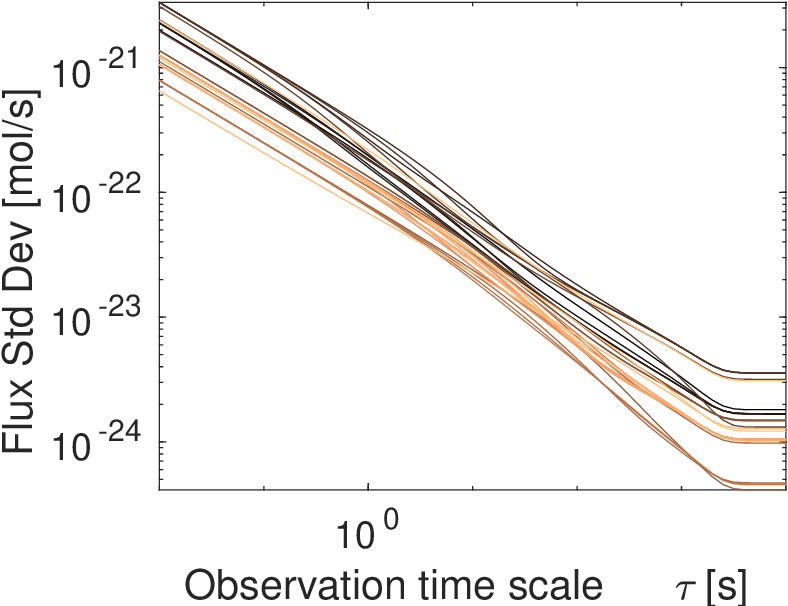}}&
 \includegraphics[height=5.5cm]{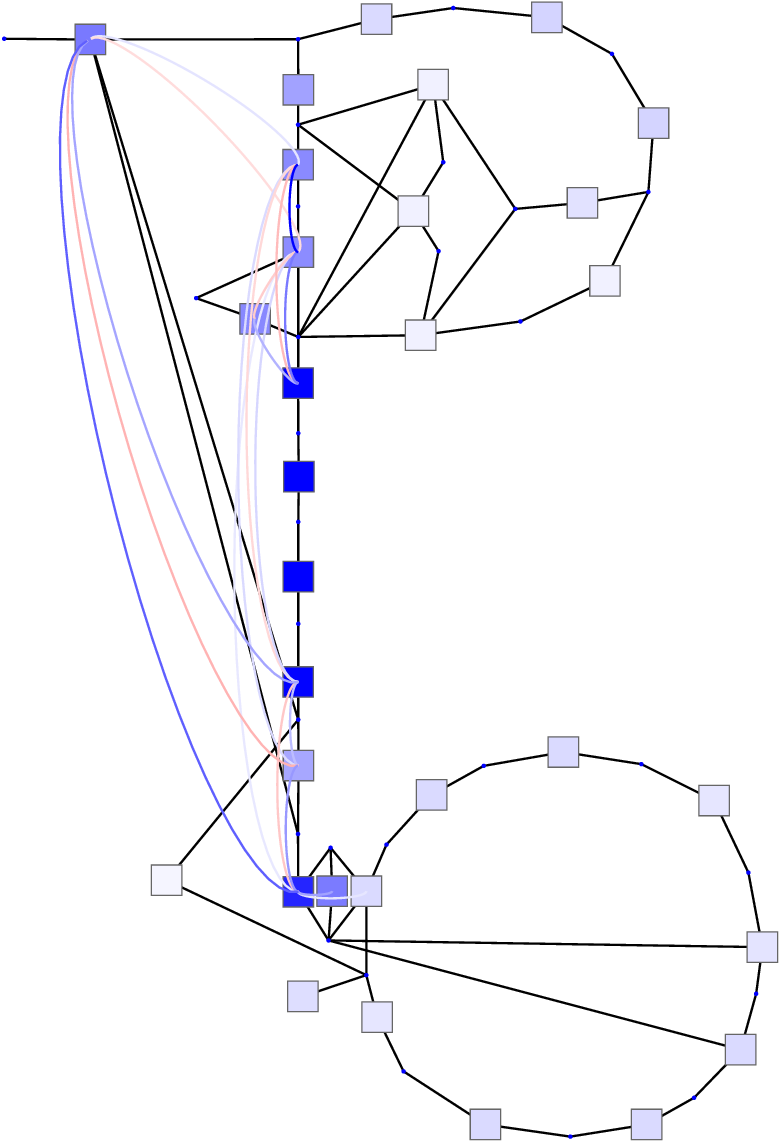}
 \end{tabular}
\end{center}
\caption{(a) Fluctuations caused by chemical noise. (a) Smoothing of stochastic time courses (e.g.~of metabolite concentrations are fluxes) with a Gaussian kernel function leads to ``observed'' stochastic time courses at a given time resolution $\tau$ (reflecting the time resolution of measurements). The variability (red arrows on the right) depends on the time resolution chosen. (b) Flux fluctuations (compare Figure~\ref{fig:chemicalnoise} in the main article).}\ \\[-5mm]
\label{fig:SIchemicalnoiseFlux}
\end{figure}


\subsection{Role of thermodynamics and enzyme saturation in metabolic control and 
 fluctuations}
\label{sec:SIEffectOfThermodynamic}

\begin{figure}[h!]
(a) Variation of thermodynamic forces \\
\begin{center}
\begin{tabular}{rccc}
 & Chemical potentials & &Thermodynamic forces \\
  $\theta=$0.1 RT & 
   \includegraphics[width=5cm]{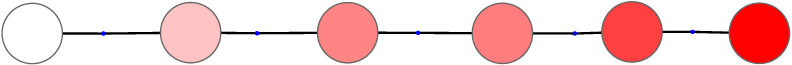} &&
   \includegraphics[width=5cm]{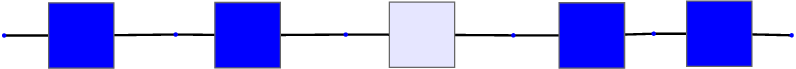}\\
  $\theta=$1 RT & 
   \includegraphics[width=5cm]{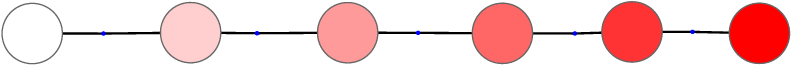}&&
   \includegraphics[width=5cm]{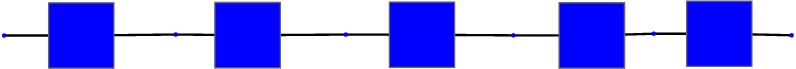}\\
  $\theta=$10 RT & 
   \includegraphics[width=5cm]{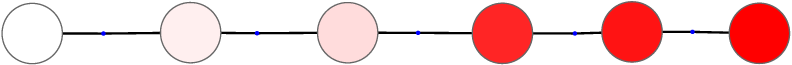}&&
   \includegraphics[width=5cm]{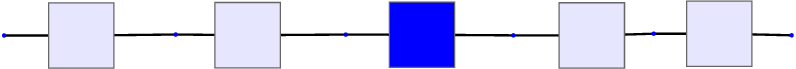}
\end{tabular}\\[7mm]
\end{center}
(b) Control coefficients \\[2mm]  
 \begin{tabular}{cc}
Concentrations & Fluxes \\
 \begin{tabular}{rccc}
   & $\beta=$ 0.1 & $\beta=$ 0.5 & $\beta=$ 0.9 \\
 {\scriptsize $\theta=$0.1 RT} & 
   \includegraphics[width=2cm]{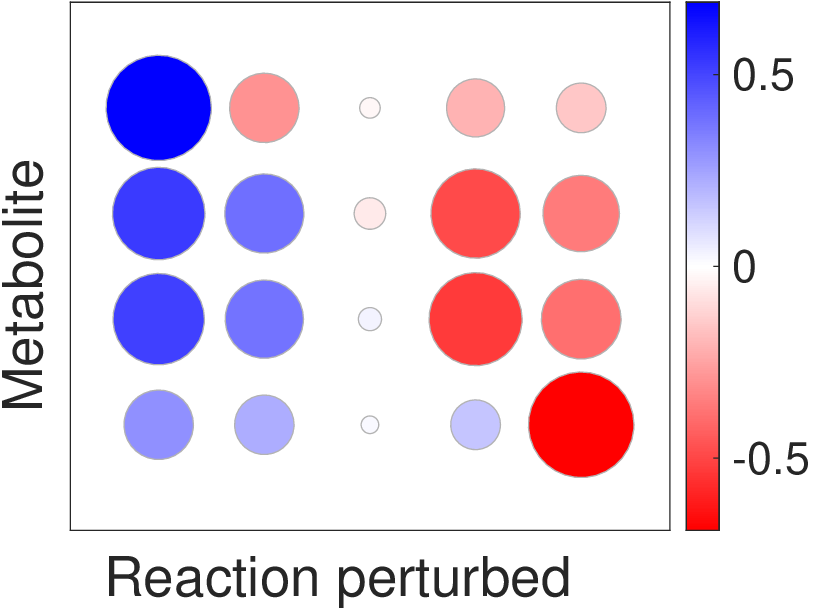}&
   \includegraphics[width=2cm]{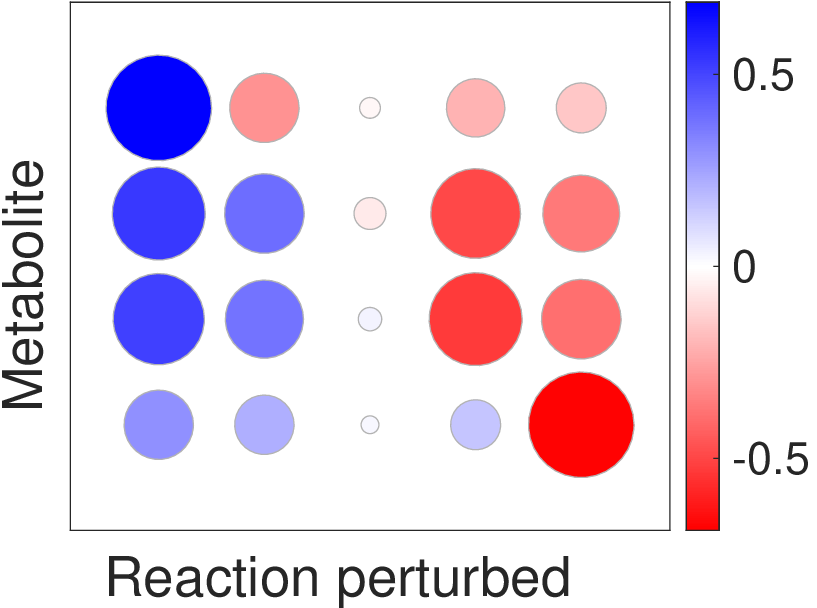}&
   \includegraphics[width=2cm]{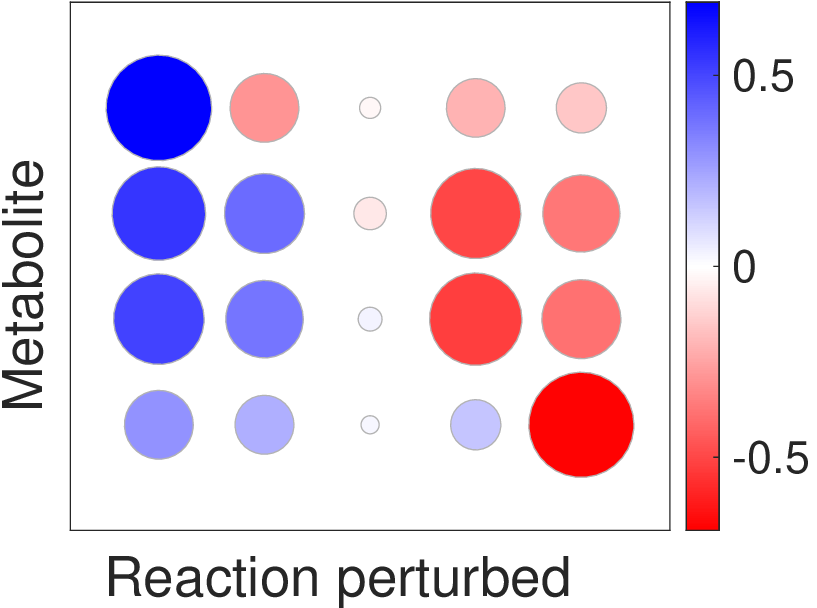}\\
 {\scriptsize $\theta=$1 RT} & 
   \includegraphics[width=2cm]{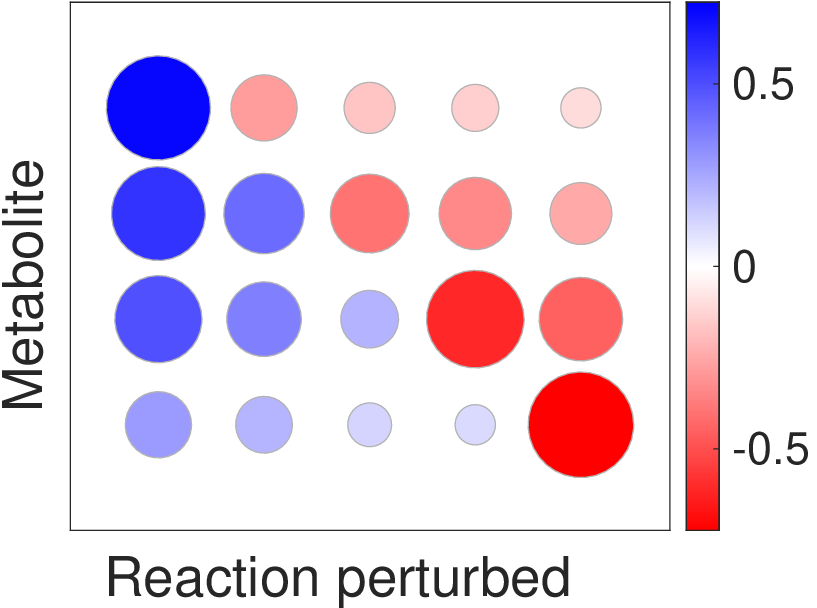}&
   \includegraphics[width=2cm]{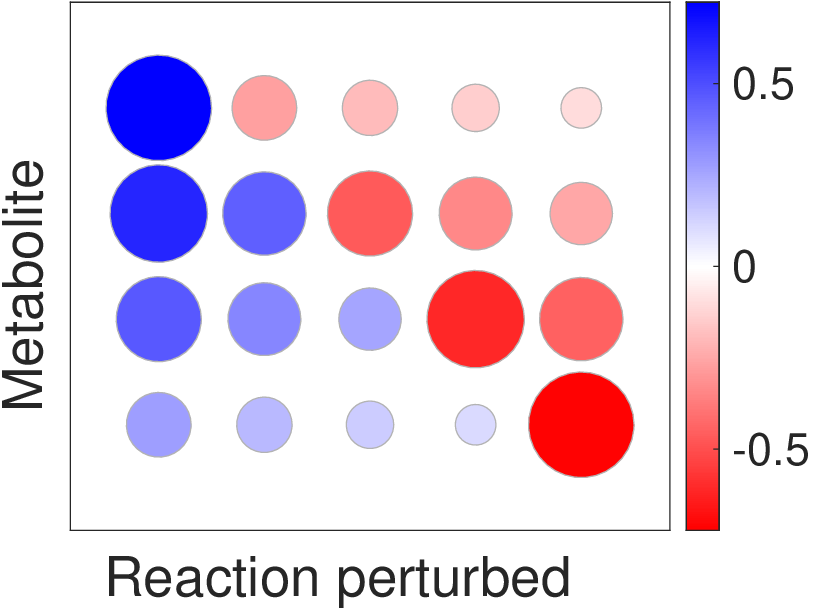}&
   \includegraphics[width=2cm]{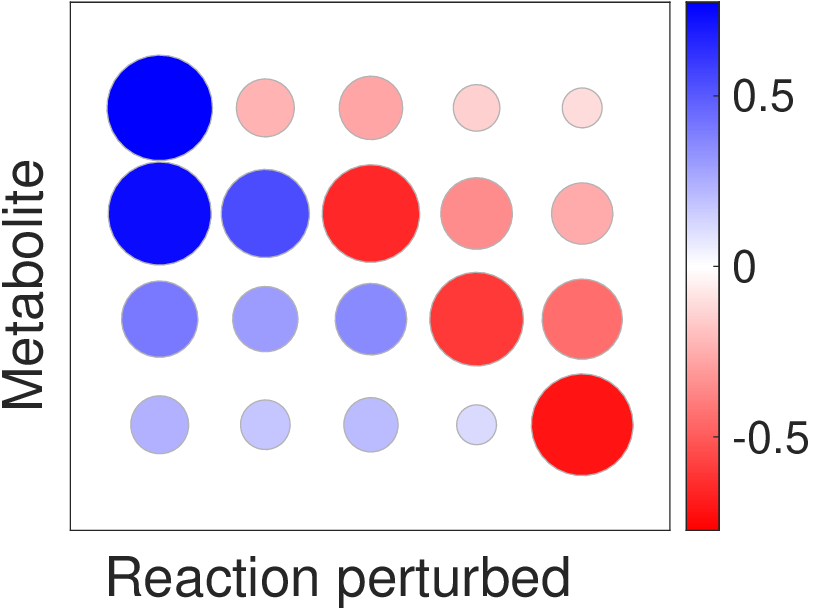}\\
 {\scriptsize $\theta=$10 RT} & 
   \includegraphics[width=2cm]{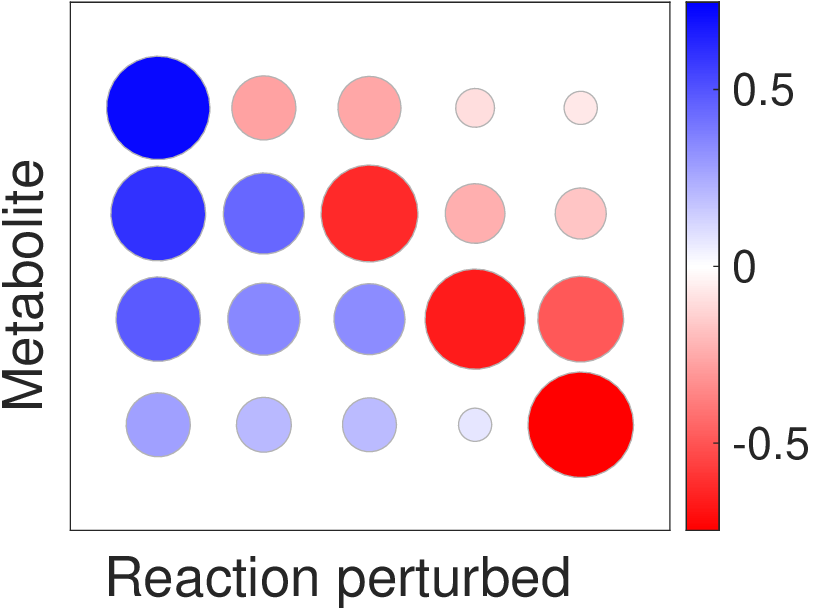}&
   \includegraphics[width=2cm]{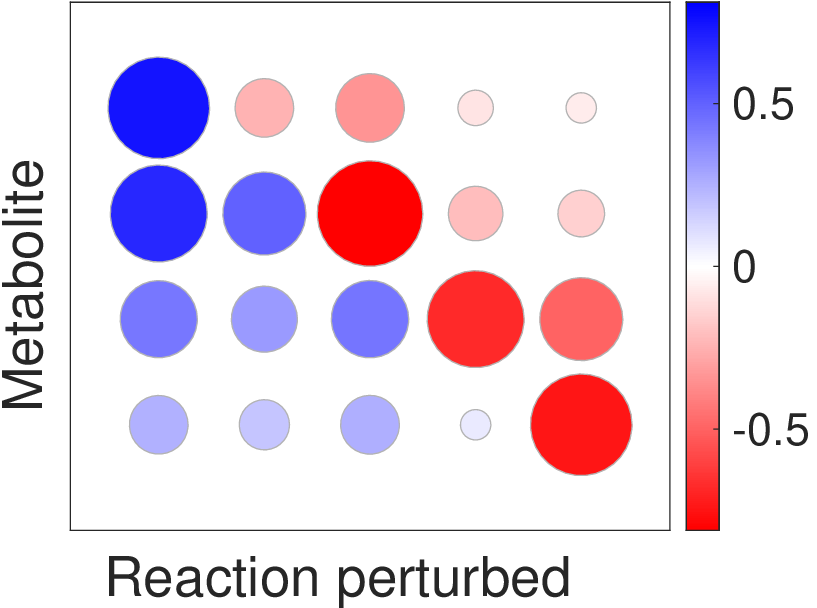}&
   \includegraphics[width=2cm]{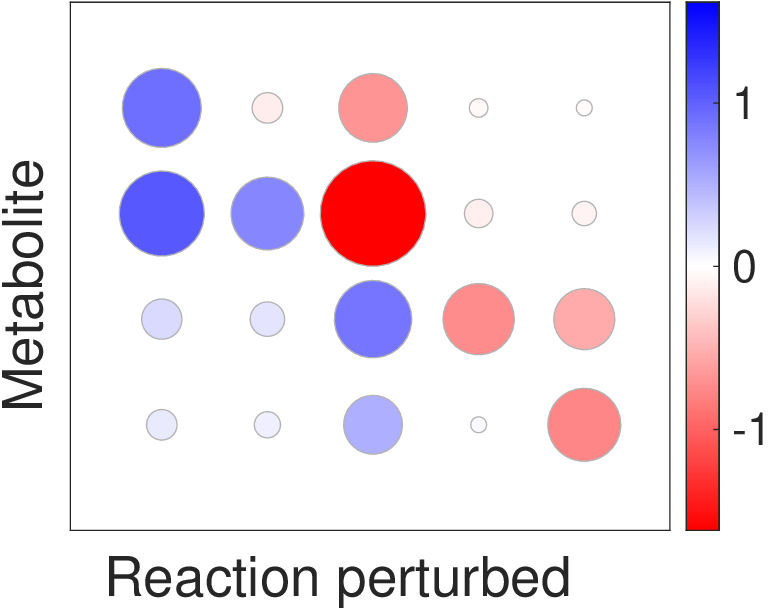}
\end{tabular} & 
 \begin{tabular}{ccc}
  $\beta=$ 0.1 & $\beta=$ 0.5 & $\beta=$ 0.9 \\
  \includegraphics[width=2cm]{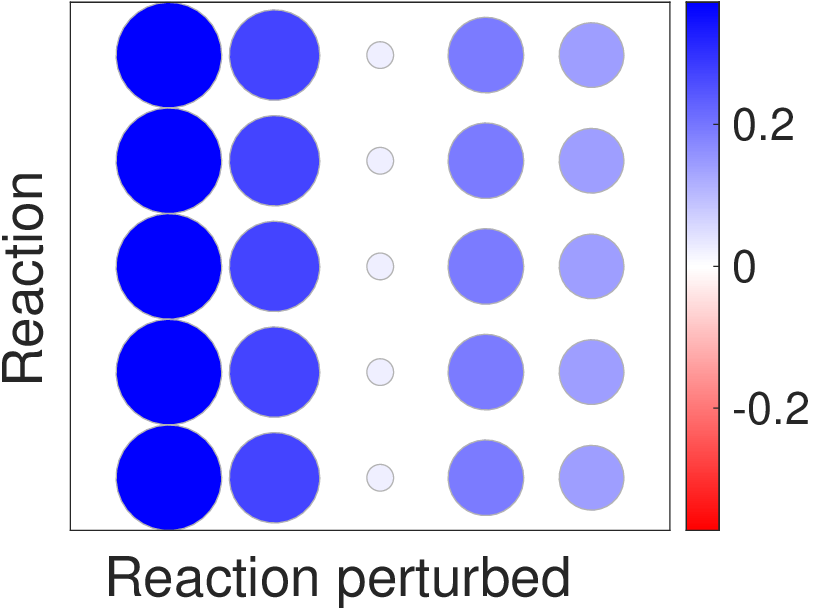}&
  \includegraphics[width=2cm]{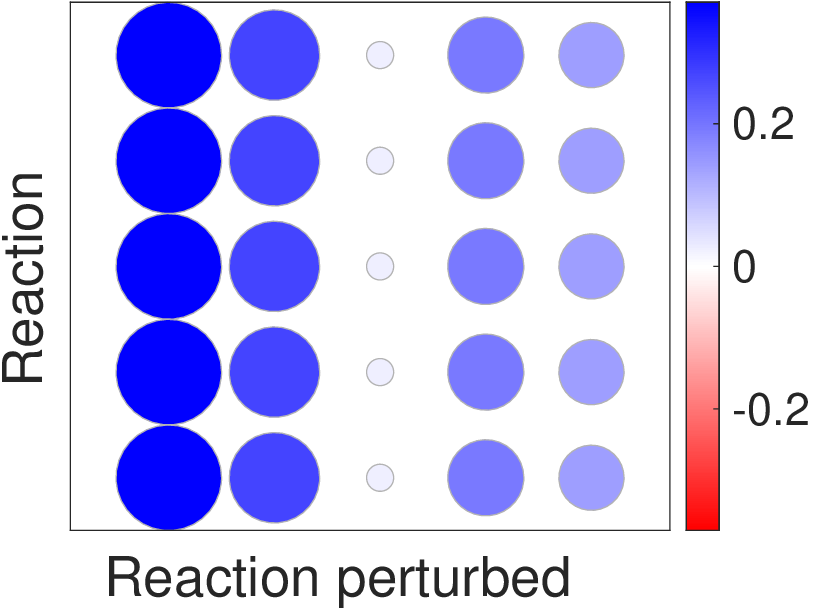}&
  \includegraphics[width=2cm]{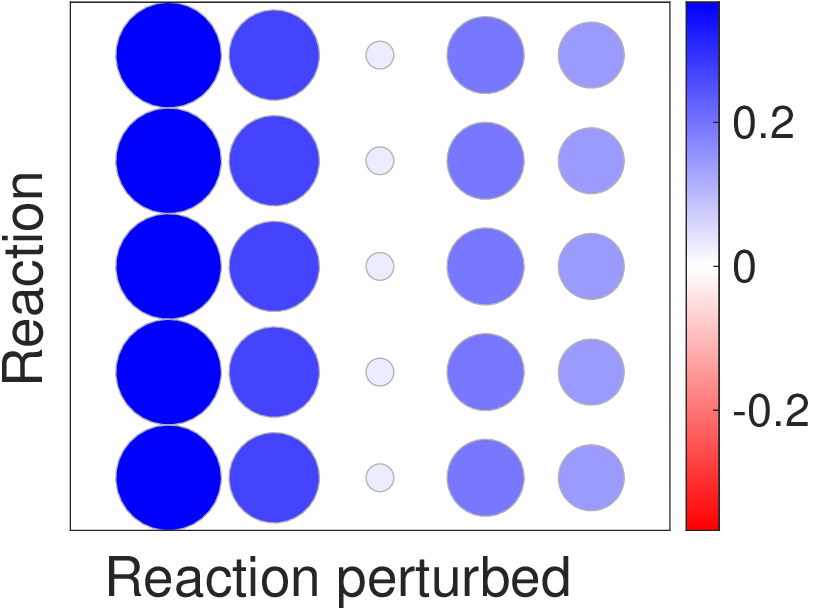}\\
  \includegraphics[width=2cm]{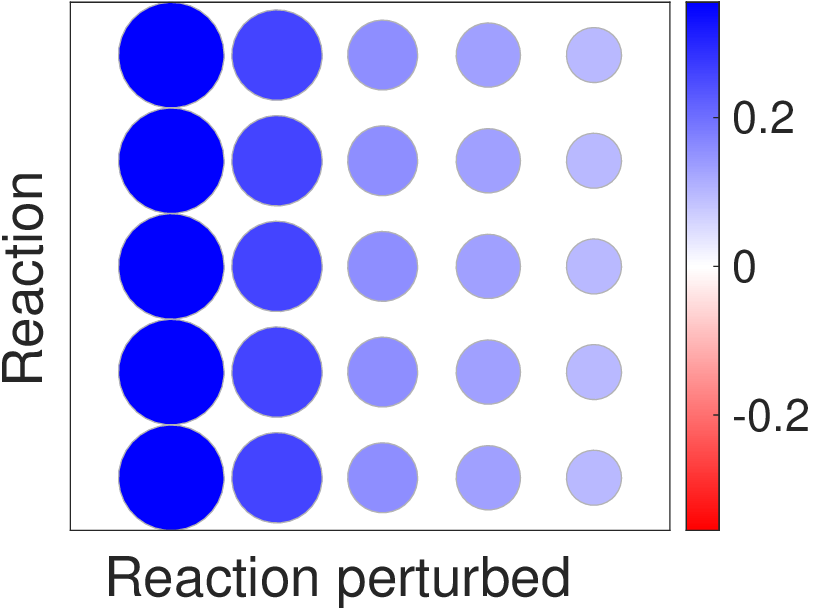}&
  \includegraphics[width=2cm]{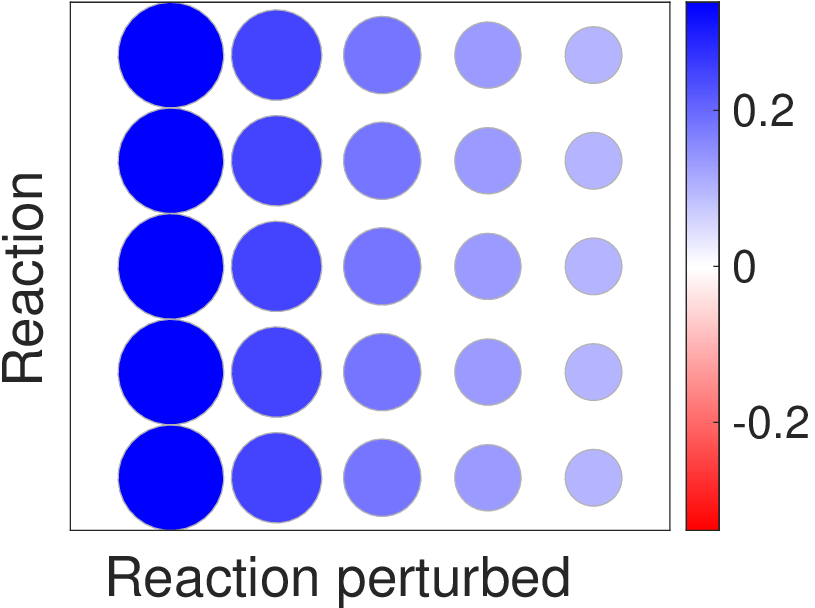}&
  \includegraphics[width=2cm]{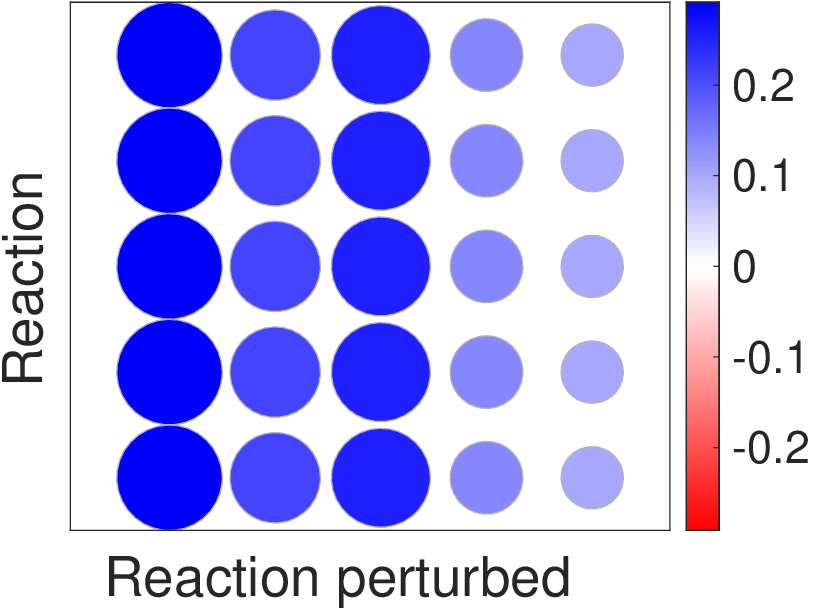}\\
  \includegraphics[width=2cm]{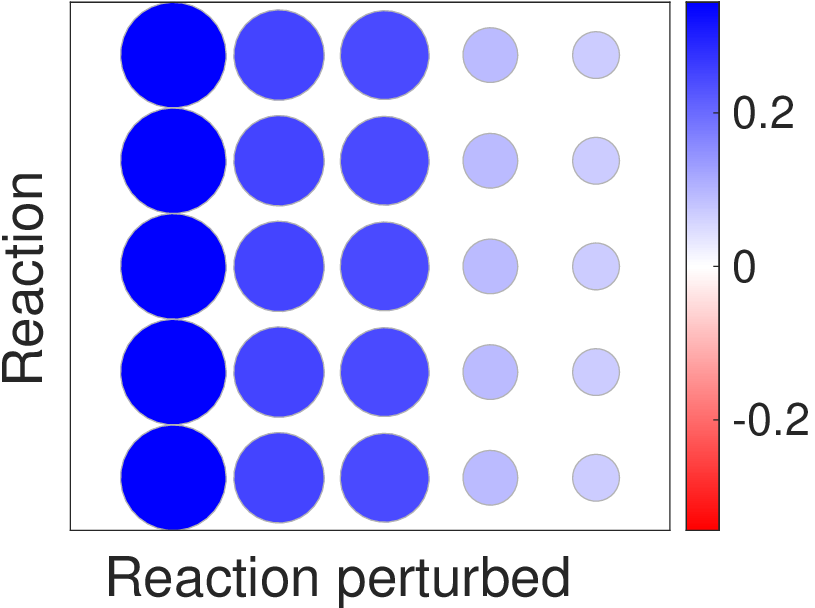}&
  \includegraphics[width=2cm]{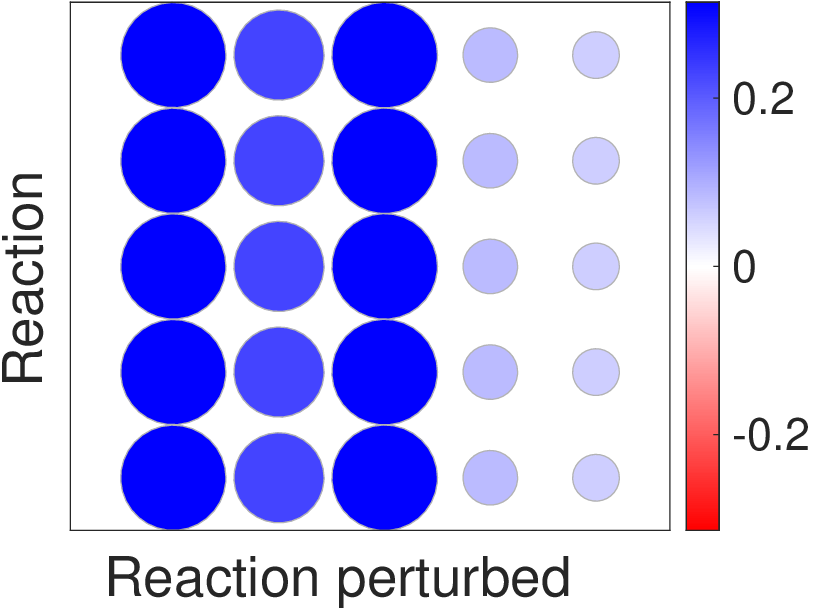}&
  \includegraphics[width=2cm]{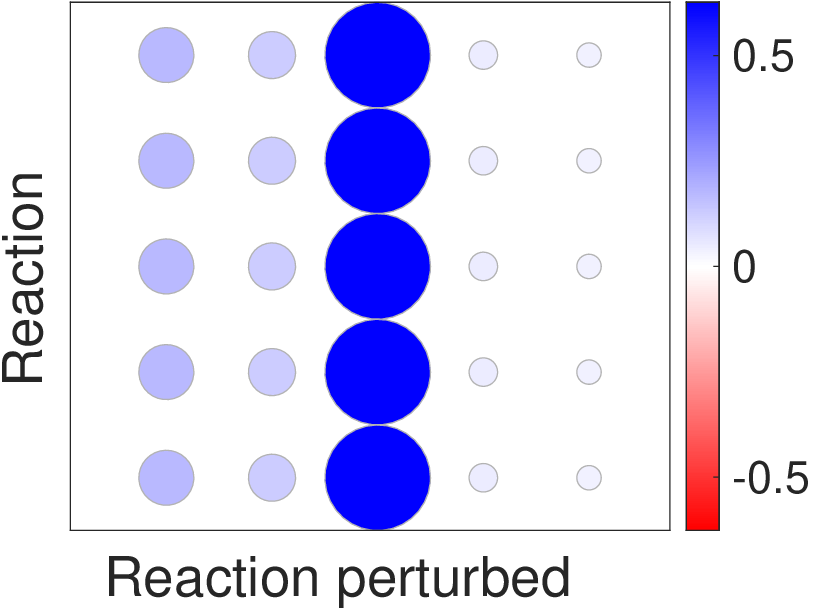}
\end{tabular}
\end{tabular}
\caption{\textbf{Thermodynamic forces in a linear chain and their
    effects on metabolic control.}  Variants of a pathway model with
  different thermodynamic forces (0.1 RT, RT, 10 RT) and saturation
  values (0.1, 0.5, 0.9) in the central reaction. (a) The
  thermodynamic force of the central reaction was made smaller, equal
  or larger than in the other reactions by varying the chemical
  potentials (left; white: high; red: low).  (b) Static control
  coefficients (for concentrations and fluxes) obtained from different
  model variants.
}
\label{fig:linchain1a}
\end{figure}

\myparagraph{\ \\Effect of varying thermodynamic forces} Thermodynamic
forces and saturation values have effects on metabolic control and
fluctuations. Figure \ref{fig:linchain1a} shows these effects for a
simple model, a pathway of five uni-uni reactions without
regulation. In the standard model version, all thermodynamic forces
are set to RT and all saturation values are set to 1/2. To assess the
effects of parameter changes, the parameters in the third reaction
were systematically varied, setting the force to values of 0.1 RT, RT,
and 10 RT, while keeping the metabolite concentrations close to their
original values (see section
\ref{sec:SIvaryingTheAffinities}). Similarly, the (substrate and
product) saturation values were set to values of 0.1, 0.5, and
0.9. The figure shows static control coefficients and spectral power
densities of chemical noise (for metabolite levels and fluxes). There
are some clear patterns: if the third reaction is close to equilibrium
(0.1 RT), it has little control on concentrations and fluxes, and also
little influence on the control exerted by other reactions. In
contrast, if the reaction is strongly driven, it exerts a larger
control, while the control exerted by downstream enzymes, and all
control on downstream metabolites decreases. If the reaction is close
to equilibrium, the substrate saturation does not play a role; as the
reaction is driven strongly, the substrate saturation additionally
increases the control exerted by the reaction. Thus, for a high flux
control, reaction must be strongly driven and the enzyme must be
saturated with substrate.

\section{Mathematical derivations}

\subsection{Probability densities for saturation values and dissociation constants}
\label{sec:SIprobabilitydistributions}

The fraction of enzyme molecules (of one sort of enzyme) bound by a
certain metabolite depends on the metabolite concentration $c$ and on
the dissociation constant $k$  (indices
omitted for simplicity). It can be described by saturation values
$\alpha = \frac{1}{1+c/k} = \frac{k}{k+c}$ or
$\beta = \frac{c}{k+c} = 1-\alpha$. The saturation values are directly
related to $c$ and $k$. If $\alpha$ is uniformly distributed in the
interval $]0,1[$, the conditional probability densities read

\begin{eqnarray}
 \label{eq:deriveddistributions}
 p(c) &=& \frac{k}{(k+c)^{2}}\qquad \mbox{for fixed}\, k \\
 p(k) &=& \frac{c}{(k+c)^{2}}\qquad \mbox{for fixed}\, c.
\end{eqnarray}
Similar formulae hold for the sampling of $\beta$.  According to this
formula, $\ln (c/\kM)$ follows a logistic distribution with location
parameter 0 and scale parameter 1. Mean, median and mode of
$\ln (c/\kM)$ are given by 0, and the variance is given by $\pi^{2}/3$
  
\textbf{Proof:} The probability density of $\alpha$ is $p(\alpha)=1$.
For fixed parameter $k$, we obtain $\partial \alpha/\partial c = -k/(c+k)^{2}$
and thus
\begin{eqnarray}
 \label{eq:dd1}
 p(c) &=& C\,p(\alpha) \, \left\vert \frac{\partial \alpha}{\partial c} \right\vert
 = C \frac{k}{(k+c)^{2}}.
\end{eqnarray}
The normalisation constant $C$ has a value of 1 because
\begin{eqnarray}
 \label{eq:dd2}
 1/C = \int_{0}^{\infty} \frac{k}{(c+k)^{2}} \md c 
 =  \left. \frac{-k}{k+c} \right\vert_{0}^{\infty} = 1.
\end{eqnarray}
For fixed concentration $c$, we compute $\partial \alpha/\partial k = k/(c+k)^{2}$ and
obtain
\begin{eqnarray}
 \label{eq:dd3}
 p(k(\alpha)) &=& C\,p(\alpha) \, \left\vert \frac{\partial \alpha}{\partial k} \right\vert
 = C \frac{c}{(k+c)^{2}},
\end{eqnarray}
again with normalisation constant $C=1$ because
\begin{eqnarray}
  \label{eq:dd4}
 1/C = \int_{0}^{\infty} \frac{c}{(k+c)^{2}} \md k 
 = \left. \frac{k}{c+k} \right\vert_{0}^{\infty} = 1.
\end{eqnarray}

\subsection{Metabolic control and response coefficients}
\label{sec:SIMCAcalculation}

\myparagraph{\ \\Metabolic control and response coefficients (first
  order)} Control and response coefficients describe the effects of
small parameter changes on state variables (metabolite concentrations
$c_{i}$ and reaction rates $v_{l}$) in a first- or second-order
approximation \cite{hesc:96,hofm:01}. The unscaled response and
control coefficients can be computed from the unscaled elasticities
and the stoichiometric matrix \cite{lieb:2005}. In systems without
conservation relations, the first-order unscaled control matrices read
\begin{eqnarray}
\label{eq:conccontrol}
 \CSun &=& - (\Nint \Ecun)\inv \Nint \\
 \CJun &=& \Imat + \Ecun\, \CSun.
\end{eqnarray}
In models with linear conservation relations, the Jacobian
$\Nint \Ecun$ is rank-deficient and not invertible, but we can still
compute the control coefficients \cite{rede:88}: we select a set of
independent internal metabolites (such that the reduced matrix $\Nr$
has full row rank, and the same rank as $\Nint$). Then the
stoichiometric matrix can be split into a matrix product
$\Nint = \Lmat\, \Nr$, and Eq.~(\ref{eq:conccontrol}) is replaced by
\begin{eqnarray}
 \CSun = - \Lmat (\Nr \Ecun \Lmat)\inv \Nr.
\end{eqnarray}

The response matrices with respect to system parameters
$p_{m}$ read
\begin{eqnarray}
 \RSunp = \frac{\partial \sv}{\partial \pv} = \CSun\, \Epun, \qquad
 \RJunp = \frac{\partial j}{\partial p} = \CJun\, \Epun. 
\end{eqnarray}
Scaled control and response matrices,
e.g.~$\Rsc^{c_{i}}_{u} = \partial \ln c_{i}/\partial \ln u$, are defined
 similar to the scaled elasticities. Since the
enzyme concentrations appear as prefactors in the rate laws, it turns out that scaled response
coefficients and scaled control coefficients are identical. With
other perturbation parameters (e.g.~external metabolite
concentrations), this will not be the case.

\myparagraph{Metabolic synergy  coefficients (second-order response coefficients)}
\label{sec:SIsecondOrderResponse}
For general perturbation parameters $u_{p}$ and $u_{q}$ (i.e.~not
necessarily enzyme levels), we can write the unscaled synergy tensors \cite{hohe:93} as
\begin{eqnarray}
\label{eq:responsetensors}
 \Run^{c_{i}}_{u_{p} u_{q}} &=& \sum\limits_{k} \Cun^{c_{i}}_{v_{l}}\, \Gamma^{v_{l}}_{u_{p}u_{q}},\qquad
 \Run^{v_{j}}_{u_{p} u_{q}} = \sum\limits_{k} \Cun^{v_{j}}_{v_{l}}\, \Gamma^{v_{l}}_{u_{p}u_{q}}
\end{eqnarray}
with a tensor $\Gamma$ defined as 
\begin{eqnarray}
 \Gamma^{v_{l}}_{u_{p}u_{q}} &=& 
 \sum\limits_{ij} \Euns^{v_{l}}_{c_{i} c_{j}} 
 \Run^{c_{i}}_{u_{p}} \Run^{c_{j}}_{u_{q}} 
 + \sum\limits_{j} \Euns^{v_{l}}_{c_{j} u_{p}} \Run^{c_{j}}_{u_{q}} 
 + \sum\limits_{i} \Euns^{v_{l}}_{c_{i} u_{q}} \Run^{c_{i}}_{u_{p}}
 + \Euns^{v_{l}}_{u_{p} u_{q}}.
\end{eqnarray}
Enzyme levels are parameters with specific properties: there is only one
enzyme per reaction, and enzyme levels appear as prefactors in the rate
law. Thus, for two enzyme levels $e_{p}$ and $e_{q}$ we obtain the
unscaled response coefficients (see Eq.~(\ref{eq:calculateRuu1a}))
\begin{eqnarray}
 \Run^{y}_{e_{l}e_{j}} 
&=& \Cun^{y}_{v_{k}} \left[ 
 \Euns^{v_{k}}_{c_{q}c_{r}}\, \Cun^{c_{q}}_{v_{l}}\,\Cun^{c_{r}}_{v_{j}}\,v_{l}\,v_{j}
+ \delta_{kj}\,\Euns^{v_{k}}_{c_{q}}\, \Cun^{c_{q}}_{v_{l}}\,v_{l}
+ \delta_{kl}\,\Euns^{v_{k}}_{c_{r}}\, \Cun^{c_{r}}_{v_{j}}\,v_{j}
\right] \frac{1}{e_{l}\,e_{j}}.
\end{eqnarray}
The scaled synergy coefficients between a steady-state  variable $y$
and  enzyme levels $e_{l}$ and $e_{j}$ read (see
Eq.~(\ref{eq:calculateRuu2}))
\begin{eqnarray}
 \label{eq:calculateRuufinal}
  \Rsc^{y}_{e_{l}e_{j}} 
&=& \sum_{kqr} (\Csc^{y}_{v_{k}} \Escs^{v_{k}}_{c_{q}c_{r}}\, 
 \Csc^{c_{q}}_{v_{l}}\, \Csc^{c_{r}}_{v_{j}})
 + \sum_{q} (\Csc^{y}_{v_{j}} \,\Escs^{v_{j}}_{c_{q}}\, \Csc^{c_{q}}_{v_{l}}) \nonumber \\
 && + \sum_{r} (\Csc^{y}_{v_{l}} \,\Escs^{v_{l}}_{c_{r}}\, \Csc^{c_{r}}_{v_{j}})
- (\Csc^{y}_{v_{l}} \Csc^{y}_{v_{j}}) + (\delta_{lj}\, \Csc^{y}_{v_{l}}).
\end{eqnarray}
The synergy coefficients  describe second-order
effects of an enzyme (indices $l=j$) or synergies of enzyme pairs
(indices $l \ne j$). For enzyme pairs, the Kronecker $\delta_{lj}$ vanishes and we
can set $\Escs^{v_{l}}_{c_{i}}\, \Csc^{c_{i}}_{v_{j}} = \Csc^{v_{l}}_{v_{j}}$
(see Eq.~\ref{eq:conccontrol}). By rewriting the second-order
elasticities $\Escs^{v_{k}}_{c_{q}c_{r}}$ as in
Eq.~\ref{eq:secondorderelasticitysplitting} and rearranging
Eq.~(\ref{eq:calculateRuufinal}), we obtain
\begin{eqnarray}
 \label{eq:calculateRuufinal2}
  R^{y}_{e_{l}e_{j}} 
  &=& 
  \Csc^{y}_{v_{l}} \,\Csc^{v_{l}}_{v_{j}} + \Csc^{y}_{v_{j}} \,\Csc^{v_{j}}_{v_{l}}
  - \Csc^{y}_{v_{l}} \Csc^{y}_{v_{j}}
+ \sum_{kqr} \Csc^{y}_{v_{k}} \vartheta^{k}_{qr}\, \Csc^{v_{k}}_{v_{l}}\, \Csc^{v_{k}}_{v_{j}}.
\end{eqnarray}
For mass-action rate laws  close to equilibrium and without regulation, the term $\vartheta^{k}_{qr}$
is approximately -1. 

\subsection{Spectral power density and temporal variability due to chemical noise}
\label{sec:timescalevariation}

To describe metabolic fluctuations caused by chemical noise, we
use the chemical Langevin equation 
\begin{eqnarray}
 \frac{\md \xv}{\md t} = \Nint\,[\av^{+}-\av^{-}] + \Nint\,\left[\diag\left(\sqrt{\av^{+}}\right)\,{\boldsymbol \xi}^{+} - \diag\left(\sqrt{\av^{-}}\right)\,{\boldsymbol \xi}^{-}\right]
\end{eqnarray}
for particle numbers $x_{i}$,The reaction propensities $a^{\pm}_{l}$
denote the probabilities per time (in s$\inv$) for events of reaction
$l$ in forward (+) or backward (-) direction, and
${\boldsymbol \xi}^{+}$ and ${\boldsymbol \xi}^{-}$ are vectors of
standard Gaussian white noise\footnote{The white noise appears in the
 formula as the derivative of a standard Wiener process. It
 has the covariance function
 $\cov_{\xi}(\tau) = \langle \xi(t)\,\xi(t+\tau)\rangle_{t} =
 \delta(\tau)$
 (in s$^{-1}$) and a spectral power density
 $S_{\xi}(\omega) = \frac{1}{2 \pi}$ (unitless). Note the prefactor 
  convention used for Fourier transforms:
 $x(t) = \int_{-\infty}^{\infty} \tilde x(\omega) \,\e^{i\,\omega\,t}
 \md \omega$
 and
 $x(\omega) = \frac{1}{2 \pi} \int_{-\infty}^{\infty} x(t)
 \,\e^{-i\,\omega\,t} \md t$.}
(in units of s$^{-1/2}$). In a cell volume $\Omega$, molecule
numbers and propensities are related to concentrations and
reaction rates as
\begin{eqnarray}
 x_{i} = N_{\rm A}\,\Omega\,c_{i}, \qquad a^{\pm}_{l} = N_{\rm A}\,v^{\pm}_{l}
\end{eqnarray}
with Avogadro's constant $N_{\rm A} \approx 6\cdot 10^{23}$ mol$\inv$. 
using these variables, we can rewrite the chemical Langevin equation as 
\begin{eqnarray}
 \frac{\md \cv}{\md t} = \Nint\,\frac{1}{\Omega}\, \vv 
 + \Nint\,\frac{1}{\Omega}\,[
 \diag\left(\sqrt{\frac{\vv_{+}}{N_{\rm A}}}\right)\,{\boldsymbol \xi}^{+} -
 \diag\left(\sqrt{\frac{\vv_{-}}{N_{\rm A}}}\right)\,{\boldsymbol \xi}^{-}].
\end{eqnarray}
For deviations $\Delta c_{i}$ from a stationary state, and setting
${\boldsymbol \xi} = { {\boldsymbol \xi}^{+} \choose {\boldsymbol
  \xi}^{-}}$, we can approximate it by
\begin{eqnarray}
\label{eq:chemlanglinear}
 \frac{\md \Delta \cv}{\md t} = 
  \Nint\, \Euns^{v}_{c}\,\Delta \cv
 + \Nint\, \Euns^{v}_{\xi} \, {\boldsymbol \xi}
\end{eqnarray}
with the unscaled elasticity matrices
\begin{eqnarray}
 \Euns^{v}_{c} = \frac{1}{\Omega}\, \frac{\partial \vv}{\partial \cv}, \quad
 \Euns^{v}_{\xi} = \frac{1}{\Omega\,\sqrt{N_{\rm A}}}\, \left(\diag\left(\sqrt{\vv_{+}}\right),\, -\diag\left(\sqrt{\vv^{-}}\right) \right)
\end{eqnarray}
in units of s$\inv$ and mM s$^{-1/2}$, respectively.
Eq.~(\ref{eq:chemlanglinear}) has the form of a standard linear model
with perturbation parameters in a vector ${\boldsymbol \xi}$. For this
model, we can compute the frequency-response matrices (see
\cite{lieb:2005})
\begin{eqnarray}
 \RS_{\xi}(\omega) &=& - \Lmat\,(\Nr\,\Euns^{v}_{c}\Lmat- i\,\omega \Imat)\inv\,\Nr \,\Euns^{v}_{\xi} \nonumber \\
 \Run^{j}_{\xi}(\omega) &=& \Omega \left[ \Euns^{v}_{\xi} + \Euns^{v}_{c}\, \RS_{\xi}(\omega) \right]
\end{eqnarray}
in units of mM\,s$^{1/2}$ and mol s$^{-1/2}$, respectively. The
concentration fluctuations have the spectral power density matrices
\begin{eqnarray}
 S_{c}(\omega) = \RS_{\xi}(\omega)\, S_{c}(\omega)\, \RSdag_{\xi}(\omega)
        = \frac{1}{2\,\pi}\,\RS_{\xi}(\omega)\,\RSdag_{\xi}(\omega)
\end{eqnarray}
(in mM$^{2}$ s). An analogous formula holds for flux fluctuations (in
mol$^{2}$ s$^{-1}$). To study fluctuations on a specific time scale
$\sigma$ (in units of seconds), we consider the fluctuating
concentration curve and smoothen it by convolving it with a
(normalised) Gaussian function of width $\sigma$. The resulting
function has the spectral power density
\begin{eqnarray}
 S^{(\sigma)}_{c}(\omega) = (\e^{-\frac{1}{2}\omega^{2}\,\sigma^{2}})^{2}\,S_{c}(\omega).
\end{eqnarray}
 The function in brackets is the Fourier transform of our
Gaussian function. The covariance function of the smoothed curve is
given by the reverse Fourier transform of the spectral power density
\begin{eqnarray}
 \cov^{(\sigma)}_c(\tau) = \int_{-\infty}^{\infty} S^{(\sigma)}_{c}(\omega)\, \e^{i\, \omega\,\tau}\, \md \omega.
\end{eqnarray}
 The variance (the covariance function for time shift $\tau=0$) is therefore given by 
\begin{eqnarray}
 \cov^{(\sigma)}_c
 = \int_{-\infty}^{\infty}  S^{(\sigma)}_{c}(\omega)\, \md \omega
 = \frac{1}{2\,\pi}\, \int_{-\infty}^{\infty}
\e^{-\omega^{2}\,\sigma^{2}}\, 
\RS_{\xi}(\omega)\,\RSdag_{\xi}(\omega)\, \md \omega
\end{eqnarray}
An analogous formula holds for the covariance of flux fluctuations.

\subsection{Synergy effects}
\label{sec:SIproofinterference}

Here we derive the Eq.~(\ref{eq:apprepistasis2} for synergy effects.
The effect of a double enzyme perturbation on an target variable $y$
can be seen as the sum of three terms: the single-inhibition effects
plus a synergy effect $\etaun_{\rm ab}$:
\begin{eqnarray}
 \Delta y^{\rm ab} = \Delta y_a + \Delta y_{b} + \etaun_{\rm ab}.
\end{eqnarray}
The synergy effect $\etaun_{\rm ab}$, defined as the difference
$\etaun_{\rm ab} = \Delta y^{\rm ab} - \Delta y_a - \Delta y_{b}$, can
be approximately determined from the metabolic response
coefficients. Given a vector $\Delta \ev$ of enzyme changes, a
second-order Taylor expansion yields
\begin{eqnarray}
\label{eq:proofepi1}
 y(\ev+\Delta \ev) &\approx& y(\ev) + \Rmatun^{y}_{\ev} \Delta \ev 
+ \frac{1}{2} \Delta \ev\trans \Rmatun^{y}_{uu} \Delta \ev.
\end{eqnarray}
Therefore, if two enzyme concentrations $e_{a}$ and $e_{b}$ are
decreased by $\Delta e_{a}$ and $\Delta e_{b}$, 
the target changes by
\begin{eqnarray}
\label{eq:proofepi2}
 \Delta y^{\rm ab} \approx - \Run^{y}_{e_{a}} \Delta e_{a} 
 - \Run^{y}_{e_{b}} \Delta e_{b}
 + \frac{1}{2} \Run^{y}_{e_{a}e_{a}} \Delta e_{a}^{2}
 + \Run^{y}_{e_{a}e_{b}} \Delta e_{a} \Delta e_{b}
 + \frac{1}{2} \Run^{y}_{e_{b}e_{b}} \Delta e_{b}^{2}.
\end{eqnarray}
 The single perturbations yield 
\begin{eqnarray}
 \Delta y_a &\approx& - \Run^{y}_{e_{a}} \Delta e_{a} 
 + \frac{1}{2} \Run^{y}_{e_{a}e_{a}} \Delta e_{a}^{2} \nonumber \\
 \Delta y_{b} &\approx& - \Run^{y}_{e_{b}} \Delta e_{b} 
 + \frac{1}{2} \Run^{y}_{e_{b}e_{b}} \Delta e_{b}^{2}.
\end{eqnarray}
With the second-order approximation Eqs~(\ref{eq:proofepi1}) and
(\ref{eq:proofepi2}), the synergy effect is given by the unscaled
synergy coefficient $\Run^{y}_{e_{a}e_{b}}$ multiplied by the
perturbations:
\begin{eqnarray}
\etaun_{\rm ab} \approx \Delta y^{\rm ab} - \Delta y_a - \Delta y_{b} &\approx& \Run^{y}_{e_{a}e_{b}} \Delta e_{a} \Delta e_{b}.
\end{eqnarray}
If all quantities (enzyme levels and target variable) are measured on
logarithmic scale, it is natural to consider the splitting 
\begin{eqnarray}
 \Delta \log y^{\rm ab} = \Delta \log y_a + \Delta \log y_{b} + \etasc_{\rm ab}.
\end{eqnarray}
It corresponds to a ``null hypothesis'' of multiplicative effects,
contains the scaled synergy $\etasc_{\rm ab}$, and in the corresponding
formulae the scaled response coefficients are used.

\subsection{Response and synergy coefficients}
\label{sec:SIderivationRuu}

\myparagraph{\ \\Synergy cofficients for stationary concentrations and
  fluxes} here we derive Eq.~(\ref{eq:calculateRuufinal} for scaled
synergy coefficients.  The unscaled synergy cofficients (second-order
response coefficients) between general parameters $u_{l}$ and $u_{j}$
and state variables $y$ (stationary concentrations and fluxes) read
(as in Eq.~(\ref{eq:responsetensors}) and using Einstein's sum
convention):
\begin{eqnarray}
 \label{eq:calculateRuu}
 \Run^{y}_{u_{l}u_{j}} = \Cun^{y}_{v_{k}} \Gamma^{v_{k}}_{u_{l}u_{j}}
= \Cun^{y}_{v_{k}} \left[ 
 \Euns^{v_{k}}_{c_{q}c_{r}}\, \Run^{c_{q}}_{u_{l}}\, \Run^{c_{r}}_{u_{j}}
+ \Euns^{v_{k}}_{c_{q}u_{j}}\, \Run^{c_{q}}_{u_{l}}
+ \Euns^{v_{k}}_{u_{l}c_{r}}\, \Run^{c_{r}}_{u_{j}}
+ \Euns^{v_{k}}_{u_{l}u_{j}}
\right].
\end{eqnarray}
If the perturbation parameters are enzyme levels $e_{l}$ and $e_{j}$,
we can write the elasticities as
$\Euns^{v_{k}}_{c_{q}e_{j}} = \delta_{kj}\,\frac{1}{e_{j}}\Euns^{v_{k}}_{c_{q}}$,
$\Euns^{v_{k}}_{c_{r}e_{l}} = \delta_{kl}\,\frac{1}{e_{l}}\Euns^{v_{k}}_{c_{r}}$,
$\Euns^{v_{k}}_{e_{l}e_{j}}=0$, and set
$\Run^{c_{i}}_{e_{l}} = \Cun^{c_{i}}_{v_{l}} \frac{v_{l}}{e_{l}}$.
By inserting this into Eq.~(\ref{eq:calculateRuu}), we obtain
\begin{eqnarray}
 \label{eq:calculateRuu1a}
 \Run^{y}_{e_{l}e_{j}} 
&=& \Cun^{y}_{v_{k}} \left[ 
 \Euns^{v_{k}}_{c_{q}c_{r}}\, \Run^{c_{q}}_{e_{l}}\, \Run^{c_{r}}_{e_{j}}
+ \delta_{kj}\,\frac{1}{e_{j}}\Euns^{v_{k}}_{c_{q}}\, \Run^{c_{q}}_{e_{l}}
+ \delta_{kl}\,\frac{1}{e_{l}}\Euns^{v_{k}}_{c_{r}}\, \Run^{c_{r}}_{e_{j}}
\right] \nonumber \\
&=& \Cun^{y}_{v_{k}} \left[ 
 \Euns^{v_{k}}_{c_{q}c_{r}}\, \Cun^{c_{q}}_{v_{l}}\,\Cun^{c_{r}}_{v_{j}}\,v_{l}\,v_{j}
+ \delta_{kj}\,\Euns^{v_{k}}_{c_{q}}\, \Cun^{c_{q}}_{v_{l}}\,v_{l}
+ \delta_{kl}\,\Euns^{v_{k}}_{c_{r}}\, \Cun^{c_{r}}_{v_{j}}\,v_{j}
\right] \frac{1}{e_{l}\,e_{j}},
\end{eqnarray}
which is equivalent to the formula given in \cite{hohe:93}. The
scaled synergy coefficients read, in analogy to
Eq.~(\ref{eq:elasticityscaling}),
\begin{eqnarray}
 \label{eq:calculateRuu2}
  \Rsc^{y}_{e_{l}e_{j}} 
&=&  \frac{e_{l}\,e_{j}}{y} \Run^{y}_{e_{l}e_{j}}
  - \frac{e_{l}\,e_{j}}{y^{2}} \Run^{y}_{e_{l}} \Run^{y}_{e_{j}}
  + \delta_{lj} \frac{e_{l}}{y} \Run^{y}_{e_{l}} \nonumber \\
&=& \frac{1}{y} \Cun^{y}_{v_{k}} \left[ 
 \Euns^{v_{k}}_{c_{q}c_{r}}\, \Cun^{c_{q}}_{v_{l}}\, \Cun^{c_{r}}_{v_{j}}\,v_{l}\,v_{j}
 + \delta_{kj} \Euns^{v_{k}}_{c_{q}}\, \Cun^{c_{q}}_{v_{l}}\,v_{l}
 + \delta_{kl} \Euns^{v_{k}}_{c_{r}}\, \Cun^{c_{r}}_{v_{j}}\,v_{j} \right]
- \frac{1}{y^{2}} \Cun^{y}_{v_{l}} \Cun^{y}_{v_{j}} v_{l} \,v_{j}
+ \delta_{lj} \frac{1}{y} \Cun^{y}_{v_{l}} v_{l}
\end{eqnarray}
They can be written -- again with sum symbols -- as
\begin{eqnarray}
 \label{eq:calculateRuu3}
  \Rsc^{y}_{e_{l}e_{j}} 
&=& \sum_{krq} (\Csc^{y}_{v_{k}} \Escs^{v_{k}}_{c_{q}c_{r}}\, \Csc^{c_{q}}_{v_{l}}\, \Csc^{c_{r}}_{v_{j}})
 + \sum_{q} (\Csc^{y}_{v_{j}} \,\Escs^{v_{j}}_{c_{q}}\, \Csc^{c_{q}}_{v_{l}})
 + \sum_{r} (\Csc^{y}_{v_{l}} \,\Escs^{v_{l}}_{c_{r}}\, \Csc^{c_{r}}_{v_{j}})
- (\Csc^{y}_{v_{l}} \Csc^{y}_{v_{j}}) + (\delta_{lj}\, \Csc^{y}_{v_{l}}).
\end{eqnarray}

\myparagraph{Response coefficients for general state variables} The
previous formulae holds for targets $y$ that are stationary
concentrations \coout{symbols??} $c_{i}$ or fluxes $v_{l}$. How can we
generalise them to other target variables $z(\cv, \vv)$, which are
functions of these state variables $y_{p}$? \coout{WO:?} Let the unscaled
derivatives be called $\zun_{y_{p}}$ and $\zun_{y_{p}y_{q}}$. We
shall first first compute the unscaled, and then the scaled response
coefficients of $z$. The unscaled response coefficients read (with sum
convention)
\begin{eqnarray}
\label{eq:schubidubiduhh}
 \Run^{z}_{e_{l}} = \frac{\partial z}{\partial e_{l}} 
=
 \frac{\partial z}{\partial y_{p}}\, \frac{\partial y_{p}}{\partial e_{l}}
=
 \zun_{y_{p}}\, \Run^{y_{p}}_{e_{l}}.
\end{eqnarray}
For the next step, we introduce the scaled
derivatives $\hat z_{y_{p}} = \frac{\partial \ln y}{\partial \ln |y_{p}|}$ and
$\hat z_{y_{p}y_{q}} = \frac{\partial^{2} \ln y}{\partial \ln |y_{p}|
\partial \ln |y_{q}|}$. With their help, we write the unscaled derivatives as 
\begin{eqnarray}
 \label{eq:calculatezyy}
\zun_{y_{p}} &=& \frac{z}{y_{p}}\,\hat z_{y_{p}} \nonumber \\
\zun_{y_{p}y_{q}} &=& 
\frac{z}{y_{p}\,y_{q}}\, \left[
\hat z_{y_{p}y_{q}} + \hat z_{y_{p}}\,\hat z_{y_{q}} - \delta_{pq} \hat z_{y_{p}} \right].
\end{eqnarray}

The  unscaled synergy coefficients for our target $z$ read
\begin{eqnarray}
 \Run^{z}_{e_{l}e_{j}} &=& \frac{\partial^{2} z}{\partial e_{_{l}} \partial e_{j}}
 = \frac{\partial}{\partial e_{j}}  \Run^{z}_{e_{l}}
= \frac{\partial}{\partial e_{j}} \left[
 \frac{\partial z}{\partial y_{p}}\, \Run^{y_{p}}_{e_{l}} 
\right] 
= 
\left[\frac{\partial}{\partial e_{j}} \frac{\partial z}{\partial y_{p}} \right]
 \, \Run^{y_{p}}_{e_{l}}
+ \zun_{y_{p}}\, \Run^{y_{p}}_{e_{l}e_{j}} \nonumber \\
&=&
\left[\frac{\partial^{2} z}{\partial y_{p}y_{q}} \frac{\partial y_{q}}{\partial e_{j}} 
 \right]
 \, \Run^{y_{p}}_{e_{l}}
+ \zun_{y_{p}}\, \Run^{y_{p}}_{e_{l}e_{j}}
=
\left[\zun_{y_{p}y_{q}}\, \Run^{y_{p}}_{e_{j}} 
 \right]\, \Run^{y_{p}}_{e_{l}}
+ \zun_{y_{p}}\, \Run^{y_{p}}_{e_{l}e_{j}}
 \nonumber\\
&=&
\zun_{y_{p}y_{q}}\, \Run^{y_{p}}_{e_{l}} \, \Run^{y_{q}}_{e_{j}} 
+ \zun_{y_{p}}\, \Run^{y_{p}}_{e_{l}e_{j}}.
\end{eqnarray}
Let us now consider the scaled response coefficients. The first-order
scaled response coefficients read
\begin{eqnarray}
 \Rsc^{z}_{e_{l}} = \frac{e_{l}}{z}\, \Run^{z}_{e_{l}}
 = \frac{e_{l}}{z}\,  \zun_{y_{p}}\, \Run^{y_{p}}_{e_{l}} 
 = \hat z_{y_{p}}\, \Rsc^{y_{p}}_{e_{l}}.
\end{eqnarray}
The  scaled synergy coefficients read (compare Eq.~\ref{eq:schubidubiduhh})
\begin{eqnarray}
 \label{eq:calculateRuu2a}
 \Rsc^{z}_{e_{l}e_{j}} 
 &=&  \frac{e_{l}\,e_{j}}{z} \Run^{z}_{e_{l}e_{j}}
 - \frac{e_{l}\,e_{j}}{z^{2}} \Run^{z}_{e_{l}} \Run^{z}_{e_{j}}
 + \delta_{lj} \frac{e_{l}}{z} \Run^{z}_{e_{l}} \nonumber \\
 &=&  \frac{e_{l}\,e_{j}}{z} 
 \left[
  \zun_{y_{p}y_{q}}\, \Run^{y_{p}}_{e_{l}} \, \Run^{y_{q}}_{e_{j}} 
  + \zun_{y_{p}}\, \Run^{y_{p}}_{e_{l}e_{j}} \right]  
 - R^{z}_{e_{l}} R^{z}_{e_{j}} + \delta_{lj} R^{z}_{e_{l}} \nonumber \\
 &=& 
\left[
\hat z_{y_{p}y_{q}} + \hat z_{y_{p}}\,\hat z_{y_{q}} - \delta_{pq} \hat z_{y_{p}} \right]
\, R^{y_{p}}_{e_{l}} \, R^{y_{q}}_{e_{j}} 
 + \hat z_{y_{p}}\, R^{y_{p}}_{e_{l}e_{j}}
 -\hat z_{y_{p}}\, R^{y_{p}}_{e_{l}} \,
 \hat z_{y_{q}}\, R^{y_{q}}_{e_{j}} 
 + \delta_{lj} \hat z_{y_{p}}\, R^{y_{p}}_{e_{l}}
 \nonumber \\
 &=& 
 \hat z_{y_{p}y_{q}}\, R^{y_{p}}_{e_{l}} \, R^{y_{q}}_{e_{j}} 
 + \hat z_{y_{p}}\, R^{y_{p}}_{e_{l}e_{j}}
 - \delta_{pq} \hat z_{y_{p}}
\, R^{y_{p}}_{e_{l}} \, R^{y_{q}}_{e_{j}} 
 + \delta_{lj} \hat z_{y_{p}}\, R^{y_{p}}_{e_{l}}
\end{eqnarray}
For functions $z(\cv,\vv)$ that are multiplicative in fluxes and
concentrations (i.e.~linear if all quantities are given on logarithmic
scale), the first term vanishes.

\end{appendix}

\end{document}